\documentclass{article}

\usepackage{arxiv}

\usepackage{algorithm}
\usepackage{algpseudocode}
\def\bSig\mathbf{\Sigma}

\usepackage{amsmath,amssymb}
\DeclareMathOperator{\E}{\mathbb{E}}
\usepackage{natbib}
\newcommand{\indep}{\perp \!\!\! \perp}
\usepackage{bbm}
\usepackage{graphicx}
\usepackage{rotating}
\usepackage{url}
\usepackage{verbatim}
\usepackage{hyperref}
\usepackage{rotating} 
\usepackage{caption}

\usepackage{prodint}
\usepackage{xcolor}
\usepackage{paralist}
\usepackage{amsthm}

\title{Targeted learning of heterogeneous treatment effect curves for right censored or left truncated time-to-event data}

\author{Matthew Pryce$^{1}$, Karla Diaz-Ordaz$^{1,*}$, Ruth H. Keogh$^{2}$, Stijn Vansteelandt$^{3}$ \vspace{0.5cm}\\
$^{1}$Department of Statistical Science, University College London, London, United Kingdom \\
$^{2}$Department of Medical Statistics, London School of Hygiene \& Tropical Medicine, London, WC1E 7HT, United Kingdom\\
$^{3}$Department of Mathematics, Computer Science, and Statistics, 
Ghent University, \\Ghent, Belgium \vspace{0.5cm}\\
*Correspondence address: karla.diaz-ordaz@ucl.ac.uk}

\begin{document}
\maketitle
\begin{abstract}
In recent years, there has been growing interest in causal machine learning estimators for quantifying subject-specific effects of a binary treatment on time-to-event outcomes. Estimation approaches have been proposed which attenuate the inherent regularisation bias in machine learning predictions, with each of these estimators addressing measured confounding, right censoring, and in some cases, left truncation. However, the existing approaches are found to exhibit suboptimal finite-sample performance, with none of the existing estimators fully leveraging the temporal structure of the data, yielding non-smooth treatment effects over time. We address these limitations by introducing surv-iTMLE, a targeted learning procedure for estimating the difference in the conditional survival probabilities under two treatments. Unlike existing estimators, surv-iTMLE accommodates both left truncation and right censoring while enforcing smoothness and boundedness of the estimated treatment effect curve over time. Through extensive simulation studies under both right censoring and left truncation scenarios, we demonstrate that surv-iTMLE outperforms existing methods in terms of bias and smoothness of time-varying effect estimates in finite samples. We then illustrate surv-iTMLE’s practical utility by exploring heterogeneity in the effects of immunotherapy on survival among non-small cell lung cancer (NSCLC) patients, revealing clinically meaningful temporal patterns that existing estimators may obscure.
\end{abstract}



\section{Introduction}

Understanding treatment effect heterogeneity has become a vital component in advancing personalized medicine, promoting health equity, and informing policy decisions. Treatment effect heterogeneity is often studied using observational data, yet when outcomes are time-to-event, estimating these effects becomes challenging. Beyond the confounding inherent in observational data, one is typically faced with right censoring (e.g., when individuals are lost to follow-up prior to having an event) and sometimes left truncation (e.g., when individuals enter the study post treatment initiation). We explore how these factors impact the identification and estimation of the difference in conditional survival probabilities under two treatments for a given patient.

The work we present is motivated by a real-world application, evaluating immunotherapy effectiveness on mortality in non–small cell lung cancer (NSCLC) patients. The efficacy of immunotherapies is known to depend on patient-level factors, such as the presence of genomic mutations in the tumour, and we explore this using the Flatiron–Foundation Medicine clinico-genomic database, described in Section \ref{Motiv_ex}. As the clinico-genomic database offers a rich array of demographic, clinical, and genomic covariates, we focus on how flexible, machine learning methods can be employed to investigate heterogeneity in treatment effects across patient characteristics. However, as patients' event times are subject to left truncation, due to patients only joining the clinico-genomic database after genomic testing, and right censoring, as not all patients die before their end of follow-up, these complexities must be addressed when using this data.

A tempting strategy when utilizing machine learning to estimate treatment effect heterogeneity is to apply off-the-shelf survival machine learning (ML) algorithms, such as survival random forests, to estimate the conditional survival probability separately in each treatment group as a function of covariates (including pre-treatment confounders and predictors of censoring/truncation). The treatment effect as a function of covariates could then be obtained by taking their difference from the separate estimates. However, such indirect approaches inherit much of the regularisation bias that affects the separate learners --- see \cite{Morz} for a general review of this problem. Alternatively, some estimators directly target the contrasts between conditional survival probabilities under two treatments in such a way that this regularisation bias is attenuated. A notable example is causal survival forests (CSFs) \citep{RN9}, designed to handle right censored (RC) data, and more recently, the so-called ltrc-DR and ltrc-R learners \citep{RN3}, which handle left truncated and right censored (LTRC) data. However, each of these estimators have limitations. CSFs restrict the estimation techniques to tree ensembles, while the ltrc learners inherit issues from their loss functions, with the ltrc-DR learner prone to producing unbounded estimates of the survival contrasts, and the ltrc-R learner requiring the contrasts to be estimated conditional on all of the variables used to adjust for confounding. Additionally, these estimators do not fully exploit the temporal structure of the data, requiring the survival contrasts be estimated separately at each time point of interest, which is inefficient and can result in non-smooth and biologically implausible treatment effect curves over time.


To overcome these limitations, we introduce surv-iTMLE, a non-parametric targeted learning framework that accommodates RC or LTRC data. This framework leverages sieve-based targeted learning, known as infinite-dimensional targeted minimum loss-based estimation (iTMLE) \citep{RN68, RN4, RN15} within a two-step pseudo-outcome construction, allowing it to yield estimates of the difference in conditional survival probabilities under two treatments for a given patient that are bounded, stable and smooth over time. We explore these properties using extensive simulations in Section \ref{Sims}, and implement surv-iTMLE and its comparators on our motivating example in Section \ref{NSCLC_example}.

\section{Motivating example} \label{Motiv_ex}

\subsection{Background}
Our motivating example focuses on the exploration of treatment effect heterogeneity when considering the use of immunotherapies to treat NSCLC. Immunotherapies work by enhancing the body’s own immune response against cancer cells. However, therapeutic response can vary substantially between patients and is known to differ depending on individual factors, including the presence of genomic mutations \citep{RN24}. Since immunotherapies are typically prescribed instead of more traditional treatments, such as chemotherapies, we explore the difference in conditional survival probabilities if patients were to receive an immunotherapy or a chemotherapy, where the focus is on treatments initiated within 30 days of a NSCLC diagnosis. 

To conduct this study, we use data from the US-based de-identified Flatiron Health-Foundation Medicine NSCLC Clinico-Genomic Database. Clinical data from the Flatiron Health Research Database \citep{RN32} are linked to genomic data, derived from Flatiron Medicine's comprehensive genomic profiling tests (FoundationOne®CDx, FoundationOne®) using deterministic matching to provide a de-identified dataset \citep{RN35,RN33,RN34}. The data comes from approximately 280 US cancer clinics (800 sites of care) and contains over 20 pre-treatment demographic and clinical variables, extracted from electronic health records, as well as an array of genomic data, obtained from a tumour biopsy taken at a date after diagnosis. It presents a rich resource for exploring our research question, but also presents practical challenges, as patients can only enter the clinico-genomic database once they have undergone genomic testing (i.e., after treatment initiation). This means patients who die prior to genomic testing are not observed in the database, and due to the delayed entry, the observed patient event times can be left truncated, as well as right censored (when they do not die within the observed follow-up period). In this work we explore how causal ML estimators overcome these issues.

\subsection{Notation and data structure}  

Let $A$ be the binary treatment, with $A=1$ indicating treatment in a class of immunotherapies ($A=0$ otherwise), and let us define the start of follow up for a patient as the time of treatment initiation, $t=0$. Further, let $Z$ be a set of pre-treatment covariates, with $X\subseteq Z$ a subset of these for which heterogeneity is of interest, for instance the presence of an EGFR mutation. Using potential outcomes \citep{RN1}, we denote $T^a$ as the potential event time (since the start of treatment) under treatment level $a$, $a\in\{0,1\}$, and define $O_{0}=(Z,A,T^0,T^1)$ as the ideal data unit, with distribution $P_{0}$. Let $T$ denote the event time under the observed treatment. With right censored data, $T$ is unobserved for censored individuals, and we observe $\tilde{T}=min\{T,C\}$, where $C$ is the (partially observed) censoring time. We also define $\Delta=\mathbbm{1}(T\leq C)$ as an event indicator and $Q$ as an individual's study entry time. In our immunotherapy example, this is either a patient's time of genomic testing, or their time of treatment initiation, whichever occurs second. Additionally, as no information is available in the clinico-genomic database for patients who die after treatment initiation but before study entry ($0<T<Q$), we only observe data for those with $T\geq Q$. Finally, since patients cannot be censored prior to undergoing genomic testing or receiving a treatment, we assume censoring only occurs on individuals once they have entered the dataset, $C\geq Q$. Thus, we define the observed data as $O=(Z,A,\tilde{T},\Delta,Q)$ for individuals with $T\geq Q$, with $P$ its corresponding distribution.

\section{Identification} \label{Identification}

We explore the difference in conditional survival probabilities under two treatments at times $t\in(0,\tau]$, conditional on a subset of covariates, $X$:
\begin{align} \label{estimand_diff}
  \theta(t|x)=P_0(T^1>t|X=x) - P_0(T^0>t|X=x).  
\end{align}

Identifiability assumptions for time-to-event estimands have been previously discussed, with \cite{RN2} presenting a set of assumptions for identifying the conditional mean survival function using RC data, while \cite{RN5} and \cite{RN3} provide assumptions for a variety of time-to-event estimands using LTRC data. Because our motivating example is LTRC, we focus on assumptions required for identification in the LTRC setting, but for assumptions for the RC setting, see Section S8.1 of the Supplementary Material. 

Our identification assumptions for \eqref{estimand_diff} follow the notation of \cite{RN2}, but are extended to the LTRC setting using the original time scale, with time 0 defined as the time of treatment initiation, as done by \cite{RN5}. Weaker assumptions stated on the residual time scale, which set time 0 to be at study entry, can be found in \cite{RN3}, and a discussion of the benefits and drawbacks of these two frameworks is provided by \cite{RN23}. We require four types of assumption; positivity assumptions (A1): $P(A=a|Z)>0$, $a\in\{0,1\}$ and $P$-almost surely, and (A2): $P(C \geq \tau| Z)>0$, $P$-almost surely, and (A3): $P(Q<\tau| Z)>0$, $P$-almost surely. A consistency assumption (A4): When $A=a$, $T = T^{a}$. Conditional exchangeability assumptions (A5): $T^aI(T^a\leq \tau) \indep A |Z$, (A6): $T^aI(T^a\leq \tau) \indep QI(Q<\tau) |A=a,Z$, and (A7): $T^aI(T^a\leq \tau) \indep CI(C\leq \tau) |Q,A=a,Z,T^a\geq Q$. And finally, a no interference assumption (A8): $T_i^a$ does not vary with $a_j$, for $j\neq i$. 

Under (A1)-(A8), and using the product integral notation which allows events to occur in either discrete or continuous time \citep{RN8}, our causal estimand can be identified and written in terms of observable data functions as:
\begin{align}\label{est_diff_ident_LTRC}
    \theta(t|x)=\E\left[\left.\Prodi_{(0,\tau]}\{1-\Lambda(du|A=1,z,u\geq Q)\} - \Prodi_{(0,\tau]}\{1-\Lambda(du|A=0,z,u\geq Q)\}\right|X=x\right],
\end{align}
\noindent where $\Lambda(t|A=a,z,t\geq Q)=\int_{(0,t]}\frac{F(du|A=a,z,u\geq Q)}{R(u-|A=a,z,u\geq Q)}$ is the conditional cumulative hazard, $R(u-|A=a,z,u\geq Q)=P(u \leq \tilde{T}|A=a,Z=z,u\geq Q)$, and $F(u|A=a,z,u\geq Q)=P(\tilde{T} \leq u,\Delta=1|A=a,Z=z,u\geq Q)$. For the identification proof in the LTRC setting, see Section S1 of the Supplementary Material. Additionally, below we highlight the considerations which should be made when assessing the practical viability of these assumptions.

Assumption (A1) requires each individual in the target population to be eligible for either treatment, and (A2) and (A3) require that each individual has a positive probability of being under observation at any given time during the follow-up of interest. These assumptions should be considered when specifying the target population, removing subgroups for whom the treatment is contraindicated, or limiting the follow-up period to a window in which event times are observed. In our NSCLC study, we aimed to avoid violations of the positivity assumptions by restricting follow-up times to three years, and by implementing exclusion criteria similar to those seen in immunotherapy trials, such as excluding immunocompromised patients as they are ineligible for immunotherapy. As treatment effects in these subgroups are often of limited clinical relevance, such exclusions are often not problematic.

The validity of the consistency assumption (A4) can be considered by reviewing the definition/recording of the treatment and event times. For treatments, this includes considering the impact that potential dosage/adherence variations may have on the event times; while for the event times, this may require considering how a patients treatment pathway may impact the data recording mechanism. In our NSCLC study, as the event times are captured from electronic health records, the measurement of outcomes should not differ between patients, irrespective of their treatment pathways. However, as treatment protocols are left to a doctor's discretion, the treatment dosages received by patients do vary. This issue is very common in observational data. One option for overcoming it is to take a pragmatic approach, defining treatment as treatment initiation, rather than as a specific treatment dose or adherence. We deem this reasonable in our NSCLC study as immunotherapy/chemotherapy dosages are standardized and as prescribing doctors should be following specified guidelines \citep{RN27}.

This leaves the no interference assumption (A8) and the conditional exchangeabilty assumptions (A5)-(A7). No interference depends on the outcome and treatment being considered, and in the NSCLC example it is highly plausible as NSCLC is not an infectious disease. Meanwhile, the conditional exchangeability assumptions, (A5)-(A7) are crucial when considering RC or LTRC observational data, as each of these mechanisms can shift the covariate profile of the observed patients at each time point away from that of the target population, or can introduce imbalances in the covariate profiles between the treatment groups. These assumptions rely on a sufficient set of pre-treatment covariates being recorded, however, when using LTRC data this can be problematic, as typically some covariate information is recorded only at study entry, $t=Q$. For example, in our data, the genomic information is only available at the time of genomic testing $t=Q$, which can be after treatment initiation. When this occurs, the validity of (A5)-(A7) should be assessed in light of how this information may impact treatment assignment, study entry and censoring times. In our example, the genomic information is not considered to impact the health markers of a patient prior to treatment, and is not known to the treating clinician, so is not expected to influence treatment decision (not to violate (A5)). In such scenarios, one should consider whether the information recorded at study entry, $t=Q$ can be used as a proxy for the pre-treatment variable. We do so by classifying our genomic data into two categories; information which is likely or unlikely to change as a result of treatment assignment and timing of recording. For the information which is unlikely to change (e.g., an EGFR mutation) we use the data at $t=Q$ as a proxy for the pre-treatment data. Crucially, variables which may change as a result of treatment (e.g., a patient’s PD-L1 expression levels) should not be adjusted for, as they are potential mediators. However, this may result in a greater risk of unmeasured confounding.

\section{Existing causal ML estimators} \label{existing_ests}

We now discuss the existing causal ML estimators for our estimand. The use of ML is desirable as it allows users to fully leverage complex data structures, such as those found in the clinico-genomic database, and allows the researcher to avoid unrealistic modelling assumptions. A simple way of implementing ML to estimate the difference in survival probabilities is via a so-called T-learner \citep{RN12} which estimates the difference by plugging in ML based estimates of the two conditional treatment-specific survival curves in eq. (\ref{est_diff_ident_LTRC}). However, the T-learner is highly prone to bias when using ML, as it tends to over-smooth the treatment-specific survival curve estimates in subgroups which are under-represented, with this regularisation bias propagating through to the estimates of the difference. Additionally, as it does not optimize the difference directly it can fail to draw upon the fact that this contrast may be smoother than the treatment-specific survival curves. Hence, even if the difference can be learnt at a faster rate than the treatment specific survival curves, the T-learner will not benefit from this property. For these reasons, estimators which directly target the estimand and which are insensitive to errors in nuisance function estimation (functions which are not the target function, such as the conditional treatment-specific survival curves) are typically preferred \citep{cher}. 

A class of estimators which achieve these properties are known as de-biased estimators and are typically derived using the efficient influence function (EIF) of the given estimand, where the EIF quantifies the sensitivity of the estimand to perturbations in the data-generating distribution \citep{Hines}. This said, when estimating heterogenous treatment effects, estimands can be infinite dimensional (when any variable in $X$ is continuous) and their EIFs may not be well defined. Instead, one must draw upon the EIF of an appropriately chosen expected loss function \citep{Morz}. By doing so, estimators are constructed such that this expected loss function experiences a relative insensitivity to the estimation of nuisance functions. 

Key examples of EIF-based estimators for heterogenous treatment effects using time-to-event data are causal survival forests (CSFs) \citep{RN9}, which handle RC event times, and more recently, the ltrc-R and ltrc-DR learners of \citep{RN3}, which handle LTRC event times. All three estimators target the mean difference in a deterministic transformation of survival times under different treatments, which includes the conditional mean difference in survival probabilities when the transformation is defined as $\mathbbm{1}\{T>t\}$. However, they are derived using different approaches. Both CSFs and the ltrc-R learner estimate treatment effect heterogeneity by drawing upon Robinson's outcome decomposition \citep{RN81}, constructing a residual-on-residual regression using the EIF presented by \cite{RN82}, noting the ltrc-R learner does this in the LTRC setting. Meanwhile, the ltrc-DR learner constructs a loss function using a ``one-step'' construction \citep{Hines}, adding a de-biasing term (which comes from the EIF of the MSE) to the plug-in estimate of the difference in survival probabilities. 

All three estimators discussed above draw upon EIFs to construct their estimators, and consequently make use of inverse probability weights (IPWs), along with outcome regressions to address imbalances in the covariate distribution introduced by treatment assignment and by right censoring and left truncation (in the case of the ltrc-R and ltrc-DR learners). Meanwhile, each estimator ensures that the influence of nuisance parameter estimation (such as for the outcome models/IPWs) is only of second-order, enabling the estimators to achieve oracle efficiency, that is, to perform as well as if the nuisance functions were known, provided that the nuisance estimators converge sufficiently quickly to the true functions \citep{RN9,RN3}. This considered, each estimator's practical performance differs. The ltrc-DR learner sometimes yields unbounded estimates of the target function, as the estimator does not mitigate extreme inverse probability of treatment, censoring or truncation weights which can occur when their respective functions near their bounds, i.e., 0 or 1. Further, CSFs and the ltrc-R learner, both of which use a residual-on-residual regression, restrict treatment effect heterogeneity to $X\subseteq Z$, delivering bias when there is heterogeneity in components of $Z$ that are not in $X$. 

In addition to this, CSFs and the ltrc-R and ltrc-DR estimators take different approaches when minimizing their expected loss functions, with CSFs estimating heterogeneity using a forest-based weighting scheme, while the ltrc estimators do not depend on a forest-based structure, but can be run using any ML algorithm that accommodates custom loss functions, i.e, \textit{xgboost} \citep{RN19}. This means CSFs are prone to bias when its forest based framework does not suit the data, i.e., when the data is high dimensional or when the covariates are strongly correlated. On the other hand, both ltrc estimators encounter a different issue at this stage, with the weights being possibly negative, which makes them non-interpretable (even when the nuisance functions are correctly bounded). This introduces bias, makes the loss function non-convex (complicating the minimization of the loss), and restricts algorithm choice.

The limitations discussed above pose practical challenges for the CSF and ltrc estimators. However, there is also a more fundamental issue affecting all of these methods, in that none of these estimators fully leverage the temporal structure of the data. In order to generate estimates of a target function across multiple time points (i.e. a treatment effect curve), each of these existing learners optimizes the estimates of the difference at each time point in $(0,\tau]$ separately. This can lead to non-smooth treatment effect estimates over time, which may not be biologically plausible. We aim to address these issues, presenting an alternative estimation technique, surv-iTMLE, which directly targets the difference in conditional survival probabilities, while also obtaining estimates which are smooth over time and conditional on a subset of covariates, $X\subseteq Z$.

\section{surv-iTMLE} \label{Method}
 
Our proposal, surv-iTMLE, involves a two step algorithm to estimate (\ref{est_diff_ident_LTRC}). The first step constructs de-biased plug-in estimates of the difference in survival probabilities at a series of time points, which are conditional on $Z$, referred to as pseudo-outcomes. The second step runs a pseudo-outcome regression, obtaining estimates of the difference in survival probabilities which are smooth over time and conditional on $X\subseteq Z$, i.e, are not restricted to the full set of adjustment variables, $Z$. This section provides details on these steps for the LTRC setting (Algorithms \ref{alg:part1} and \ref{alg:part2}), and details for the RC setting can be found in Section S8 of the Supplementary Material.

\subsection{Step 1: Generating pseudo-outcomes using iTMLE} \label{Alg_Part1}

To derive a de-biased estimator of the difference in conditional survival probabilities, we first define our expected loss as the components of the MSE (of $\theta(t|X)$) which depend on $\theta(t|X)$, noting that the remaining components do not affect its minimization.
\begin{gather} \label{Diff_risk}
    \psi(P_0) =  \E_0\left[\theta(t|X)^2 -2\theta(t|X)\E_0\left[\left.\left(P_0(T^1>t|Z) - P_0(T^0>t|Z)\right)\right|X\right]\right].
\end{gather}
For a fixed $\theta(t|X)$, we write the uncentered EIF of equation (\ref{Diff_risk}) in the LTRC setting as:
\begin{allowdisplaybreaks}
  \begin{align} \label{General_IF}
  \nonumber  \phi&(P_0) = \theta(t|X)^2 - 2\theta(t|X)\left(S(t|A=1,Z,t\geq Q)-S(t|A=0,Z,t\geq Q)\right) + \\ 
    \nonumber &~~~~~~~~~~~~~~~~~~2\theta(t|X)\frac{\mathbbm{1}(t\geq Q)(A-\pi(Z))}{(1-\pi(Z))\pi(Z)}S(t|A,Z,t\geq Q)\times\\
    &~~~~~~~~~~~~~~~~~~~~~~~~~~~~~\int_{(0,t]}\frac{\left\{\mathbbm{1}(\tilde{T}=u, \Delta=1) - \mathbbm{1}(u \leq \tilde{T})\Lambda(du|A,Z,u\geq Q)\right\}}{S(u|A,Z,u\geq Q)\int_{(0,u]}G(u-|A,Q,Z)H(dq|A,Z)},
\end{align} 
\end{allowdisplaybreaks}
\noindent where $S(t|A=a,Z,t\geq Q)=P(\tilde{T}>t,\Delta=1|A=a,Z,t\geq Q)$, $a\in\{0,1\}$ are the probabilities of surviving up to time $t$ under each treatment conditional on surviving to the study entry time $Q$, $G(t|A,Q,Z)=P(C>t|A,Q,Z)$ is the probability of being censored after time $t$, $H(t|A,Z)=P(Q\leq t|Z)$ is the probability of entering the study at or before time $t$, and $\pi(Z)=P(A=1|Z)$ the probability of being treated. The derivation for this EIF can be found in Section S2 in the Supplementary Materials.

The EIF in equation (\ref{General_IF}) includes IPWs for treatment assignment, $1/\pi(Z)$, right censoring, $1/G(t|A,Q,Z)$, and left truncation, $1/H(t|A,Z)$, and consists of two key components: the plug-in loss function, $S(t|A=1,Z,t\geq Q) - S(t|A=0,Z,t\geq Q)$, and a de-biasing term, written in equation (\ref{General_IF_LTRC_debiasing_term}) below,
\begin{allowdisplaybreaks}
  \begin{align} \label{General_IF_LTRC_debiasing_term}
  2\theta(t|X)&\frac{\mathbbm{1}(t\geq Q)(A-\pi(Z))}{(1-\pi(Z))\pi(Z)}S(t|A,Z,t\geq Q)\int_{(0,t]}\frac{\left\{\mathbbm{1}(\tilde{T}=u, \Delta=1) - \mathbbm{1}(u \leq \tilde{T})\Lambda(du|A,Z,u\geq Q)\right\}}{S(u|A,Z,u\geq Q)\int_{(0,u]}G(u-|A,Q,Z)H(dq|A,Z)}.
\end{align} 
\end{allowdisplaybreaks}

The de-biasing term in equation (\ref{General_IF_LTRC_debiasing_term}) captures the bias introduced by estimating rather than knowing the nuisance parameters. Moreover, the EIF in equation (\ref{General_IF}) can be used to derive de-biased estimators by constructing pseudo-outcomes. In this paper we present a novel targeted learning approach, combining two existing methods; the EP-learner \citep{RN15}, which uses iTMLE to estimate heterogeneous treatment effects, and a survival TMLE algorithm, presented by \cite{Moore}, using a pre-defined grid of times. 

In general, the goal of targeted learning is to update the plug-in estimates of a function, in our case $\theta(t|X)$, such that the sample average of the de-biasing term in equation (\ref{General_IF_LTRC_debiasing_term}) converges to zero sufficiently quickly, ensuring that it does not impact the overall convergence rate of the expected loss. Typically, this is done in two steps: first, the EIF is used to define a weighted regression; and second, the estimated linear predictor from this model is used to update the plug-in estimates. However, when constructing a targeting process for our estimand, we encounter additional challenges. One of these challenges is that the de-biasing term contains $\theta(t|X)$, which is unknown and infinite-dimensional in $t$ and $X$ (when any variables in $X$ are continuous). This means that in order to remove bias from the plug-in estimator, the targeting step must set the sample average of infinitely many terms equal to 0. An estimator which presented a solution to this problem was the EP-learner \citep{RN15}, which overcame this problem in the context of the CATE by introducing a sieve-based targeting procedure, regressing the outcomes against the sieve basis $\varphi(X)$ in a weighted regression. We draw upon this method, known as iTMLE \citep{RN68, RN4, RN15}, extending it to the time-to-event setting.

In the time to event setting, we focus on the targeted learning procedure proposed by \cite{Moore} which considers a pre-defined grid of times, $\{t_1,...,\tau\}$. It updates the survival functions, $S(t|A,Z,t\geq Q)$, $A\in\{0,1\}$ by first updating the hazard functions, $\lambda(k|A,Z,k\geq Q)$, for $A\in\{0,1\}$ at each time point up to and including time point of interest, $k\in\{t_1,...,t\}$. The plug-in hazard estimates, $\hat{\lambda}(k|A,Z,k\geq Q)$ are obtained from plug-in survival function estimates, $\hat{S}(t|A,Z,t\geq Q)$), and this algorithm requires separate targeting models for each $k\in\{t_1,...,t\}$. Further, it must be repeated for each $t\in\{t_1,...,\tau\}$, as the de-biasing term we wish to set to 0 will differ across each time, $t$. By updating the treatment-specific hazard function estimates first, we can maintain the bounds and monotonicity of the survival functions. Additionally, as we consider discrete pre-defined time intervals, we write the hazards as $\hat{\lambda}(k|A,Z,k\geq Q)$ (rather than in terms of the cumulative hazard/product integral notation), noting that they can be obtained from estimates of the treatment specific survival functions. 

\begin{figure}[!htb]
\begin{center}
\includegraphics[scale=0.45]{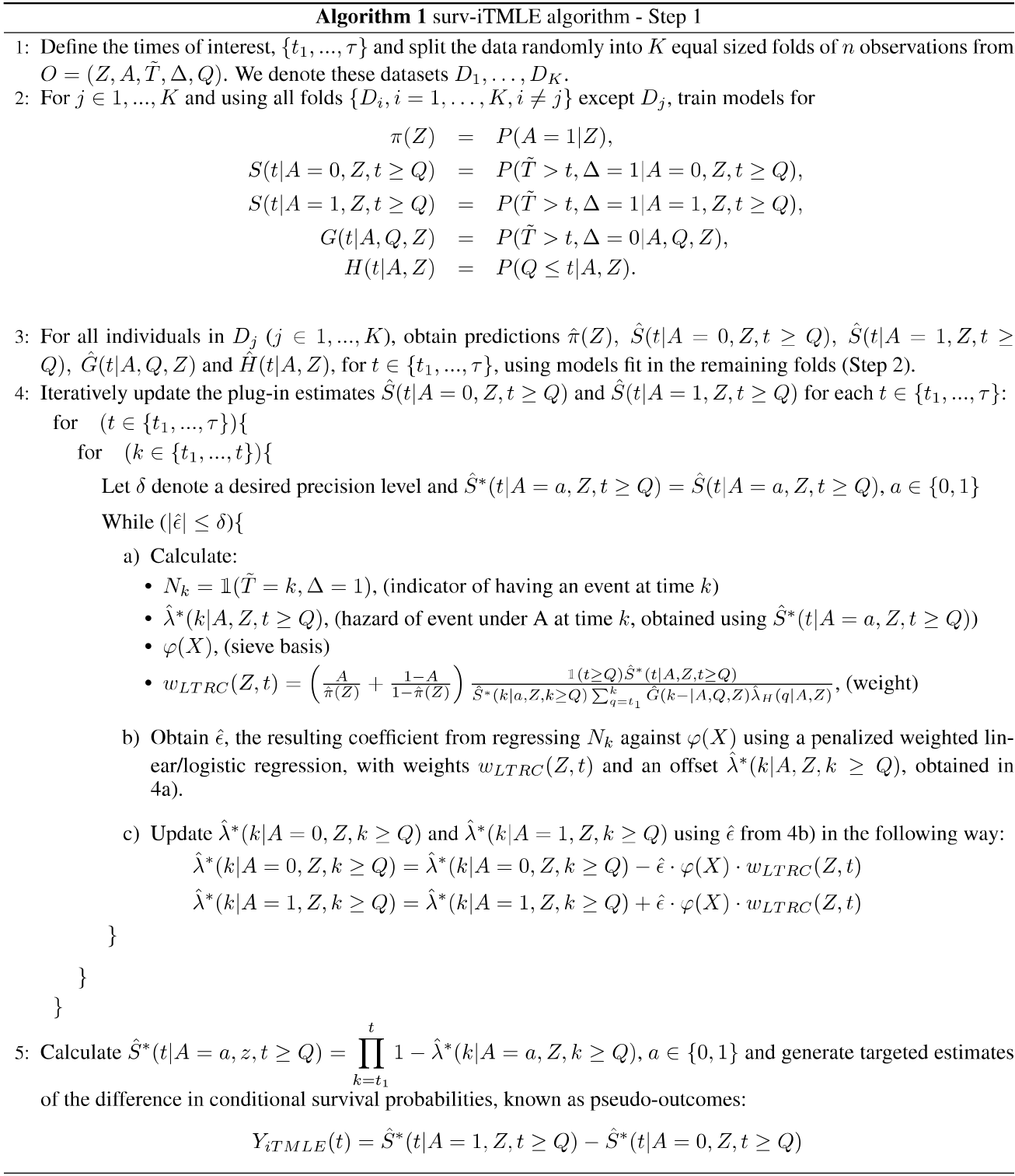} 
\end{center}
\caption{surv-iTMLE algorithm - Step 1 - Obtaining pseudo-outcomes.}
\label{alg:part1}
\end{figure}

To adapt this algorithm to our infinite dimensional estimand, we now incorporate a sieve basis in the targeting step. The adapted algorithm incorporating this is summarised in Algorithm \ref{alg:part1}, which fits a weighted logistic regression (or alternatively linear or log-linear regression - see below) for each $k\in\{t_1,...,t\}$ and $t\in\{t_1,...,\tau\}$, regressing the indicator of having an event during a given interval, $\mathbbm{1}(\tilde{T}=k,\Delta=1)$, against the sieve basis, $\varphi(X)$, with the plug-in hazard function estimates, $\hat{\lambda}(k|A,Z,k\geq Q)$, included as an offset, and the weights defined as 
$$w_{LTRC}(Z,t) = \left(\frac{A}{\hat{\pi}(Z)} + \frac{1-A}{1-\hat{\pi}(Z)}\right)\frac{\mathbbm{1}(t\geq Q)\hat{S}(t|a,Z,t\geq Q)}{\hat{S}(k|a,Z,k\geq Q)\sum_{q=t_1}^{k}\hat{G}(k-|A,q,Z)\hat{\lambda}_H(q|A,Z)}.$$ 
Here, $\hat{\lambda}_H(q|A,Z)$ represents the hazard of study entry, which can be obtained using $\hat{H}(q|A,Z)$, and we write the integral of the censoring/truncation probabilities as a sum over the respective intervals. Once the hazards are updated, these can be used to reconstruct the survival functions, generating updated survival functions $S^{*}(t|A=a,Z)$, for $a\in\{0,1\}$. By making this subtle change to the survival TMLE algorithm, now using the sieve basis regression, the model coefficients can now vary by $X$, and the targeting process (when iterated) will set the sample average of the de-biasing term to 0. This being said, the inclusion of sieves introduces new practical challenges. When defining a sieve basis, it is useful to remember that sieves, by definition, grow with sample size. This means that when used in practice, the sieve basis, $\varphi(X)$, can grow to be high-dimensional, and hence its inclusion in parametric models can result in model convergence issues. Further, as the event of interest in this setting is binary (and potentially rare, depending on the time grid), logistic regressions can experience further convergence issues. Together, these two issues prevent reliable convergence of the targeting models in this algorithm, and hence we suggest two alterations to the estimation process. 

Firstly, one can use linear regressions, or log-linear regressions in the targeting steps. Both of these options experience fewer convergence issues than logistic regressions, and in theory achieve the same targeting goal. However, the resulting hazards are no longer guaranteed to respect the bounds $[0,1]$. Secondly, it can be useful to implement each of these regressions using penalization (e.g. LASSO) \citep{RN4} when the sieve is high dimensional. Together, these techniques aid the convergence of the hazard targeting steps, and the resulting updated hazards can be used to construct pseudo-outcomes. 

Figure \ref{alg:part1} presents this algorithm, including the need for cross-fitting. Additionally, as the pseudo-outcomes are obtained using targeted learning, which mitigates the impact of extreme IPWs, the pseudo-outcomes can be less volatile than those generated by one-step methods (e.g., the ltrc-DR learner).

\subsection{Step 2: Generating causal contrast estimates using a pseudo-outcome regression} \label{Alg_Part2_text}

We now consider how to obtain treatment effect curves which not only minimize the expected loss function in equation (\ref{Diff_risk}), but which are also smooth over time. A simple solution would be to separately regress the pseudo-outcomes at each time, $\hat{Y}_{iTMLE}$(t), for $t\in\{t_1,...,\tau\}$, against $X$, 
$$\hat{\theta}_{iTMLE-Naive}(t|X)=\E\left[\left.\hat{Y}_{iTMLE}(t)\right|X\right]~~~\text{for }t\in\{t_1,...,\tau\}.$$
However, this would generate non-smooth treatment effect curves over time, similar to those produced by CSFs. For this reason, we present an alternative solution, proposing the use of a pooled regression which allows for the estimates of $\theta_0(t|X)$ to be estimated jointly across time. It begins by generating a long dataset, where each person contributes one row per time. The outcome is then defined as the incremental change in an individual's pseudo-outcome since the previous time point, $\Delta Y_{iTMLE}(k)=Y_{iTMLE}(j)-Y_{iTMLE}(k-1)$, for $k\in\{t_1,...,t\}$, using the pseudo-outcomes generated in Step 1. We use the incremental changes in the pseudo-outcomes, rather than the pseudo-outcomes themselves in order to dampen correlations between rows. Estimates of $\theta_0(t|X)$ can then be obtained by regressing $\Delta Y_{iTMLE}(k)$ against $X$, the covariates in which treatment effect heterogeneity is of interest, and $time$, with $time$ included as a continuous predictor. This produces estimates $\Delta \hat{\theta}_{iTMLE}(k|X)=\hat{\theta}(k|X)-\hat{\theta}(k-1|X)$ for $k\in\{t_1,...,\tau\}$, which represent the incremental increase in the difference in conditional survival probabilities (given $X$) at each time, and estimates of $\hat{\theta}_{iTMLE}(t|X)$ can be obtained by summing these incremental predictions up to time $t$, $t\in\{t_1,...,\tau\}$. This algorithm is summarised in Figure \ref{alg:part2}, and in Section S3 in the Supplementary Materials, we provide an overview of why this regression can be run on the incremental outcomes, even though it changes the loss function being minimized.
$$\hat{\theta}_{iTMLE}(t|X) = \sum_{k=t_1}^{t}\Delta\hat{\theta}_{iTMLE}(k|X).$$

\begin{figure}[!htb]
\begin{center}
\includegraphics[scale=0.45]{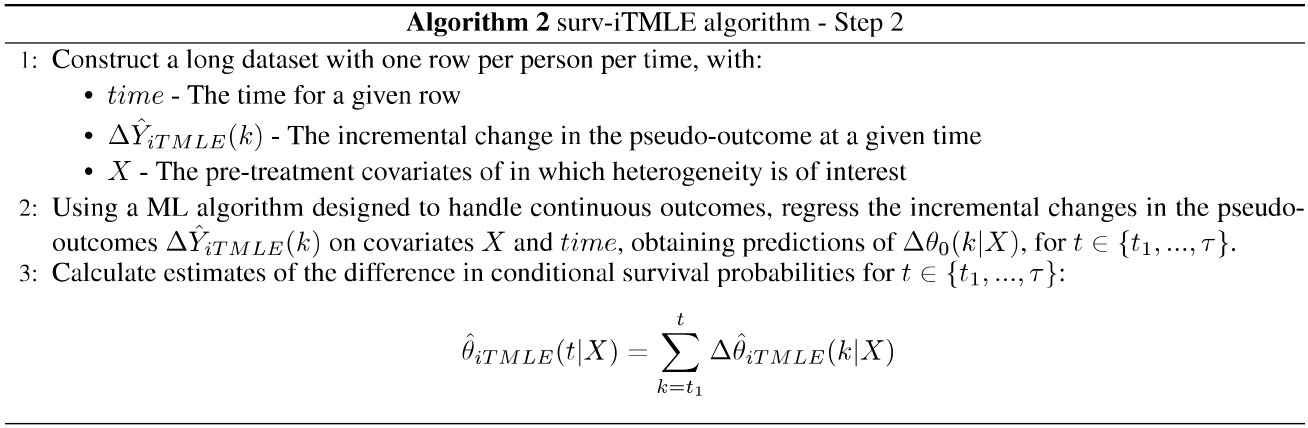} 
\end{center}
\caption{surv-iTMLE algorithm - Step 2 - Estimating the difference in conditional survival probabilities under two treatments for a given patient.}
\label{alg:part2}
\end{figure}

\subsection{Implementation guidance} \label{imp}

The application of surv-iTMLE involves several important user-specified choices that influence the estimation procedure. The first is how to estimate each nuisance function found in equation (\ref{General_IF}). As seen in Algorithm \ref{alg:part1}, steps 1-3, cross-fitting is required when estimating the nuisance functions, assuring that independence between the nuisance function estimates and the estimation of the target function is maintained. In Algorithm \ref{alg:part1} we achieve this using a $K$-fold cross-fitting procedure. 

We next consider how each nuisance function can be estimated in the LTRC setting. Both the survival functions, $S(t|A=a,Z,t\geq Q)$, $a\in\{0,1\}$, and the censoring function, $G(t|A,Q,Z)$, can be estimated using ML algorithms, as long as the algorithms account for LTRC and RC event times respectively (See Section S4 of the Supplementary Materials for a discussion of available options). However, when estimating the truncation probabilities, $H(t|A,Z)$, or the propensity score, $\pi(Z)$, additional caution is required. Both functions are defined conditional on observing the full target population, but in practice, we only observe data on those who enter the study prior to having an event, $T>Q$. To address this disparity in populations, one can re-weight the population, up-weighting the subgroups who are likely to have an event prior to truncation. These inverse probability weights, $1/P(T>Q|A,Z)$, can be estimated using the same algorithms used to estimate the survival functions and can be used within the truncation/treatment models to address this bias. Once estimated, the truncation probabilities and propensity score can then be estimated using algorithms which handle time-to-event and binary outcomes respectively. Yet as errors in these weights can propagate through to the estimates of $H(t|A,Z)$ and $\pi(Z)$, we note this process can slow the rate of convergence of these nuisance estimators, which may impact the overall rate of convergence of the estimator. This issue is discussed by \cite{RN3}. 


Once the nuisance functions have been estimated, one needs to decide how to define the sieve basis and how to carry out the weighted regressions in the targeting step (Algorithm \ref{alg:part1}, step 4). Our implementations found in Sections \ref{Sims} and \ref{NSCLC_example} define the sieve basis using a univariate cosine polynomial basis, as it offers strong approximation guarantees under smoothness assumptions \citep{RN21}. However, as this sieve basis can be high dimensional, we recommend carrying out the targeting step by using penalized regressions, as done by \cite{RN4}. This can be penalized linear regressions as found in Sections \ref{Sims} and \ref{NSCLC_example}, but could also be penalised logistic or log-linear regressions, depending on the trade-off between maintaining the hazard estimate bounds (i.e. $[0,1]$) and achieving model convergence.

Finally, the Step 2 regression (Algorithm \ref{alg:part2}) can be performed using any algorithm which handles continuous outcomes. In Section \ref{Sims}, we use the \textit{SuperLearner}, but to obtain smooth treatment effect curves over time, one may wish to use an algorithm which can capture the level of smoothness expected in that particular setting. In Section \ref{NSCLC_example}, we use a generalized additive model (GAM), including factor–smooth interactions between $time$ and $X$, as they allow treatment effect curves to vary in shape across $X$ values. However, if using GAMs, one must limit the maximum basis dimension for each term, such that the algorithm generates smooth curves which can still depart from strict linearity. In the NSCLC example, a maximum of four degrees of freedom ($K=4$) offered flexible but smooth treatment effect curves, however, other algorithm/tuning parameters may be more appropriate in other settings.

\section{Simulations} \label{Sims}

\subsection{Methods} \label{sims_methods}
We demonstrate the empirical performance of surv-iTMLE and compare it to existing methods across three data generating processes (DGPs). The first DGP is taken from \cite{RN9}, but with additional left truncation introduced to the event times; the second closely resembles the event/censoring/truncation patterns seen in our NSCLC analysis, and the third DGP provides an example with no treatment effect and high levels of right censoring ($\sim 50\%$). In each of these DGPs, we generate 20 independent uniformly distributed covariates, $Z$, a binary treatment, $A$, and generate event and censoring times, $T$ and $C$, along with two levels of left truncation, $Q$, allowing $\sim 25\%$ and $\sim 50\%$ of individuals to have an event prior to their truncation time. Individuals with an event time prior to their truncation time are excluded from analyses. Details of the data generating functions are in Section S5 of the Supplementary Material.  

Using surv-iTMLE, we estimate the conditional survival probabilities under the two treatment levels conditional on all covariates, $Z$. In each DGP, we review performance across three training sample sizes; $n=800$, $1600$ and $2400$, and for each DGP/sample size combination, 250 simulated datasets are generated. All nuisance models are fit using 10-fold cross-fitting, and the survival, censoring and truncation models in surv-iTMLE are estimated using a local survival stacking (LSS) approach \citep{RN17} (Section S5 of the Supplementary Materials) which includes variations of \textit{glmnet} and \textit{ranger}. Meanwhile, the propensity score and pseudo-outcome regressions are estimated using the \textit{SuperLearner} (for binary/continuous outcomes respectively). Details of the libraries used for each model are reported in Section S6 of the Supplementary Material. We compare our estimates to those obtained using a T-learner, fit using the same local survival stacking approach; CSFs, fit using the default estimation options in the \textit{grf} package on R; and the ltrc-DR and ltrc-R learners, fit using two nuisance parameter estimation options, (1) local survival stacking/\textit{SuperLearner}, and (2) penalized Cox regressions/generalized boosted regressions, as presented by \cite{RN3}. 

To evaluate estimator performance, we generated a single test dataset (which remains the same across simulations) with $n=10,000$ individuals per DGP. For each individual in the test dataset, the true difference in conditional survival probabilities between treatments was obtained via Monte Carlo simulation: we generated 1,000,000 survival times under each treatment, estimated survival probabilities at each time point as the proportion of simulated times exceeding that time, and then computed their difference between treatments. Each model was the trained on each of the 250 training datasets, generating estimates of the treatment effect for each individual in the test dataset, given their covariates Z, and at 10 time points. At each time point we obtain the squared differences between the estimated and true effects for each individual, and averaged these squared differences across individuals within a given simulation, taking the square root to give a RMSE. The RMSEs for each time point are are then averaged across the simulations, calculating the mean RMSE at each time point, and also averaged across time points, to calculate the overall mean RMSE. An analogous RC-only simulation study is provided in Section 8.3 of the Supplementary Material.

\subsection{Findings}

\begin{figure}[!htb]
\begin{center}
\includegraphics[scale=0.4]{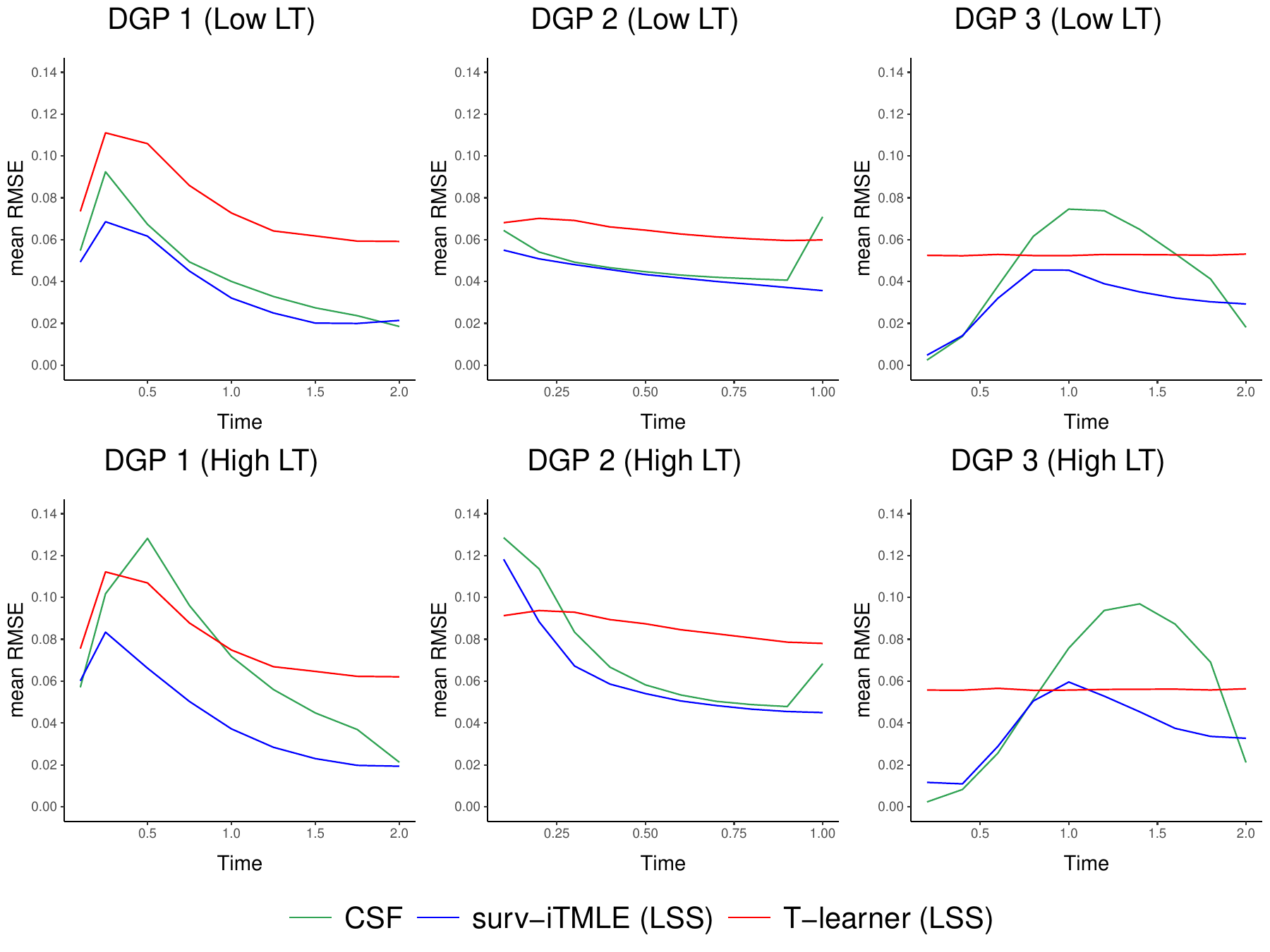} 
\end{center}
\caption{Mean root mean squared error (RMSE) by time point for surv-iTMLE, CSFs and the T-learner when estimating the difference in conditional survival probabilities under two treatments using training data with sample size $n=2400$, with varying proportions of left truncation (LT). LSS: local survival stacking.}
\label{Sim_plot2}
\end{figure}

Figure \ref{Sim_plot2} shows that surv-iTMLE generally outperformed both the T-learner and CSFs when the mean RMSE was reviewed at each time point individually. The performance of surv-iTMLE and CSFs varies depending on the degree of truncation and censoring which occurs at each time point, whereas the T-learner was less dependent on censoring/truncation rates, but generally exhibited poorer performance over time. Moreover, as surv-iTMLE accounts for left truncation, it outperforms CSFs in the early, heavily left truncated time periods (DGP1), with this trend most prominent in settings with high left truncation rates. We also note surv-iTMLE outperformed CSFs and the T-learner across all settings when reviewing the overall mean RMSE with surv-iTMLE obtaining estimates between 5\% and 36\% lower than CSFs, and between 12\% and 46\% lower than the T-learner across each of our DGP/sample size combinations. The largest benefits are seen in settings with high levels of left truncation (See Figure 1 in Section S7 of the Supplementary Materials). Additionally, both ltrc-R and ltrc-DR learners performed poorly, likely due to the generation of negative weights in their estimation process. This was most prominent when left truncation rates were high. For plot readability, we only include the corresponding results in the Supplementary Materials Section S7, Figures 1-5. 

\begin{figure}[!htb]
\begin{center}
\includegraphics[scale=0.4]{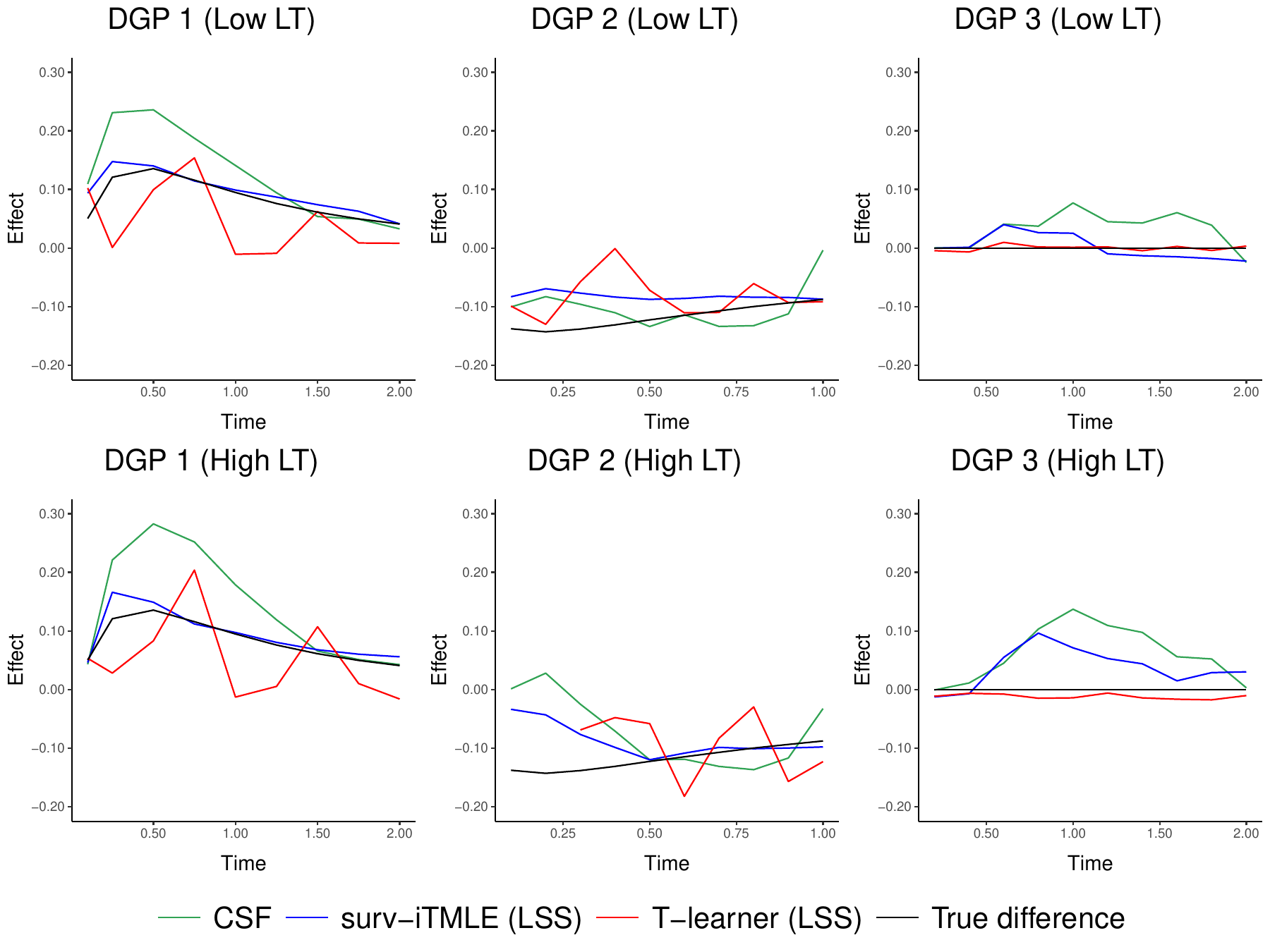} 
\end{center}
\caption{Individual estimates of the difference in conditional survival probabilities under two treatments plotted for one individual over time, with estimates obtained by surv-iTMLE, CSFs and the T-learner using training data of samples size $n=2400$. }
\label{Sim_plot3}
\end{figure}

Figure \ref{Sim_plot3} shows individual estimates of the difference in conditional survival probabilities, corresponding to one randomly chosen individual in one simulated set for each DGP. This illustrates how surv-iTMLE's estimates more closely track the true curve, providing smoother estimates than those produced by CSFs, the T-learner and the ltrc-R and ltrc-DR learners. This is a consequence of surv-iTMLE's second step, which estimates the difference jointly across all times, and can be seen for all the individuals in the test dataset (See Figures 3-5 in Section S7 of the Supplementary Materials, which also presents ltrc-R and ltrc-DR learner predictions). In the following section (Section \ref{NSCLC_example}),  we use GAMs for this step to obtain smooth treatment effect curves. 

\section{NSCLC analysis} \label{NSCLC_example}

\subsection{Background and Methods}
We now return to our motivating example and illustrate how surv-iTMLE, as well as the T-learner, CSFs and the ltrc-R and ltrc-DR learners can be used to estimate the difference in conditional survival probabilities if patients with NSCLC were to receive an immunotherapy or a
chemotherapy. As discussed in Section \ref{Motiv_ex}, the treatment strategy considered is treatment initiation (such that the consistency assumption (A4) holds). We focus on treatments initiated within 30 days of a NSCLC diagnosis and explore three years of follow-up for patients. To improve the plausibility of the positivity assumptions, (A1)-(A3), we restrict our target population using criteria outlined in previous immunotherapy trials \citep{RN22}; requiring patients be aged over 18 at diagnosis, have advanced NSCLC at the time of diagnosis, and not have an autoimmune disease at time of diagnosis. We also restrict our population to patients who do not have systemic steroid use in the 90 days prior to treatment initiation and further exclude patients who initiate a chemotherapy but have prior immunotherapy use. Under these eligibility criteria, there is a total of 3086 patients, of which 1368 (44\%) initiated immunotherapy and 1718 (56\%) initiated chemotherapy. 

We apply each estimator, adjusting for all pre-treatment covariates with less than 30\% missingness, which includes demographic and clinical measurements, as well genomic data which is considered to not vary over time (e.g., presence of EGFR and KRAS mutations). A full list of the available pre-treatment covariates can be found in Section S9.1 of the Supplementary Materials. A single dataset with no missing covariate information was obtained by imputing missing values for each covariate with less than 30\% missingness. This was obtained using Multiple Imputation by Chained Equations. A summary of this process, along with details on how each of these estimators were fit can be found in Sections S9.2 and S9.3 of the Supplementary Materials respectively.

The initial estimand of interest is the difference in survival probabilities conditional on the sufficient adjustment set of pre-treatment covariates, $Z$. This is estimated using all methods. Further, since surv-iTMLE can estimate the difference conditional on a subset $X$ of the sufficient set $Z$, we target two extra estimands. One conditional on only EGFR mutations, and one conditional on only age, both of which are of clinical interest. Finally, to draw upon the fact surv-iTMLE can obtain smooth treatment effect curves over time, we use a GAM, as described in Section \ref{imp} to estimate the pseudo-outcome model in surv-iTMLE.


\subsection{Findings}

\begin{figure}[!htb]
\begin{center}
\includegraphics[scale=0.4]{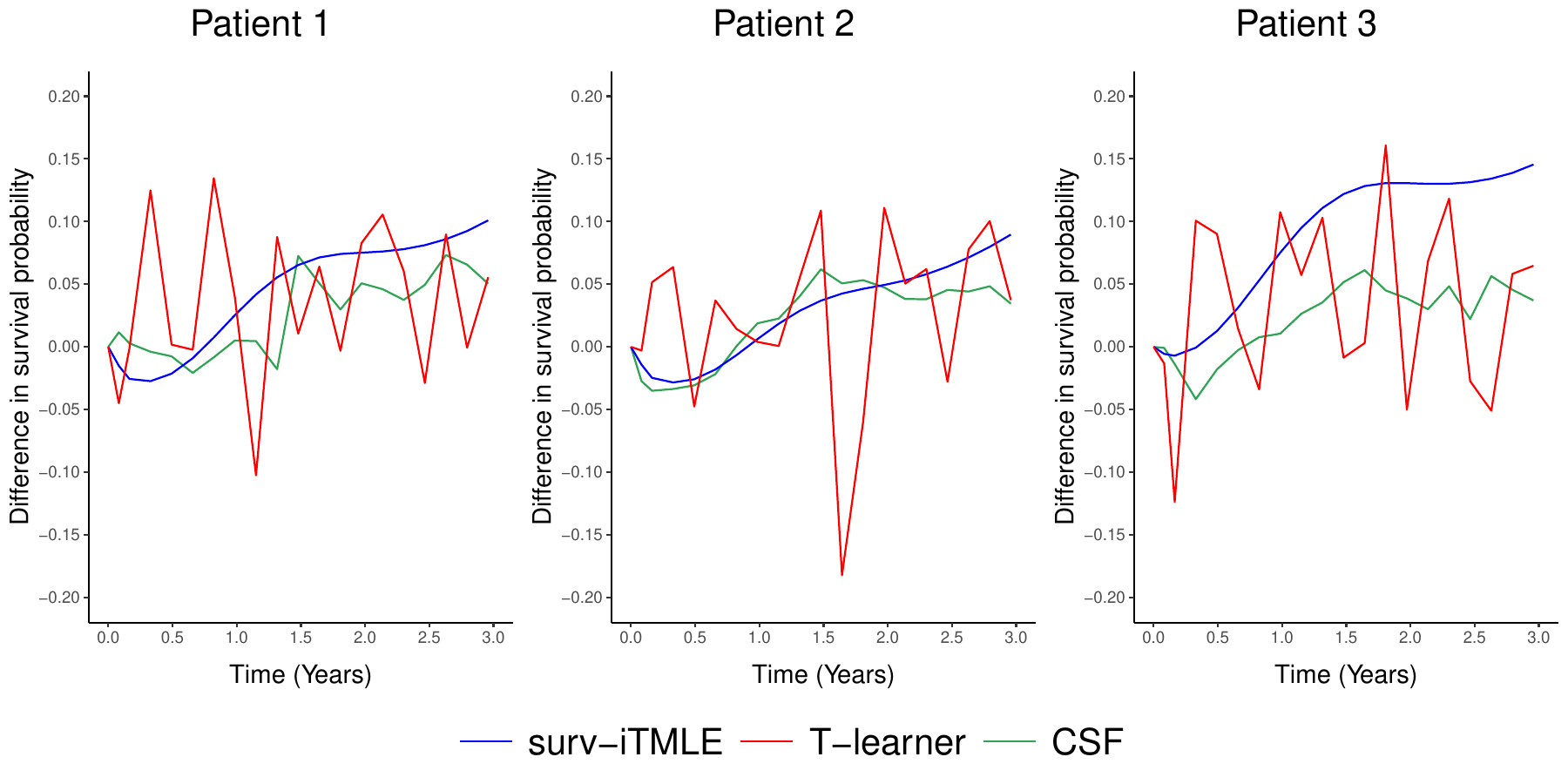} 
\end{center}
\caption{Individual difference in conditional survival probabilities over time for three example NSCLC patients if they were to initiate immunotherapy/chemotherapy after a NSCLC diagnosis, with treatment effects estimated by surv-iTMLE, CSFs and the T-learner. }
\label{NSCLC_plot1}
\end{figure}

Figure \ref{NSCLC_plot1} shows plots of individual treatment effect curves for three randomly chosen example patients, with different gender, genomic and demographic profiles. We see that both surv-iTMLE and CSFs predict a limited response to the immunotherapies over the first 6 months, with immunotherapies offering increased effectiveness relative to chemotherapy over time. On the other hand, the T-learner, which does not directly target the estimand of interest, produces highly non-smooth curves over time in which trends are difficult to detect. Similarly, the curves produced by CSFs and ltrc-R and ltrc-DR learners are non-smooth, while the curves generated by the surv-iTMLE are smooth over time, reflecting more intuitive treatment effect patterns. Results from the ltrc-R and ltrc-DR learners are only presented in Section S10 of the supplementary materials to aid readability.

\begin{figure}[!htb]
\begin{center}
\includegraphics[scale=0.45]{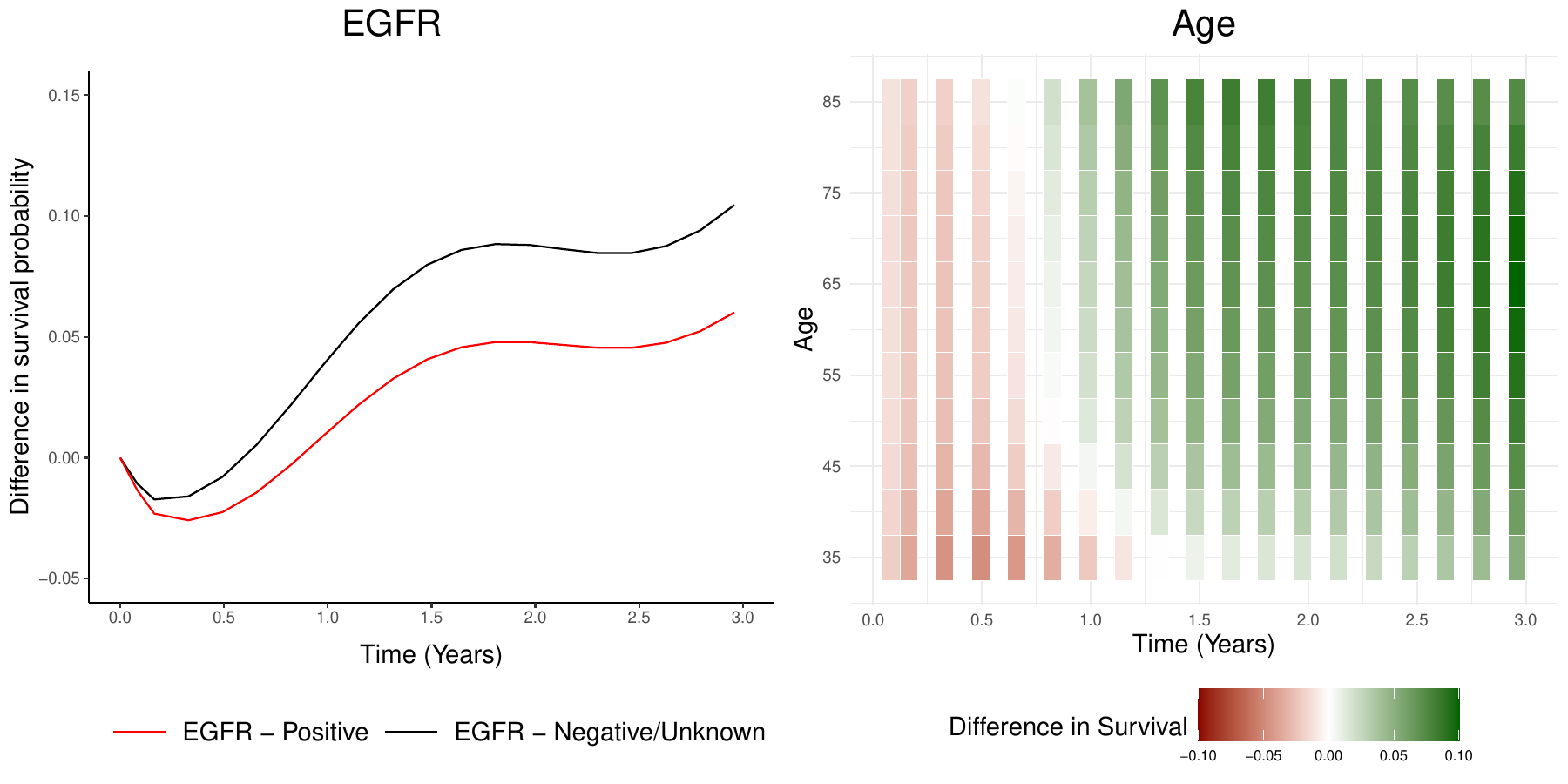} 
\end{center}
\caption{Estimates from surv-iTMLE of the difference in the survival probabilities if patients were to receive an immunotherapy or a
chemotherapy after a NSCLC diagnosis. Left panel:  estimates conditional on having an EGFR mutation (binary). Right panel: conditional on age (continuous). }
\label{NSCLC_plot2}
\end{figure}

Figure \ref{NSCLC_plot2} shows the estimated difference in survival probabilities conditional on only a subset of covariates obtained by surv-iTMLE. On the left panel, the resulting estimates suggest that individuals who have an EGFR mutation tend to benefit less from immunotherapies than those who do not have the mutation (or whose status is unknown). On the right hand panel, we use a heat-map to display the estimated differences conditional on continuous age. Here, we observe that older patients tend to benefit more from receiving immunotherapies over chemotherapies, however this is not necessarily true for very elderly patients, with effectiveness dropping amongst patients aged 85 and above.

\section{Discussion} \label{Discussion}

In this paper we have shown how the difference in conditional survival probabilities can be estimated using surv-iTMLE, a fully non-parametric estimator which produces bounded and stable treatment effect estimates that are smooth over time. We have shown how it can accommodate both RC and LTRC event times, demonstrated its performance through simulations, and illustrated how it can provide meaningful insights in the NSCLC application. However, we believe further improvements and extensions to both surv-iTMLE, and our comparators are possible.

Firstly, when used in the LTRC setting, surv-iTMLE requires additional weights in order to estimate the truncation or propensity score models. These are also required by the ltrc-R and ltrc-DR learners. Orthogonal estimation for these nuisance parameters could in principle lead to improved performance of each of these estimators. Secondly, while the use of sieves for infinite-dimensional targeting has been formally studied in other settings \citep{RN15}, including with the use of lasso \citep{RN4}, we did not quantify the resulting approximation error 
and regularisation bias. Thirdly, we do not present an approach for obtaining measures of uncertainty for surv-iTMLE. This is inherently challenging due to surv-iTMLE's non-parametric design. Extending the recent work by \cite{Zhang} to the time-to-event setting, which uses highly adaptive lasso to provide confidence intervals for infinite dimensional parameters, could be a possible strategy which we leave to further work. In addition, we note that neither the ltrc-R or ltrc-DR learners performed well in our simulation studies (Section S7 of Supplementary Materials). This may be due to the negative weights generated when constructing their loss functions, but further exploration of their performance is needed and proposals that avoid the negative weights would also be of interest. 

Finally, we highlight that the surv-iTMLE algorithm could be used to estimate treatment specific survival curves, offering TMLE alternatives to the methods presented by \cite{RN2} and \cite{RN5}. 





\section*{Acknowledgments}
The authors would like to thank Chris Harbron (Roche) for facilitating data access, and for his valuable input regarding the data structure and scientific  problem formulation. 
This work was supported by the Medical Research Council [grant number MR/N013638/1]. KDO was funded by a Royal Society-Welcome Trust Sir Henry Dale fellowship, grant number 218554/Z/19/Z. RHK was funded by UK Research and Innovation (Future Leaders Fellowship MR/X015017/1). SV was supported by Advanced ERC grant ACME (101141305).

\bibliographystyle{unsrtnat}

\bibliography{Paper_endnote_lib.bib}

\newpage
\appendix
\section*{S1. Identification - Left truncated, right censored time-to-event data} \label{App_LTRC_ident_proof}

We show that our estimand, $\theta(t|x)$ is identifiable by first showing that $S_0(t|A=a,Z=z)=P_0(T^a> t|Z=z)$, $a\in\{0,1\}$ is identifiable under (A1)-(A8), extending the equivalent proof provided by \cite{RN2} to the LTRC setting. 

To simplify notation, we write $S_0(t|A=a,Z=z)=S_0(t|a,z)=P_0(T^a> t|Z=z)$, $a\in\{0,1\}$. We then show that this is identifiable using left truncated, right censored time-to-event data under (A1)-(A8) by first writing $S_0(t|a,z)$ using product integral notation \citep{RN8}:
\begin{align} \label{surv_haz_LT}
    S_0(t|a,z) =  \Prodi_{(0,t]}{\{1-\Lambda_0(du|a,z)\}},\tag{1}
\end{align}
where $\Lambda_0(t|a,z)=-\int_{(0,t]}\frac{S_0(du|a,z)}{S_0(u-|a,z)}$, with $S_0(u-|a,z) = P_0(T^a \geq u|Z=z)$. We then show that $\Lambda_0(t|a,z)$ is identifiable as $\Lambda(t|a,z,t\geq Q)=\int_{(0,t]}\frac{F(du|a,z,u\geq Q)}{R(u-|a,z,u\geq Q)}$, where $R(u-|a,z,u\geq Q)=P(u \leq \tilde{T}|A=a,Z=z,u\geq Q)$ and $F(u|a,z,u\geq Q)=P(\tilde{T} \leq u,\Delta=1|A=a,Z=z,u\geq Q)$ are the observable functions. We also note that for notational ease, we write functions of $T$ or $C$ using the distribution $P$, as they are partially observed. 

\begin{proof}
\noindent We begin by noting that under assumptions (A1) and (A5), $S_0(t|a,z)$ can be written as:
\begin{align*}
  S_0(t|a,z) &= P_0(T^a>t|Z=z) \\
  &= P_0(T^a>t|A=a,Z=z), 
\end{align*} 

\noindent for all $t\in(0,\tau]$ and $P_0$-almost every $z$. This holds as $\mathbbm{1}(T^a>t)$ is a measurable function of $T^a\mathbbm{1}(T^a\leq\tau)$ for $t\leq\tau$. Additionally, we define $G(t|a,q,z)=P(C>t|A=a,Q=q,Z=z)$ and $H(t|a,z)=P(Q\leq t|A=a,Z=z)$ and then demonstrate the equivalence between $\Lambda_0(t|a,z)$ and $\Lambda(t|a,z,u\geq Q)$ by re-writing $R(t|a,z,t\geq Q)$ and $F(dt|a,z,t\geq Q)$ in terms of $S_0(t|a,z)$, $G(t|a,q,z)$ and $H(t|a,z)$. Let us begin by considering $R(t|a,z,t\geq Q)$:

\begin{allowdisplaybreaks}
\begin{align*}
    R(t|a,z,&t\geq Q) = P(t \leq \tilde{T}|A=a,Z=z,t\geq Q) \\
    &= \frac{P(Q\leq t,t \leq T,t\leq C|A=a,Z=z)}{P(t\geq Q|A=a,Z=z)} \\
    &= \frac{P_0(Q\leq t,t \leq T^a,t\leq C|A=a,Z=z)}{P(t\geq Q|A=a,Z=z)} ~~~\text{By (A4)}\\ 
    &= \frac{\int_{(0,t]}P_0(t \leq T^a,t\leq C|Q=q,A=a,Z=z)P(Q\in dq|A=a,Z=z)}{P(t\geq Q|A=a,Z=z)}~~~\text{By (A6)} \\
    &= \frac{\int_{(0,t]}P_0(t \leq T^a|Q=q,A=a,Z=z)P(t\leq C|Q=q,A=a,Z=z)P(Q\in dq|A=a,Z=z)}{P(t\geq Q|A=a,Z=z)} \\
    &~~~~~~~\text{By (A7)} \\
    &= \frac{\int_{(0,t]}P_0(t \leq T^a|A=a,Z=z)P(t\leq C|Q=q,A=a,Z=z)P(Q\in dq|A=a,Z=z)}{P(t\geq Q|A=a,Z=z)} \\ 
    &= S_0(t-|a,z)\frac{\int_{(0,t]}G(t-|a,q,z)H(dq|a,z)}{P(t\geq Q|A=a,Z=z)} 
\end{align*}
\end{allowdisplaybreaks}

\noindent Next, we consider $F(t|a,z,t\geq Q)$: 
\begin{allowdisplaybreaks}
\begin{align*}
    F(t|a,z,t\geq Q) &= P(\tilde{T}\leq t,\Delta=1|A=a,Z=z,t\geq Q) \\
    &= \frac{P(T\leq t,T\leq C,t\geq Q|A=a,Z=z)}{P(t\geq Q|A=a,Z=z)} \\
    &= \frac{\int_{(0,t]}\int_{[u,\infty)}\int_{(0,u]}P(du,dc,dq|A=a,Z=z)}{P(t\geq Q|A=a,Z=z)}, \\
\end{align*}
\end{allowdisplaybreaks}
\noindent where $P(u,c,q|A=a,Z=z)$ is the joint distribution function of $T$, $C$ and $Q$ given $A=a$ and $Z=z$. Similarly, under assumptions (A4), (A6) and (A7) we can re-write this in terms of $S_0(u|a,z)$, $G(u|a,q,z)$ and $H(u|a,z)$, as
\begin{allowdisplaybreaks}
\begin{align*}
    P&(u,c,q|A=a,Z=z) = P(T< u,C< c,Q< q|A=a,Z=z) \\ 
    &= P_0(T^a< u,C< c,Q< q|A=a,Z=z) \\ 
    &= \int_{(0,q)}P_0(T^a< u,C< c|Q=q',A=a,Z=z)P(Q\in dq'|A=a,Z=z) \\ 
    &= \int_{(0,q)}P_0(T^a< u|Q=q',A=a,Z=z)P(C< c|Q=q',A=a,Z=z)P(Q\in dq'|A=a,Z=z) \\ 
    &= \int_{(0,q)}P_0(T^a< u|A=a,Z=z)P(C< c|Q=q',A=a,Z=z)P(Q\in dq'|A=a,Z=z) \\ 
    &= [1-S_0(u-|a,z)]\int_{(0,q)}[1-G(c-|a,q',z)]H(dq'|a,z) 
\end{align*}
\end{allowdisplaybreaks}

\noindent We can then use that 
$P(du,dc,dq|A=a,Z=z) = S_0(du|a,z)G(dc|a,q,z)H(dq|a,z)$, substituting this into our definition of $F(t|a,z,t\geq Q)$, to give us
\begin{allowdisplaybreaks}
\begin{align*}
    F(t|a,z,t\geq Q)&= \frac{\int_{(0,t]}\int_{(0,u]}\int_{[u,\infty)}S_0(du|a,z)H(dq|a,z)G(dc|a,q,z)}{P(t\geq Q|A=a,Z=z)} \\
    &= \frac{\int_{(0,t]}S_0(du|a,z)\int_{(0,u]}H(dq|a,z)\int_{[u,\infty)}G(dc|a,q,z)}{P(t\geq Q|A=a,Z=z)} \\
    &= \frac{\int_{(0,t]}S_0(du|a,z)\int_{(0,u]}H(dq|a,z)[G(\infty|a,q,z) - G(u-|a,q,z)]}{P(t\geq Q|A=a,Z=z)} \\
    &= -\frac{\int_{(0,t]}S_0(du|a,z)\int_{(0,u]}G(u-|a,q,z)H(dq|a,z)}{P(t\geq Q|A=a,Z=z)} 
\end{align*}
\end{allowdisplaybreaks}

\noindent Consequently, we can write $F(dt|a,z,t\geq Q) = -\frac{S_0(dt|a,z)\int_{(0,t]}G(t-|a,q,z)H(dq|a,z)}{P(t\geq Q|A=a,Z=z)}$, and under (A2) and (A3) we see
\begin{align*}
    \Lambda(t|a,z,t\geq Q) = \int_{(0,t]}\frac{F(du|a,z,t\geq Q)}{R(u-|a,z,t\geq Q)} = -\int_{(0,t]}\frac{S_0(du|a,z)}{S_0(u-|a,z)} = 
    \Lambda_0(t|a,z)
\end{align*}

\noindent for all $t\in(0,\tau]$, demonstrating that $S_0(t|a,z)$ can be identified as $S(t|A=a,Z=z,t\geq Q) = S(t|a,z,t\geq Q) = \Prodi_{(0,t]}\{1-\Lambda(du|A=1,Z,u\geq Q)\}$ using the observed data. Consequently, $\theta(t|x)$ is identifiable as: 
\begin{align*}
    \theta(t|x) &= \E\left[\left.\Prodi_{(0,t]}\{1-\Lambda(du|A=1,Z,u\geq Q)\} - \Prodi_{(0,t]}\{1-\Lambda(du|A=0,Z,u\geq Q)\}\right|X=x\right]
\end{align*}

\noindent for all $t\in(0,\tau]$, completing the proof. 
\end{proof}

\newpage
\section*{S2. EIF derivation - Left truncated, right censored time-to-event data} \label{LTRC_EIFs}

\noindent Let us consider the following risk function for a given $\theta(t|X)$:
\begin{align*}
  \psi(P_0) = \E_0\left[\theta(t|X)^2 -2\theta(t|X)\E_0\left[\left.\left(S_0(t|A=1,Z)-S_0(t|A=0,Z)\right)\right|X\right]\right],  
\end{align*}

\noindent which under (A1)-(A8) can be written as:
\begin{align*}
    &\psi(P_0) = \psi(P) \\
    &= \E\left[\theta(t|X)^2 -2\theta(t|X)\E\left[\left.\Prodi_{(0,t]}\{1-\Lambda(du|A=1,Z,u\geq Q)\} - \Prodi_{(0,t]}\{1-\Lambda(du|A=0,Z,u\geq Q)\}\right|X\right]\right] \\
    &= \E\left[\E\left[\theta(t|X)^2 -2\theta(t|X)\left(\left.\Prodi_{(0,t]}\{1-\Lambda(du|A=1,Z,u\geq Q)\} - \Prodi_{(0,t]}\{1-\Lambda(du|A=0,Z,u\geq Q)\}\right)\right|X\right]\right] \\
    &= \E\left[\theta(t|X)^2 -2\theta(t|X)\left(\Prodi_{(0,t]}\{1-\Lambda(du|A=1,Z,u\geq Q)\} - \Prodi_{(0,t]}\{1-\Lambda(du|A=0,Z,u\geq Q)\}\right)\right] .
\end{align*}

\noindent We can then write the outer expectation as an integral as,
\begin{align} \label{LTRC_MSE}
    &\psi(P) = \int f(z)\left[\theta(t|x)^2 -2\theta(t|x)\left(\Prodi_{(0,t]}\{1-\Lambda(du|A=1,Z=z,u\geq Q)\} - \right.\right. \tag{4}\\
    &~~~~~~~~~~~~~~~~~~~~~~~~~~~~~~~~~~~~~~~~~~~~~~\left.\left.\Prodi_{(0,t]}\{1-\Lambda(du|A=0,Z=z,u\geq Q)\}\right)\right]dz, \nonumber
\end{align}

\noindent and derive this EIF for (\ref{LTRC_MSE}) by perturbing $P$ in the direction parameterized via the one-dimensional mixture model
$$P_s = s\Tilde{P} + (1-s)P,$$
where $\Tilde{P}$ is a fixed,
deterministic distribution with its support contained in the support of $P$. By perturbing $P$
in the direction of a point mass at $(\Tilde{z},\Tilde{a},\Tilde{c},\Tilde{v},\tilde{q})$, we get:
\begin{align*}
\psi(P_s) &= \int f_s(z)\left[\theta(t|x)^2 -2\theta(t|x)\left(\Prodi_{(0,t]}\{1-\Lambda_s(du|A=1,Z=z,u\geq Q)\} - \right.\right. \\
&~~~~~~~~~~~~~~~~~~~~~~~~~~~~~~~~~~~~~~~~\left.\left.\Prodi_{(0,t]}\{1-\Lambda_s(du|A=0,Z=z,u\geq Q)\}\right)\right]dz, 
\end{align*}
\noindent where, $f_s(z) = s\mathbbm{1}_{\Tilde{z}}(z) + (1-s)f(z)$ and $\Lambda_s(u|a,z,u\geq Q) = \int_{(0,t]}\frac{F_s(du|a,z,u\geq Q)}{R_s(u-|a,z,u\geq Q)}$. \\

\noindent We then calculate the Gateaux derivative as:
\begin{allowdisplaybreaks}
  \begin{align*}
    \frac{d\Psi(P_s)}{ds}&\Bigr|_{s=0} = \frac{d}{ds}\int \left\{f_s(z)\left[\theta(t|x)^2 - 2\theta(t|x)\left(\Prodi_{(0,t]}\{1-\Lambda_s(du|A=1,Z=z,u\geq Q)\} - \right.\right.\right.\\ &~~~~~~~~~~~~~~~~~~~~~~~~~~~~~~~~~~~~~~~~~~~~~~~~~~~~\left.\left.\left.\Prodi_{(0,t]}\{1-\Lambda_s(du|A=0,Z=z,u\geq Q)\}\right)\right]\right\}dz \Bigr|_{s=0} \\
    &= \int \frac{d}{ds}\left\{f_s(z)\left[\theta(t|x)^2 - 2\theta(t|x)\left(\Prodi_{(0,t]}\{1-\Lambda_s(du|A=1,Z=z,u\geq Q)\} - \right.\right.\right.\\ &~~~~~~~~~~~~~~~~~~~~~~~~~~~~~~~~~~~~~~~~~~~~~~~\left.\left.\left.\Prodi_{(0,t]}\{1-\Lambda_s(du|A=0,Z=z,u\geq Q)\}\right)\right]\right\}\Bigr|_{s=0}dz  \\
    &= \int \left[\mathbbm{1}_{\Tilde{z}}(z) -f(z)\right]\left\{\theta(t|x)^2 - 2\theta(t|x)\left(\Prodi_{(0,t]}\{1-\Lambda(du|A=1,Z=z,u\geq Q)\} - \right.\right.\\ &~~~~~~~~~~~~~~~~~~~~~~~~~~~~~~~~~~~~~~~~~~~~~~~~~~~~\left.\left.\Prodi_{(0,t]}\{1-\Lambda(du|A=0,Z=z,u\geq Q)\}\right)\right\} + \\ 
    &~~~~~~~~~ f(z)\frac{d}{ds}\left\{\theta(t|x)^2 - 2\theta(t|x)\left(\Prodi_{(0,t]}\{1-\Lambda_s(du|A=1,Z=z,u\geq Q)\} - \right.\right.\\ &~~~~~~~~~~~~~~~~~~~~~~~~~~~~~~~~~~~~~~~~~~~~~\left.\left.\Prodi_{(0,t]}\{1-\Lambda_s(du|A=0,Z=z,u\geq Q)\}\right)\right\}\Bigr|_{s=0}dz  \\
    &= \int \left[\mathbbm{1}_{\Tilde{z}}(z) -f(z)\right]\left\{\theta(t|x)^2 - 2\theta(t|x)\left(\Prodi_{(0,t]}\{1-\Lambda(du|A=1,Z=z,u\geq Q)\} - \right.\right.\\ &~~~~~~~~~~~~~~~~~~~~~~~~~~~~~~~~~~~~~~~~~~~~~~~~~~~~\left.\left.\Prodi_{(0,t]}\{1-\Lambda(du|A=0,Z=z,u\geq Q)\}\right)\right\} - \\ 
    & 2f(z)\theta(t|x)\frac{d}{ds}\left(\Prodi_{(0,t]}\{1-\Lambda_s(du|A=1,Z=z,u\geq Q)\} - \Prodi_{(0,t]}\{1-\Lambda_s(du|A=0,Z=z,u\geq Q)\}\right)\Bigr|_{s=0}dz  
\end{align*} 
\end{allowdisplaybreaks}

\noindent By using the chain rule and using  Theorem 8 from \cite{RN8}, we can show that,
\begin{align*}
    \frac{d}{ds}\Prodi_{(0,t]}\{1-\Lambda_s(du|A=a,Z=z,u\geq Q)\}\Bigr|_{s=0} = -S(t|a,z,t\geq Q)\int_{(0,t]}\frac{S(u-|a,z,u\geq Q)}{S(u|a,z,u\geq Q)}\frac{d}{ds}\Lambda_s(du|a,z,u\geq Q)\Bigr|_{s=0}.
\end{align*}

\noindent We then use our definition of $\Lambda_s(t|a,z,t\geq Q)$ and see
\begin{align*}
    \frac{d}{ds}\Lambda_s(t|a,z,t\geq Q)\Bigr|_{s=0} &= \frac{d}{ds}\int_{(0,t]}F_s(du|a,z,u\geq Q)R_s(u|a,z,u\geq Q)^{-1}\Bigr|_{s=0} \\
    &= \int_{(0,t]}\frac{d}{ds}F_s(du|a,z,u\geq Q)\Bigr|_{s=0}R(u|a,z,u\geq Q)^{-1} - \\
    &~~~~\int_{(0,t]}F(du|a,z,u\geq Q)R(u|a,z,u\geq Q)^{-2}\frac{d}{ds}R_s(u|a,z,u\geq Q)\Bigr|_{s=0}.
\end{align*}

\noindent Hence we have,
\begin{align*}
    \frac{d}{ds}\Lambda_s(du|a,z,u\geq Q)\Bigr|_{s=0} &= \frac{\frac{d}{ds}F_s(du|a,z,u\geq Q)\Bigr|_{s=0}}{R(u|a,z,u\geq Q)} - \frac{F(du|a,z,u\geq Q)\frac{d}{ds}R_s(u|a,z,u\geq Q)\Bigr|_{s=0}}{R(u|a,z,u\geq Q)^2}.
\end{align*}

\noindent Additionally, we have 
\begin{allowdisplaybreaks}
\begin{align*}
    \frac{d}{ds}F_s(u|a,z,u\geq Q)\Bigr|_{s=0} &= \frac{d}{ds}P_s(\tilde{T} \leq u,\Delta=1|A=a,Z=z,u\geq Q)\Bigr|_{s=0} \\ 
    &= \frac{d}{ds}P_s(T \leq u,C \leq u,T\leq C|A=a,Z=z,t\geq Q)\Bigr|_{s=0} \\
    &= \frac{d}{ds}\frac{P_s(T \leq u,C \leq u,T\leq C,A=a,Z=z,t\geq Q)}{P_s(A=a,Z=z,t\geq Q)}\Bigr|_{s=0} \\
    &= \frac{\frac{d}{ds}P_s(T \leq u,C \leq u,T\leq C,A=a,Z=z,t\geq Q)\Bigr|_{s=0}}{P(A=a,Z=z,t\geq Q)} - \\
    &~~~~~\frac{\frac{d}{ds}P_s(A=a,Z=z,t\geq Q)\Bigr|_{s=0}P(T \leq u,C \leq u,T\leq C,A=a,Z=z,t\geq Q)}{P(A=a,Z=z,t\geq Q)^2},
\end{align*}    
\end{allowdisplaybreaks}

\noindent where 
\begin{allowdisplaybreaks}
\begin{align*}
   \frac{d}{ds}P_s(&T \leq u,C \leq u,T\leq C,A=a,Z=z,u\geq Q)\Bigr|_{s=0} \\
   &= \frac{d}{ds}\int_{(0,u]}\int_{[v,\infty)}\int_{(0,u]}f_s(v,c,q,A=a,Z=z)dqdcdv\Bigr|_{s=0} \\
   &= \int_{(0,u]}\int_{[v,\infty)}\int_{(0,u]}\frac{d}{ds}f_s(v,c,q,A=a,Z=z)\Bigr|_{s=0}dqdcdv
   \\
   &= \int_{(0,u]}\int_{[v,\infty)}\int_{(0,u]}\left\{\mathbbm{1}_{\Tilde{v},\Tilde{c},\tilde{q},\Tilde{a},\Tilde{z}}(v,c,q,A=a,Z=z) - f(v,c,q,A=a,Z=z)\right\}dqdcdv \\
   &= \mathbbm{1}_{\tilde{v},\tilde{c},\tilde{q},\tilde{a},\tilde{z}}(v\leq u,c\leq u, v\leq c, u \geq q,A=a,Z=z) - \\
   &~~~~P(T \leq u,C \leq u,T\leq C,A=a,Z=z,u\geq Q),
\end{align*}
\end{allowdisplaybreaks}

\noindent and 
\begin{allowdisplaybreaks}
\begin{align*}
   \frac{d}{ds}P_s(A=a,Z=z,u\geq Q)\Bigr|_{s=0} &= \frac{d}{ds}\int_{(0,u]}f_s(q,A=a,Z=z)dq\Bigr|_{s=0} \\
   &= \int_{(0,u]}\frac{d}{ds}f_s(q,A=a,Z=z)\Bigr|_{s=0}dq
   \\
   &= \int_{(0,u]}\left\{\mathbbm{1}_{\tilde{q},\Tilde{a},\Tilde{z}}(q,A=a,Z=z) - f(q,A=a,Z=z)\right\}dq \\
   &= \mathbbm{1}_{\tilde{q},\Tilde{a},\Tilde{z}}(u \geq q,A=a,Z=z) - P(A=a,Z=z,u\geq Q).
\end{align*}
\end{allowdisplaybreaks}

\noindent Then by substituting this back into $\frac{d}{ds}F_s(u|a,z,u\geq Q)\Bigr|_{s=0}$, we get
\begin{allowdisplaybreaks}
\begin{align*}
    &\frac{d}{ds}F_s(u|a,z,u\geq Q)\Bigr|_{s=0} = \frac{\mathbbm{1}_{\Tilde{v},\Tilde{c},\tilde{q},\Tilde{a},\Tilde{z}}(v\leq u,c\leq u, v\leq c, u \geq q,A=a,Z=z)}{P(A=a,Z=z,u\geq Q)} - \\
    &~~~~~~~~~~~~~~~~~~~~~~~~~~~~~~~~~~\frac{\mathbbm{1}_{\tilde{q},\Tilde{a},\Tilde{z}}(u \geq Q,A=a,Z=z)P(T \leq u,C \leq u,T\leq C,A=a,Z=z,u\geq Q)}{P(A=a,Z=z,u\geq Q)^2} \\ 
    &= \frac{\mathbbm{1}_{\tilde{q},\Tilde{a},\Tilde{z}}(u \geq Q,A=a,Z=z)}{P(A=a,Z=z,u\geq Q)}\left\{\mathbbm{1}_{\Tilde{v},\Tilde{c}}(v\leq u,c\leq u, v\leq c) - \right.\\
    &~~~~~~~~~~~~~~~~~~~~~~~~~~~~~~~~~~~~~~~~~~\left.P(T \leq u,C \leq u,T\leq C|A=a,Z=z,u\geq Q)\right\} \\ 
    &= \frac{\mathbbm{1}_{\tilde{q},\Tilde{a},\Tilde{z}}(u \geq Q,A=a,Z=z)}{P(A=a,Z=z,u\geq Q)}\left\{\mathbbm{1}_{\Tilde{v},\Tilde{c}}(v\leq u, v\leq c) - F(u|a,z,u\geq Q)\right\},
\end{align*}    
\end{allowdisplaybreaks}

\noindent and hence 
\begin{align*}
    \frac{d}{ds}F_s(du|a,z,u\geq Q)\Bigr|_{s=0} = \frac{\mathbbm{1}_{\tilde{q},\Tilde{a},\Tilde{z}}(u \geq q,A=a,Z=z)}{P(A=a,Z=z,u\geq Q)}\left\{\mathbbm{1}_{\Tilde{v},\Tilde{c}}(v= u, v\leq c) - F(du|a,z,u\geq Q)\right\}.
\end{align*}

\noindent We also show that 
\begin{allowdisplaybreaks}
\begin{align*}
    \frac{d}{ds}R_s(u|a,z,u\geq Q)\Bigr|_{s=0} &= \frac{d}{ds}P_s(u \leq \tilde{T}|A=a,Z=z,u\geq Q)\Bigr|_{s=0} \\
    &= \frac{d}{ds}P_s(u \leq T,u\leq C|A=a,Z=z,u\geq Q)\Bigr|_{s=0} \\
    &= \frac{d}{ds}\frac{P_s(u \leq T,u\leq C,A=a,Z=z,u\geq Q)}{P_s(A=a,Z=z,u\geq Q)}\Bigr|_{s=0} \\
    &= \frac{\frac{d}{ds}P_s(u \leq T,u\leq C,A=a,Z=z,u\geq Q)\Bigr|_{s=0}}{P(A=a,Z=z,u\geq Q)} - \\
    &~~~~~\frac{\frac{d}{ds}P_s(A=a,Z=z,u\geq Q)\Bigr|_{s=0}P(u \leq T,u\leq C,A=a,Z=z,u\geq Q)}{P(A=a,Z=z,u\geq Q)^2},
\end{align*}    
\end{allowdisplaybreaks}
\noindent where
\begin{allowdisplaybreaks}
\begin{align*}
    \frac{d}{ds}&P_s(u \leq T,u\leq C,A=a,Z=z,u\geq Q)\Bigr|_{s=0} \\
   &= \frac{d}{ds}\int_{[u,\infty)}\int_{[u,\infty)}\int_{(0,u]}f_s(v,c,q,A=a,Z=z)dqdcdv\Bigr|_{s=0} \\
   &= \int_{[u,\infty)}\int_{[u,\infty)}\int_{(0,u]}\frac{d}{ds}f_s(v,c,q,A=a,Z=z)\Bigr|_{s=0}dqdcdv
   \\
   &= \int_{[u,\infty)}\int_{[u,\infty)}\int_{(0,u]}\left\{\mathbbm{1}_{\Tilde{v},\Tilde{c},\tilde{q},\Tilde{a},\Tilde{z}}(v,c,q,A=a,Z=z) - f(v,c,q,A=a,Z=z)\right\}dqdcdv \\
   &= \mathbbm{1}_{\Tilde{v},\Tilde{c},\tilde{q},\Tilde{a},\Tilde{z}}(u \leq v,u\leq c,A=a,Z=z,u\geq q) - P(u \leq T,u\leq C,A=a,Z=z,u\geq Q).
\end{align*}
\end{allowdisplaybreaks}
We can then use this to show 
\begin{allowdisplaybreaks}
\begin{align*}
    \frac{d}{ds}R_s(u|a,z,u\geq Q)\Bigr|_{s=0} &= \frac{\mathbbm{1}_{\Tilde{v},\Tilde{c},\tilde{q},\Tilde{a},\Tilde{z}}(u \leq v,u\leq c,A=a,Z=z,u\geq q)}{P(A=a,Z=z,u\geq Q)} - \\
    &~~~~~\frac{\mathbbm{1}_{\tilde{q},\Tilde{a},\Tilde{z}}(u \geq q,A=a,Z=z)P(u \leq T,u\leq C,A=a,Z=z,u\geq Q)}{P(A=a,Z=z,u\geq Q)^2}  \\
    &= \frac{\mathbbm{1}_{\tilde{q},\Tilde{a},\Tilde{z}}(u \geq q,A=a,Z=z)}{P(A=a,Z=z,u\geq Q)}\{\mathbbm{1}_{\Tilde{v},\Tilde{c}}(u \leq v,u\leq c) - \\
    &~~~~~~~~~~~~~~~~~~~~~~~~~~~~~~~~~~~~~~~~~~P(Q\leq u, u \leq T,u\leq C|A=a,Z=z,u\geq Q)\} \\
    &= \frac{\mathbbm{1}_{\tilde{q},\Tilde{a},\Tilde{z}}(u \geq q,A=a,Z=z)}{P(A=a,Z=z,u\geq Q)}\{\mathbbm{1}_{\Tilde{v},\Tilde{c}}(u \leq v,u\leq c) - R(u|a,z,u\geq Q)\}.
\end{align*}
\end{allowdisplaybreaks}

\noindent Substituting these into $\frac{d}{ds}\Lambda_s(du|a,z,u\geq Q)\Bigr|_{s=0}$ we get,
\begin{align*}
    \frac{d}{ds}\Lambda_s(du|a,z,u\geq Q)\Bigr|_{s=0} &= \frac{\frac{d}{ds}F_s(du|a,z,u\geq Q)\Bigr|_{s=0}}{R(u|a,z,u\geq Q)} - \frac{F(du|a,z,u\geq Q)\frac{d}{ds}R_s(u|a,z,u\geq Q)\Bigr|_{s=0}}{R(u|a,z,u\geq Q)^2} \\
    &= \frac{\mathbbm{1}_{\tilde{q},\Tilde{a},\Tilde{z}}(u \geq q,A=a,Z=z)}{P(A=a,Z=z,u\geq Q)R(u|a,z,u\geq Q)}\\
    &~~~\left\{\mathbbm{1}_{\Tilde{v},\Tilde{c}}(v= u, v\leq c) - \mathbbm{1}_{\Tilde{v},\Tilde{c}}(u \leq v,u\leq c)\frac{F(du|a,z,u\geq Q)}{R(u|a,z,u\geq Q)} \right\},
\end{align*}

\noindent which gives us,
\begin{align*}
    &\frac{d}{ds}\Prodi_{(0,t]}\{1-\Lambda_s(du|A=a,Z=z,u\geq Q)\}\Bigr|_{s=0} \\
    &= -S(t|a,z,t\geq Q)\int_{(0,t]}\frac{S(u-|a,z,u\geq Q)}{S(u|a,z,u\geq Q)}\frac{d}{ds}\Lambda_s(du|a,z,u\geq Q)\Bigr|_{s=0} \\
    &=-\frac{\mathbbm{1}_{\tilde{q},\Tilde{a},\Tilde{z}}(t \geq q,A=a,Z=z)S(t|a,z,t\geq Q)}{P(A=a,Z=z,t\geq Q)}\int_{(0,t]}\frac{S(u-|a,z,u\geq Q)}{S(u|a,z,u\geq Q)R(u|a,z,u\geq Q)}\\
    &~~~~~~~~~~~~~~~~~~~~~~~~~~~~~~~~~~\left\{\mathbbm{1}_{\Tilde{v},\Tilde{c}}(v= u, v\leq c) - \mathbbm{1}_{\Tilde{v},\Tilde{c}}(u \leq v,u\leq c)\frac{F(du|a,z,u\geq Q)}{R(u|a,z,u\geq Q)} \right\} \\
    &=-\frac{\mathbbm{1}_{\tilde{q},\Tilde{a},\Tilde{z}}(t \geq q,A=a,Z=z)S(t|a,z,t\geq Q)}{P(A=a,Z=z,t\geq Q)}\int_{(0,t]}\frac{S(u-|a,z,u\geq Q)P(u\geq Q|A=a,Z=z)}{S(u|a,z,u\geq Q)S_0(u-|a,z)\int_{(0,u]}G(u-|a,q,z)H(dq|a,z)}\\
    &~~~~~~~\left\{\mathbbm{1}_{\Tilde{v},\Tilde{c}}(v= u, v\leq c) - \mathbbm{1}_{\Tilde{v},\Tilde{c}}(u \leq v,u\leq c)\frac{S_0(du|a,z)\int_{(0,u]}G(u-|a,q,z)H(dq|a,z)}{S_0(u-|a,z)\int_{(0,u]}G(u-|a,q,z)H(dq|a,z)} \right\} \\
    &=-\frac{\mathbbm{1}_{\tilde{q},\Tilde{a},\Tilde{z}}(t \geq q,A=a,Z=z)S(t|a,z,t\geq Q)}{P(A=a,Z=z,t\geq Q)}\int_{(0,t]}\frac{P(u\geq Q|A=a,Z=z)}{S(u|a,z,u\geq Q)\int_{(0,u]}G(u-|a,q,z)H(dq|a,z)}\\
    &~~~~~~~\left\{\mathbbm{1}_{\Tilde{v},\Tilde{c}}(v= u, v\leq c) - \mathbbm{1}_{\Tilde{v},\Tilde{c}}(u \leq v,u\leq c)\frac{S(du|a,z,u\geq Q)\int_{(0,u]}G(u-|a,q,z)H(dq|a,z)}{S(u-|a,z,u\geq Q)\int_{(0,u]}G(u-|a,q,z)H(dq|a,z)} \right\} \\
    &=-\frac{\mathbbm{1}_{\tilde{q},\Tilde{a},\Tilde{z}}(u \geq q,A=a,Z=z)S(t|a,z,t\geq Q)P(t\geq Q|A=a,Z=z)}{P(A=a,Z=z,t\geq Q)}\\
    &~~~~~~~~~~~~~~~~~~~~~~~~~~~~~~~~~~~~~~~~~~~~\int_{(0,t]}\frac{\mathbbm{1}_{\Tilde{v},\Tilde{c}}(v= u, v\leq c) - \mathbbm{1}_{\Tilde{v},\Tilde{c}}(u \leq v,u\leq c)\frac{S(du|a,z,u\geq Q)}{S(u-|a,z,u\geq Q)}}{S(u|a,z,u\geq Q)\int_{(0,u]}G(u-|a,q,z)H(dq|a,z)}
\end{align*}

\noindent Finally, we see
\begin{allowdisplaybreaks}
  \begin{align*}
    &\frac{d\Psi(P_s)}{ds}\Bigr|_{s=0} \\
    &= \int \left[\mathbbm{1}_{\Tilde{z}}(z) -f(z)\right]\left\{\theta(t|x)^2 - 2\theta(t|x)\left(\Prodi_{(0,t]}\{1-\Lambda(du|A=1,Z=z,u\geq Q)\} - \right.\right.\\ &~~~~~~~~~~~~~~~~~~~~~~~~~~~~~~~~~~~~~~~~~~~~~~~~~~~~\left.\left.\Prodi_{(0,t]}\{1-\Lambda(du|A=0,Z=z,u\geq Q)\}\right)\right\} - \\ 
    &~~ 2f(z)\theta(t|x)\frac{d}{ds}\left(\Prodi_{(0,t]}\{1-\Lambda_s(du|A=1,Z=z,u\geq Q)\} - \Prodi_{(0,t]}\{1-\Lambda_s(du|A=0,Z=z,u\geq Q)\}\right)\Bigr|_{s=0}dz  \\
    &= \theta(t|\tilde{x})^2 - 2\theta(t|\tilde{x})\left(S(t|A=1,\tilde{z},t\geq Q) - S(t|A=0,\tilde{z},t\geq Q)\}\right) - \Psi(P) - \\ 
    &~~ \int 2f(z)\theta(t|x)\frac{d}{ds}\left(\Prodi_{(0,t]}\{1-\Lambda_s(du|A=1,Z=z,u\geq Q)\} - \Prodi_{(0,t]}\{1-\Lambda_s(du|A=0,Z=z,u\geq Q)\}\right)\Bigr|_{s=0}dz  \\ \\
    &= \theta(t|\tilde{x})^2 - 2\theta(t|\tilde{x})\left(S(t|A=1,\tilde{z},t\geq Q) - S(t|A=0,\tilde{z},t\geq Q)\}\right) - \Psi(P) + \\ 
    &~~~~~ 2\int \theta(t|x)\left\{\left(\frac{\mathbbm{1}_{\tilde{q},\Tilde{a},\Tilde{z}}(t \geq q,A=1,Z=z)S(t|A=1,z,t\geq Q)P(t\geq Q|A=1,Z=z)}{P(A=1,t\geq Q|Z=z)}\right.\right.\\
    &~~~~~~~~~~~~~~~~~~~~~~~~~~~~~~~~~~~~~~~~\left.\int_{(0,t]}\frac{\mathbbm{1}_{\Tilde{v},\Tilde{c}}(v= u, v\leq c) -\mathbbm{1}_{\Tilde{v},\Tilde{c}}(u \leq v,u\leq c)\frac{S(du|A=1,z,u\geq Q)}{S(u-|A=1,z,u\geq Q)}}{S(u|A=1,z,u\geq Q)\int_{(0,u]}G(u-|A=1,q,z)H(dq|A=1,z)}\right)- \\    &~~~~~~~~~~~~~~~~~~~~~~~~~~~\left(\frac{\mathbbm{1}_{\tilde{q},\Tilde{a},\Tilde{z}}(t \geq q,A=0,Z=z)S(t|A=0,z,t\geq Q)P(t\geq Q|A=0,Z=z)}{P(A=0,t\geq Q|Z=z)}\right.\\
    &~~~~~~~~~~~~~~~~~~~~~~~~~~~~~~~~~~~\left.\left.\int_{(0,t]}\frac{\mathbbm{1}_{\Tilde{v},\Tilde{c}}(v=u, v\leq c) -\mathbbm{1}_{\Tilde{v},\Tilde{c}}(u \leq v,u\leq c)\frac{S(du|A=0,z,u\geq Q)}{S(u-|A=0,z,u\geq Q)}}{S(u|A=0,z,u\geq Q)\int_{(0,u]}G(u-|A=0,q,z)H(dq|A=0,z)}\right)\right\}dz  \\ \\
    &= \theta(t|\tilde{x})^2 - 2\theta(t|\tilde{x})\left(S(t|A=1,\tilde{z},t\geq Q) - S(t|A=0,\tilde{z},t\geq Q)\}\right) - \Psi(P) + \\ 
    &~~~~~~~~~ 2\int \theta(t|x)\left\{\left(\frac{\mathbbm{1}_{\tilde{q},\Tilde{a},\Tilde{z}}(t \geq q,A=1,Z=z)S(t|A=1,z,t\geq Q)P(t\geq Q|A=1,Z=z)}{P(t\geq Q|A=1,Z=z)P(A=1|Z=z)}\right.\right.\\
    &~~~~~~~~~~~~~~~~~~~~~~~~~~~~~~~~\left.\int_{(0,t]}\frac{\mathbbm{1}_{\Tilde{v},\Tilde{c}}(v= u, v\leq c) -\mathbbm{1}_{\Tilde{v},\Tilde{c}}(u \leq v,u\leq c)\frac{S(du|A=1,z,u\geq Q)}{S(u-|A=1,z,u\geq Q)}}{S(u|A=1,z,u\geq Q)\int_{(0,u]}G(u-|A=1,q,z)H(dq|A=1,z)}\right)- \\    &~~~~~~~~~~~~~~~~~~~~~~~~~~\left(\frac{\mathbbm{1}_{\tilde{q},\Tilde{a},\Tilde{z}}(t \geq q,A=0,Z=z)S(t|A=0,z,t\geq Q)P(t\geq Q|A=0,Z=z)}{P(t\geq Q|A=0,Z=z)P(A=0|Z=z)}\right.\\
    &~~~~~~~~~~~~~~~~~~~~~~~~~~~~~~~~\left.\left.\int_{(0,t]}\frac{\mathbbm{1}_{\Tilde{v},\Tilde{c}}(v=u, v\leq c) -\mathbbm{1}_{\Tilde{v},\Tilde{c}}(u \leq v,u\leq c)\frac{S(du|A=0,z,u\geq Q)}{S(u-|A=0,z,u\geq Q)}}{S(u|A=0,z,u\geq Q)\int_{(0,u]}G(u-|A=0,q,z)H(dq|A=0,z)}\right)\right\}dz  \\ 
    &~~~~~~~~\text{Using (A4)} \\ \\
    &= \theta(t|\tilde{x})^2 - 2\theta(t|\tilde{x})\left(S(t|A=1,\tilde{z},t\geq Q) - S(t|A=0,\tilde{z},t\geq Q)\}\right) - \Psi(P) + \\ 
    &~~~~~~~~~ 2\int \theta(t|x)\left\{\left(\frac{\mathbbm{1}_{\tilde{q},\Tilde{a},\Tilde{z}}(t \geq q,A=1,Z=z)S(t|A=1,z,t\geq Q)}{P(A=1|Z=z)}\right.\right.\\
    &~~~~~~~~~~~~~~~~~~~~~~~~~~~~~~~~~~~~~~~~\left.\int_{(0,t]}\frac{\mathbbm{1}_{\Tilde{v},\Tilde{c}}(v= u, v\leq c) -\mathbbm{1}_{\Tilde{v},\Tilde{c}}(u \leq v,u\leq c)\frac{S(du|A=1,z,u\geq Q)}{S(u-|A=1,z,u\geq Q)}}{S(u|A=1,z,u\geq Q)\int_{(0,u]}G(u-|A=1,q,z)H(dq|A=1,z)}\right)- \\    &~~~~~~~~~~~~~~~~~~~~~~~~~~~~\left(\frac{\mathbbm{1}_{\tilde{q},\Tilde{a},\Tilde{z}}(t \geq q,A=0,Z=z)S(t|A=0,z,t\geq Q)}{P(A=0|Z=z)}\right.\\
    &~~~~~~~~~~~~~~~~~~~~~~~~~~~~~~~~~~~~~~~~\left.\left.\int_{(0,t]}\frac{\mathbbm{1}_{\Tilde{v},\Tilde{c}}(v=u, v\leq c) -\mathbbm{1}_{\Tilde{v},\Tilde{c}}(u \leq v,u\leq c)\frac{S(du|A=0,z,u\geq Q)}{S(u-|A=0,z,u\geq Q)}}{S(u|A=0,z,u\geq Q)\int_{(0,u]}G(u-|A=0,q,z)H(dq|A=0,z)}\right)\right\}dz  \\   
    &= \theta(t|\tilde{x})^2 - 2\theta(t|\tilde{x})\left(S(t|A=1,\tilde{z},t\geq Q) - S(t|A=0,\tilde{z},t\geq Q)\}\right) - \Psi(P) + \\ 
    &~~~~~~ 2\theta(t|\tilde{x})\left\{\left(\frac{\mathbbm{1}_{\Tilde{v},\tilde{q},\Tilde{a}}(v \geq q,A=1)S(t|A=1,\tilde{z},\tilde{T}\geq Q)}{P(A=1|Z=\tilde{z})}\right.\right.\\
    &~~~~~~~~~~~~~~~~~~~~~~~~~~~~~\left.\int_{(0,t]}\frac{\mathbbm{1}_{\Tilde{v},\Tilde{c}}(v= u, v\leq c) -\mathbbm{1}_{\Tilde{v},\Tilde{c}}(u \leq v,u\leq c)\frac{S(du|A=1,\tilde{z},u\geq Q)}{S(u-|A=1,\tilde{z},u\geq Q)}}{S(u|A=1,\tilde{z},u\geq Q)\int_{(0,u]}G(u-|A=1,\tilde{q},\tilde{z})H(dq|A=1,\tilde{z})}\right)- \\    &~~~~~~~~~~~~~~~~~~~~\left(\frac{\mathbbm{1}_{\tilde{q},\Tilde{a}}(t \geq q,A=0,Z=\tilde{z})S(t|A=0,\tilde{z},t\geq Q)}{P(A=0|Z=\tilde{z})}\right.\\
    &~~~~~~~~~~~~~~~~~~~~~~~~~~~~\left.\left.\int_{(0,t]}\frac{\mathbbm{1}_{\Tilde{v},\Tilde{c}}(v=u, v\leq c) -\mathbbm{1}_{\Tilde{v},\Tilde{c}}(u \leq v,u\leq c)\frac{S(du|A=0,\tilde{z},u\geq Q)}{S(u-|A=0,\tilde{z},u\geq Q)}}{S(u|A=0,\tilde{z},u\geq Q)\int_{(0,u]}G(u-|A=0,\tilde{q},\tilde{z})H(dq|A=0,\tilde{z})}\right)\right\}  \\    \\
    &= \theta(t|\tilde{x})^2 - 2\theta(t|\tilde{x})\left(S(t|A=1,\tilde{z},t\geq Q) - S(t|A=0,\tilde{z},t\geq Q)\}\right) - \Psi(P) + \\ 
    &~~~~~~~~~~ 2\theta(t|\tilde{x})\frac{\mathbbm{1}_{\tilde{q}}(t \geq q)(\mathbbm{1}_{\Tilde{a}}(A=1)-\pi(\tilde{z}))S(t|\tilde{a},\tilde{z},t\geq Q)}{(1-\pi(\tilde{z}))\pi(\tilde{z})} \\
    &~~~~~~~~~~~~~~~~~~~~~~~~~~~~~~~~~~\int_{(0,t]}\frac{\left\{\mathbbm{1}_{\Tilde{v},\Tilde{c}}(v=u, v\leq c) - \mathbbm{1}_{\Tilde{v},\Tilde{c}}(u \leq v,u\leq c)\Lambda(du|\tilde{a},\tilde{z},u\geq Q)\right\}}{S(u|\tilde{a},\tilde{z},u\geq Q)\int_{(0,u]}G(u-|\tilde{a},\tilde{q},\tilde{z})H(dq|\tilde{a},\tilde{z})},\\
\end{align*} 
\end{allowdisplaybreaks}
\noindent where, $\pi(z)=P(A=1|Z=z)$.

\section*{S3. EIF for incremental difference MSE}  \label{inc_EIF}

When carrying out Step 2 of surv-iTMLE, we note that by using $\Delta Y_{iTMLE}(t)$ as the outcome in the second stage regression, the algorithm no longer minimizes our expected loss function, but instead, minimizes the following expected loss function, 
\begin{gather*} \label{inc_loss_main}
    L_{surv-inc}(\mathcal{P}) = \E\left[\sum_{t\in\{t_1,...,\tau\}}\left(\Delta\theta(t|X) - \Delta\theta_0(t|X)\right)^2\right].
\end{gather*}
Below we outline how the targeting process used to generate surv-iTMLE's pseudo-outcomes still succeeds at setting the sample average of the de-biasing term, which arises in the increment-based EIF, to 0. 

To show that the targeted learning process defined by surv-iTMLE can be used to de-bias estimates of the incremental changes in the difference in conditional survival probabilities, we derive the EIF of the expected loss function which the pooled regression minimizes. First, consider times $\{t_0,t_1,...,t_{\tau-1},\tau\}$, where $t_0$ is time 0, $t_0=0$. Then, define the incremental change in the difference in survival probabilities as $\Delta\theta(t|X)=\theta(t|X)-\theta(t-1|X)$, and let us consider the following MSE which is minimized by the pooled regression:
\begin{align*} 
  L_{surv-inc}(\mathcal{P}) = \E\left[\sum_{t\in\{t_1,...,\tau\}}\left(\Delta\theta(t|X)-\Delta\theta_0(t|X)\right)^2\right],  
\end{align*}
Under assumptions (A1)-(A8), this expected loss function can be written as
\begin{align*}
  L_{surv-inc}(\mathcal{P}) &= \E\left[\sum_{t\in\{t_1,...,\tau\}}\left(\Delta\theta(t|X)-\E\left[\{S(t|A=1,Z,t\geq Q)-S(t|A=0,Z,t\geq Q)\} -\right.\right.\right.\\
  &~~~~~~~~~~~~~~~~~~~~~~~~~~~~~~~~~~~~~~~\left.\left.\left.\{S(t-1|A=1,Z,t\geq Q)-\left.S(t-1|A=0,Z,t\geq Q)\}\right|X\right]\right)^2\right],  
\end{align*}
and as was done in the main text, I consider the components of this expected loss function which depend on $\Delta\theta(t|X)$, written
\begin{align*}
  L_{surv-inc}(\mathcal{P}) &= \E\left[\sum_{t\in\{t_1,...,\tau\}}\Delta\theta(t|X)^2-2\Delta\theta(t|X)\left(\{S(t|A=1,Z,t\geq Q)-S(t|A=0,Z,t\geq Q)\} -\right.\right.\\
  &~~~~~~~~~~~~~~~~~~~~~~~~~~~~~~~~~~~~~~~~~~~~~~~~~~~\left.\{S(t-1|A=1,Z,t\geq Q)-\left.S(t-1|A=0,Z,t\geq Q)\}\right)\right].  
\end{align*}
By doing so I can derive this EIF for this expected loss function by perturbing $P$ in the direction parameterized via the one-dimensional mixture model
$$P_s = s\Tilde{P} + (1-s)P,$$
where $\Tilde{P}$ is a fixed,
deterministic distribution with its support contained in the support of $P$. By perturbing $P$
in the direction of a point mass at $(\Tilde{z},\Tilde{a},\Tilde{c},\Tilde{v},\Tilde{q})$, I get:
\begin{align*}
L_{surv-inc}(\mathcal{P}_s) &= \int f_s(z)\sum_{t\in\{t_1,...,\tau\}}\left[\Delta\theta(t|x)^2 -2\Delta\theta(t|x)\left(\{S_s(t|A=1,Z,t\geq Q)-S_s(t|A=0,Z,t\geq Q)\} -\right.\right.\\
  &~~~~~~~~~~~~~~~~~~~~~~~~~~~~~~~~~~~~~~~~~~~\left.\{S_s(t-1|A=1,Z,t\geq Q)-\left.S_s(t-1|A=0,Z,t\geq Q)\}\right)\right]dz, 
\end{align*}
\noindent where, $f_s(z) = s\mathbbm{1}_{\Tilde{z}}(z) + (1-s)f(z)$. \\

\noindent I then calculate the Gateaux derivative as:
{\small
\begin{allowdisplaybreaks}
\begin{align*}
&\frac{dL_{surv-inc}(\mathcal{P}_s)}{ds}\Bigr|_{s=0} \\
&= \frac{d}{ds}\int f_s(z)\sum_{t\in\{t_1,...,\tau\}}\left[\Delta\theta(t|x)^2 -2\Delta\theta(t|x)\left(\{S_s(t|A=1Z=z,t\geq Q)-S_s(t|A=0Z=z,t\geq Q)\} -\right.\right.\\
  &~~~~~~~~~~~~~~~~~~~~~~~~~~~~~~~~~~~~~~~~~~~~~~~~~~~~~~~~~~~~~\left.\{S_s(t-1|A=1Z=z,t\geq Q)-\left.S_s(t-1|A=0Z=z,t\geq Q)\}\right)\right]dz \Bigr|_{s=0} \\
  &= \sum_{t\in\{t_1,...,\tau\}}\left[\Delta\theta(t|x)^2 -2\Delta\theta(t|x)\left(\{S(t|A=1Z=z,t\geq Q)-S(t|A=0Z=z,t\geq Q)\} -\right.\right.\\
  &~~~~~~~~~~~~~~~~~~~~~~~~~~~~~~~~~~~~~~~~~~~~~\left.\left.\{S(t-1|A=1Z=z,t\geq Q)-S(t-1|A=0Z=z,t\geq Q)\}\right)\right] - L_{surv-inc}(\mathcal{P}) + \\
  &~~~\int f(z)\sum_{t\in\{t_1,...,\tau\}}\frac{d}{ds}\left[\Delta\theta(t|x)^2 -2\Delta\theta(t|x)\left(\{S_s(t|A=1Z=z,t\geq Q)-S_s(t|A=0Z=z,t\geq Q)\} -\right.\right.\\
  &~~~~~~~~~~~~~~~~~~~~~~~~~~~~~~~~~~~~~~~~~~~~~~~~~~~~~~~~~~~\left.\{S_s(t-1|A=1Z=z,t\geq Q)-\left.S_s(t-1|A=0Z=z,t\geq Q)\}\right)\right]\Bigr|_{s=0}dz \\
  &= \sum_{t\in\{t_1,...,\tau\}}\left[\Delta\theta(t|x)^2 -2\Delta\theta(t|x)\left(\{S(t|A=1Z=z,t\geq Q)-S(t|A=0Z=z,t\geq Q)\} -\right.\right.\\
  &~~~~~~~~~~~~~~~~~~~~~~~~~~~~~~~~~~~~~~~~~~~~~\left.\left.\{S(t-1|A=1Z=z,t\geq Q)-S(t-1|A=0Z=z,t\geq Q)\}\right)\right] - L_{surv-inc}(\mathcal{P}) - \\
  &~~~\int f(z)\sum_{t\in\{t_1,...,\tau\}}\left[2\Delta\theta(t|x)\frac{d}{ds}\left(\{S_s(t|A=1Z=z,t\geq Q)-S_s(t|A=0Z=z,t\geq Q)\} -\right.\right.\\
  &~~~~~~~~~~~~~~~~~~~~~~~~~~~~~~~~~~~~~~~~~~~~\left.\{S_s(t-1|A=1Z=z,t\geq Q)-\left.S_s(t-1|A=0Z=z,t\geq Q)\}\right)\Bigr|_{s=0}\right]dz \\
  &= \sum_{t\in\{t_1,...,\tau\}}\left[\Delta\theta(t|x)^2 -2\Delta\theta(t|x)\left(\{S(t|A=1Z=z,t\geq Q)-S(t|A=0Z=z,t\geq Q)\} -\right.\right.\\
  &~~~~~~~~~~~~~~~~~~~~~~~~~~~~~~~~~~~~~~~~~~~~~\left.\left.\{S(t-1|A=1Z=z,t\geq Q)-S(t-1|A=0Z=z,t\geq Q)\}\right)\right] - L_{surv-inc}(\mathcal{P}) - \\
  &~~~\int f(z)\sum_{t\in\{t_1,...,\tau\}}\left[2\Delta\theta(t|x)\left(\frac{d}{ds}\{S_s(t|A=1Z=z,t\geq Q)-S_s(t|A=0Z=z,t\geq Q)\}\Bigr|_{s=0} -\right.\right.\\
  &~~~~~~~~~~~~~~~~~~~~~~~~~~~~~~~~~~~~~~~~~\left.\frac{d}{ds}\{S_s(t-1|A=1Z=z,t\geq Q)-\left.S_s(t-1|A=0Z=z,t\geq Q)\}\Bigr|_{s=0}\right)\right]dz.
\end{align*}
\end{allowdisplaybreaks}
}
where,
\begin{allowdisplaybreaks}
\begin{align*}
&\int f(z)\sum_{t\in\{t_1,...,\tau\}}\left[2\Delta\theta(t|x)\left(\frac{d}{ds}\{S_s(t|A=1Z=z,t\geq Q)-S_s(t|A=0Z=z,t\geq Q)\}\Bigr|_{s=0} -\right.\right.\\
  &~~~~~~~~~~~~~~~~~~~~~~~~~~~~~~~~~~~~~~\left.\frac{d}{ds}\{S_s(t-1|A=1Z=z,t\geq Q)-\left.S_s(t-1|A=0Z=z,t\geq Q)\}\Bigr|_{s=0}\right)\right]dz \\
&=\sum_{t\in\{t_1,...,\tau\}}2\Delta\theta(t|x)\frac{(\mathbbm{1}_{\Tilde{a}}(A=1)-\pi(\tilde{z}))\mathbbm{1}(t\geq Q)S(t|\tilde{a},\tilde{z},t\geq Q)}{(1-\pi(\tilde{z}))\pi(\tilde{z})} \\
&~~~~~~~~~~~~~~~~~~~~~~~~~~~~~~~~\int_{(0,t]}\frac{\left\{\mathbbm{1}_{\Tilde{v},\Tilde{c}}(v=u, v\leq c) - \mathbbm{1}_{\Tilde{v},\Tilde{c}}(u \leq v,u\leq c)\Lambda(du|\tilde{a},\tilde{z},t\geq Q)\right\}}{S(u|\tilde{a},\tilde{z},t\geq Q)\int_{(0,u]}G(u-|\tilde{a},\tilde{q},\tilde{z})H(dq|\tilde{a},\tilde{z})} \\
&~~~~~~~~~~~~~~~2\Delta\theta(t-1|x)\frac{(\mathbbm{1}_{\Tilde{a}}(A=1)-\pi(\tilde{z}))\mathbbm{1}(t\geq Q)S(t-1|\tilde{a},\tilde{z})}{(1-\pi(\tilde{z}))\pi(\tilde{z})} \\
&~~~~~~~~~~~~~~~~~~~~~~~~~~~~~~~~\int_{(0,t-1]}\frac{\left\{\mathbbm{1}_{\Tilde{v},\Tilde{c}}(v=u, v\leq c) - \mathbbm{1}_{\Tilde{v},\Tilde{c}}(u \leq v,u\leq c)\Lambda(du|\tilde{a},\tilde{z},t\geq Q)\right\}}{S(u|\tilde{a},\tilde{z},t\geq Q)\int_{(0,u]}G(u-|\tilde{a},\tilde{q},\tilde{z})H(dq|\tilde{a},\tilde{z})}.
\end{align*}
\end{allowdisplaybreaks}
This allows the full EIF to be written as 
{\small
\begin{allowdisplaybreaks}
\begin{align*}
&\frac{dL_{surv-inc}(\mathcal{P}_s)}{ds}\Bigr|_{s=0} \\
&= \sum_{t\in\{t_1,...,\tau\}}\left[\Delta\theta(t|x)^2 -2\Delta\theta(t|x)\left(\{S(t|A=1Z=z,t\geq Q)-S(t|A=0Z=z,t\geq Q)\} -\right.\right.\\
&~~~~~~~~~~~~~~~~~~~~~~~~~~~~~~~~~~~~~~~~~~~~~\left.\left.\{S(t-1|A=1Z=z,t\geq Q)-S(t-1|A=0Z=z,t\geq Q)\}\right)\right] - L_{surv-inc}(\mathcal{P}) - \\
&\sum_{t\in\{t_1,...,\tau\}}2\Delta\theta(t|x)\frac{(\mathbbm{1}_{\Tilde{a}}(A=1)-\pi(\tilde{z}))\mathbbm{1}(t\geq Q)S(t|\tilde{a},\tilde{z},t\geq Q)}{(1-\pi(\tilde{z}))\pi(\tilde{z})} \\
&~~~~~~~~~~~~~~~~~~~~~~~~~~~~~~~~\int_{(0,t]}\frac{\left\{\mathbbm{1}_{\Tilde{v},\Tilde{c}}(v=u, v\leq c) - \mathbbm{1}_{\Tilde{v},\Tilde{c}}(u \leq v,u\leq c)\Lambda(du|\tilde{a},\tilde{z},t\geq Q)\right\}}{S(u|\tilde{a},\tilde{z},t\geq Q)\int_{(0,u]}G(u-|\tilde{a},\tilde{q},\tilde{z})H(dq|\tilde{a},\tilde{z})} \\
&~~~~~~~~~~~~~2\Delta\theta(t-1|x)\frac{(\mathbbm{1}_{\Tilde{a}}(A=1)-\pi(\tilde{z}))\mathbbm{1}(t\geq Q)S(t-1|\tilde{a},\tilde{z},t\geq Q)}{(1-\pi(\tilde{z}))\pi(\tilde{z})} \\
&~~~~~~~~~~~~~~~~~~~~~~~~~~~~~~~~\int_{(0,t-1]}\frac{\left\{\mathbbm{1}_{\Tilde{v},\Tilde{c}}(v=u, v\leq c) - \mathbbm{1}_{\Tilde{v},\Tilde{c}}(u \leq v,u\leq c)\Lambda(du|\tilde{a},\tilde{z},t\geq Q)\right\}}{S(u|\tilde{a},\tilde{z},t\geq Q)\int_{(0,u]}G(u-|\tilde{a},\tilde{q},\tilde{z})H(dq|\tilde{a},\tilde{z})}.
\end{align*}
\end{allowdisplaybreaks}
}

\noindent From this expected loss function I point out the de-biasing term in this setting is now defined as
\begin{allowdisplaybreaks}
\begin{align*} 
&\sum_{t\in\{t_1,...,\tau\}}2\Delta\theta(t|x)\frac{(\mathbbm{1}_{\Tilde{a}}(A=1)-\pi(\tilde{z}))\mathbbm{1}(t\geq Q)S(t|\tilde{a},\tilde{z},t\geq Q)}{(1-\pi(\tilde{z}))\pi(\tilde{z})} \\
&~~~~~~~~~~~~~~~~~~~~~~~~~~~~~~~~\int_{(0,t]}\frac{\left\{\mathbbm{1}_{\Tilde{v},\Tilde{c}}(v=u, v\leq c) - \mathbbm{1}_{\Tilde{v},\Tilde{c}}(u \leq v,u\leq c)\Lambda(du|\tilde{a},\tilde{z},t\geq Q)\right\}}{S(u|\tilde{a},\tilde{z},t\geq Q)\int_{(0,u]}G(u-|\tilde{a},\tilde{q},\tilde{z})H(dq|\tilde{a},\tilde{z})} \\
&~~~~~~~~~~~~~2\Delta\theta(t-1|x)\frac{(\mathbbm{1}_{\Tilde{a}}(A=1)-\pi(\tilde{z}))\mathbbm{1}(t\geq Q)S(t-1|\tilde{a},\tilde{z},t\geq Q)}{(1-\pi(\tilde{z}))\pi(\tilde{z})} \\
&~~~~~~~~~~~~~~~~~~~~~~~~~~~~~~~~\int_{(0,t-1]}\frac{\left\{\mathbbm{1}_{\Tilde{v},\Tilde{c}}(v=u, v\leq c) - \mathbbm{1}_{\Tilde{v},\Tilde{c}}(u \leq v,u\leq c)\Lambda(du|\tilde{a},\tilde{z},t\geq Q)\right\}}{S(u|\tilde{a},\tilde{z},t\geq Q)\int_{(0,u]}G(u-|\tilde{a},\tilde{q},\tilde{z})H(dq|\tilde{a},\tilde{z})}.
\end{align*}
\end{allowdisplaybreaks}
When constructing a targeted learning based estimator, the objective is to set the sample average of this de-biasing term to 0. However, I note that the de-biasing in the above equation is simply a sum over $t$ of the de-biasing terms which were defined by surv-iTMLE in the main text, now defined below.
\begin{align*} \label{surv-iTMLE_RC_debiasing_term2}
    2\theta(t|X)\frac{(A-\pi(Z))}{(1-\pi(Z))\pi(Z)}\mathbbm{1}(t\geq Q)S(t|A,Z,t\geq Q)\int_{(0,t]}\frac{\left\{\mathbbm{1}(\tilde{T}=u, \Delta=1) - \mathbbm{1}(u \leq \tilde{T})\Lambda(du|A,Z,t\geq Q)\right\}}{S(u|A,Z,t\geq Q)\int_{(0,u]}G(u-|A,Q,Z)H(dq|A,Z)}.
\end{align*}
Consequently, as surv-iTMLE works to set the de-biasing term above to 0 for all $t\in\{t_1,...,\tau\}$, that same targeting process could be used to de-bias the estimates of the incremental changes in the difference in conditional survival probabilities. However, I note that in the de-biasing terms, the estimand which is conditional on $X$ are different, defined as $\Delta\theta(t|X)$ in the incremental case, and defined as $\theta(t|X)$ in orginal setting. Even so, this is not a problem for surv-iTMLE as it's sieve approximation is invariant to the specific function, but simply allows the updating step to produce updates which are conditional on $X$.

\newpage
\section*{S4. Machine learning algorithm use for time-to-event data} \label{App_ML_TTE}

When estimating the survival probabilities using left truncated, or right censored event times, a range of estimation options exist. Estimation options which handle left truncated and right censored event times include individual ML algorithms, such as an adaptive Cox lasso \citep{RN16}, or ensemble approaches, such as the survival stacking options provided by \cite{RN17}. Meanwhile, a wider array of estimation options are available when event times are only right censored, with alternative ensemble learners available, such as the \textit{survSuperLearner} \citep{RN2}. In this work, for computational efficiency reasons we utilize a ``local'' survival stacking approach which breaks down the survival problem down into a series of binary outcome problems, generating estimates of the hazard at each time interval, then using these to obtain survival estimates. We utilize the \textit{SuperLearner}, an ensemble learner to estimate the binary outcome problems within this approach. This said, we highlight that ``global'' survival stacking, which jointly models the entire time horizon using a single meta-learner may offer improved stability and performance \citep{RN17}, but its existing implementation requires significant computing time, and hence was deemed infeasible for our analyses.

\section*{S5. LTRC simulation study DGP specifications} \label{App_sim_specs}

The LTRC simulation study was designed by drawing upon simulation studies from previous papers in this area, including design features from \cite{RN9} and \cite{RN5}. In each DGP, we generated 20 covariates, $Z=\{Z_1,...,Z_{20}\}$, uniformly, $Z \sim \text{Unif}(0,1)^{20}$, and treatment was generated using the propensity score $\pi(Z)$; $A|Z \sim \text{Bernoulli}(\pi(Z))$, with $\pi(Z)=(1+\beta(Z_1;2,4))/4$, where $\beta(.;a,b)$ is the density function of a Beta distribution with shape parameters $a$ and $b$. Event times, $T$, were generated for each DGP using either an accelerated failure time model (DGP1), or a Cox proportional hazards model (DG2 and DGP3). Event times for DGP1 were obtained from the following accelerated failure time model, 
$$log(T) = -1.85 -(0.8(Z_1<0.5)) + (0.7(Z_2^{1/2})) + (0.2Z_3) + \left(0.7-(0.4(Z_1<0.5))-(0.4(Z_2^{1/2}))\right)A +\epsilon_T$$
where $\epsilon_T\sim N(0,1)$. Meanwhile, event times for DGP2 and DGP3 were obtained by Cox models, with $\lambda(t|A,Z)$ defined as: 
$$\lambda(t|A,Z)=\lambda_0(t)exp[0.25 + (0.5(Z_1^{1/3})) - ((1.5((0.5-Z_2)^3))-(0.4(Z_3^{1/2})))A],~\lambda_0(t)=\frac{1}{2}t^{-1/2}~~~(\text{DGP2})$$
and
$$\lambda(t|A,Z)=\lambda_0(t)exp[-0.75 -((0.75-Z_1)^3) + (0.3(Z_2^{1/2})) + (0.2Z_3)],~\lambda_0(t)=2t~~~(\text{DGP3}).$$

For each DGP, two levels of left truncation were introduced. Truncation times, $Q$, were generated by defining $Q= U\cdot Q_{max}$, where $U$ is a Beta random variable with parameters $a$ and $b$, and $Q_{max}$ is the maximum length a truncation time can take. All three DGPs define $a=2+\left(2Z_1^2 + Z^2\right)A + \epsilon_a$, where $\epsilon_a\sim N(0,0.1)$, and define $b=15+Z_2^3+0.5(Z_3>0.5)+\epsilon_b$, where $\epsilon_b\sim N(0,0.5)$. We note here that all errors are independent of the previously introduced variables. Differing levels of truncation are then achieved by varying $Q_{max}$, with DGP1 defining $Q_{max}=0.8$ and $2$; DGP2 defining $Q_{max}=0.1$ and $0.6$; while DGP3 defines $Q_{max}=4$ and $Q_{max}=7$. Follow-up time horizons for DGP1-DGP3 were defined as 2, 1, and 2 respectively. Censoring times for each DGP were generated as $C=Q+D$, where $D$ was generated using a Cox model in DG1, define as:
$$\lambda(t|A,Z)=\lambda_0(t)exp[-1.75 - (0.5(Z_2^{1/2})) + (0.2Z_3) + \left(1.15+(0.5(Z_1<0.5))-(0.3(Z_2^{1/2}))\right)A],$$

\noindent with $\lambda_0(t)=2t$, or using accelerated failure time models, define as follows for DGP2 and DGP3 respectively: 
$$log(T) = -0.75 -(1.5(0.5-Z_1)^3) + (0.5(Z_2^{1/2})) + (0.3Z_3) + \left(0.2-((0.5-Z_1)^3)-(0.4*(Z_2^{1/2}))\right)A +\epsilon_C,$$
and 
$$log(T) = 0.25 + \left(0.2-(((0.5-Z_1)^3))-(0.8(Z_2^{1/2})) + (0.6X3)\right)A +\epsilon_C,$$
where $\epsilon_C\sim N(0,1)$. Using these measurements, the observed event times $\tilde{T}$ were calculated as $\tilde{T}=\min\{T,C\}$, with an indicator of being observed in the study defined as $\mathbbm{1}\{T>C\}$. Finally, ground truth for the difference in conditional survival probabilities was obtained by simulating 1,000,000 event times for each person in the test dataset, under each exposure. These times were then used to identify the survival probability for each person at each time, conditional on all of the baseline covariates $Z$ and over the follow-up horizon under each exposure.

\newpage
\section*{S6. Algorithm libraries for Local survival stacking, SuperLearner and xgboost models} \label{App_sim_SL_libs}

\begin{table}[ht]
\caption{Super learner algorithm libraries used for binary outcome models (including those in local survival stacking (LSS) algorithms)}
\begin{center}
\begin{tabular}{ll}
\hline
\textbf{Algorithm}   & \textbf{Tuning parameters}     \\ \hline
Mean (SL.mean)     & .     \\
& \\
Linear model (SL.glm)     & .    \\
& \\
LASSO/Elastic net (SL.glmnet with logit link) & nlambda = (50,100,250)\\ 
 & alpha = (0.5,1)       \\
 & useMin = (False,True) \\
 & \\
Random forest (SL.ranger)     & mtry = (3,5)\\ 
& min.node.size = (10,20)      \\ 
& sample.fraction = (0.2,0.4,0.6) \\ \hline                          
\end{tabular}
\end{center}
\end{table}

\begin{table}[ht]
\caption{Super learner algorithm libraries used for continuous outcome models}
\begin{center}
\begin{tabular}{ll}
\hline
\textbf{Algorithm}   & \textbf{Tuning parameters}     \\ \hline
Mean (SL.mean)     & .     \\
& \\
Linear model (SL.lm)     & .    \\
& \\
LASSO/Elastic net (SL.glmnet with identity link) & nlambda = (50,100,250)\\ 
 & alpha = (0.5,1)       \\
 & useMin = (False,True) \\
 & \\
Random forest (SL.ranger)     & mtry = (3,5)\\ 
& min.node.size = (10,20)      \\ 
& sample.fraction = (0.2,0.4,0.6) \\   \\      \hline                          
\end{tabular}
\end{center}
\end{table}

\noindent When using XGBoost for the final minimization problem in the ltrc-R and ltrc-DR learners, the cross-validated function found at \href{https://github.com/wangyuyao98/truncAC}{https://github.com/wangyuyao98/truncAC}  was used, with maximum numbers of trees set to 500 and number of search rounds of the tuning parameters set to 5. This implementation can be found at \href{https://github.com/Matt-Pryce/surv-iTMLE}{https://github.com/Matt-Pryce/surv-iTMLE}.

\newpage
\section*{S7. Additional LTRC simulation study results} \label{App_sim_results}

\begin{table}[ht]
\centering
\caption{Left truncated and right censored (LTRC) simulation study - Difference in survival probabilities - Mean root mean squared error (RMSE) for the T-learner, CSFs and surv-iTMLE by sample size} 
\vspace{1em} 
\begin{sideways}
\resizebox{1.1\textwidth}{!}{%
\begin{tabular}{ccccccccccccccc}
\hline
DGP & Truncation & N &  & T-learner & & CSFs & & surv-iTMLE &  & \multicolumn{2}{c}{ltrc-R} &  & \multicolumn{2}{c}{ltrc-DR}    \\
    &  level  &    &  &  &  &  & & &  &  \multicolumn{1}{c}{LSS/SuperLearner} & \multicolumn{1}{c}{pcox/gbm} &  & \multicolumn{1}{c}{LSS/SuperLearner} & \multicolumn{1}{c}{pcox/gbm}    \\\hline
1    & $\sim 25\%$  & 800    &   & 0.091 &  & 0.060 &  & \textbf{0.057} & & 0.118 & 0.099 & & 0.138 & 0.201    \\
     &              & 1600   &   & 0.082 &  & 0.054 &  & \textbf{0.049} & & 0.100 & 0.087 & & 0.114 & 0.138   \\
     &              & 2400   &   & 0.079 &  & 0.051 &  & \textbf{0.043} & & 0.091 & 0.082 & & 0.104 & 0.115   \\
     & $\sim 50\%$  & 800    &   & 0.101 &  & 0.083 &  & \textbf{0.068} & & 0.145 & 0.103 & & 0.161 & 0.253   \\
     &              & 1600   &   & 0.087 &  & 0.080 &  & \textbf{0.056} & & 0.129 & 0.096 & & 0.141 & 0.178   \\
     &              & 2400   &   & 0.084 &  & 0.078 &  & \textbf{0.052} & & 0.128 & 0.091 & & 0.138 & 0.144  \\
     &              &        &   &  &  &  &  &  & & & & & &   \\
2    & $\sim 25\%$  & 800    &   & 0.083 &  & 0.066 &  & \textbf{0.061} & & 0.160 & 0.138 & & 0.199 & 0.257   \\
     &              & 1600   &   & 0.068 &  & 0.056 &  & \textbf{0.050} & & 0.150 & 0.137 & & 0.176 & 0.199  \\
     &              & 2400   &   & 0.064 &  & 0.052 &  & \textbf{0.045} & & 0.143 & 0.136 & & 0.162 & 0.174   \\
     & $\sim 50\%$  & 800    &   & 0.109 &  & 0.091 &  & \textbf{0.083} & & 0.224 & 0.192 & & 0.253 & 0.348  \\
     &              & 1600   &   & 0.093 &  & 0.083 &  & \textbf{0.073} & & 0.216 & 0.200 & & 0.231 & 0.274   \\
     &              & 2400   &   & 0.086 &  & 0.078 &  & \textbf{0.067} & & 0.208 & 0.201 & & 0.217 & 0.244   \\
     &              &        &   &  &  &  &  &  & & & & & &   \\
3    & $\sim 25\%$  & 800    &   & 0.071 &  & 0.062 &  & \textbf{0.049} & & 0.174 & 0.141 & & 0.192 & 0.282   \\
     &              & 1600   &   & 0.059 &  & 0.057 &  & \textbf{0.040} & & 0.147 & 0.132 & & 0.157 & 0.199  \\
     &              & 2400   &   & 0.053 &  & 0.052 &  & \textbf{0.035} & & 0.133 & 0.120 & & 0.141 & 0.167   \\
     & $\sim 50\%$  & 800    &   & 0.076 &  & 0.074 &  & \textbf{0.066} & & 0.229 & 0.152 & & 0.234 & 0.342  \\
     &              & 1600   &   & 0.063 &  & 0.072 &  & \textbf{0.049} & & 0.210 & 0.146 & & 0.208 & 0.248  \\
     &              & 2400   &   & 0.056 &  & 0.065 &  & \textbf{0.042} & & 0.197 & 0.139 & & 0.195 & 0.212   \\
 \hline
  \multicolumn{15}{l}{\small\textit{DGP = Data generating process; N = Training sample size; CSFs = Causal Survival Forests; LSS = Local Survival Stacking;}} \\
  \multicolumn{15}{l}{\small\textit{pcox = penalized cox regression; gbm = gradient boosting machine.}}
\end{tabular}
}
\end{sideways}
\end{table}

\newpage
\begin{table}[ht]
\centering
\caption{Left truncated and right censored (LTRC) simulation study - Data generating process 1 (N=2400) - Difference in survival probabilities - Mean root mean squared error (RMSE) by time for the T-learner, CSFs and surv-iTMLE} 
\vspace{1em} 
\begin{sideways}
\resizebox{1.1\textwidth}{!}{%
\begin{tabular}{ccclcccccccccccc}
\hline
DGP & Truncation & N & Time & & T-learner & & CSFs & & surv-iTMLE &  & \multicolumn{2}{c}{ltrc-R} &  & \multicolumn{2}{c}{ltrc-DR}    \\
    &  level  &  &   &  &  &  &  & & &  &  \multicolumn{1}{c}{LSS/SuperLearner} & \multicolumn{1}{c}{pcox/gbm} &  & \multicolumn{1}{c}{LSS/SuperLearner} & \multicolumn{1}{c}{pcox/gbm}    \\\hline
1    & $\sim 25\%$  & 2400 & 0.1    &   & 0.074 &  & 0.055 &  & \textbf{0.049} & & 0.177 & 0.146 & & 0.187 & 0.201    \\
     &              &      & 0.25   &   & 0.111 &  & 0.092 &  & \textbf{0.069} & & 0.129 & 0.128 & & 0.139 & 0.167    \\
     &              &      & 0.5    &   & 0.106 &  & 0.067 &  & \textbf{0.062} & & 0.091 & 0.092 & & 0.107 & 0.127    \\
     &              &      & 0.75   &   & 0.086 &  & 0.049 &  & \textbf{0.045} & & 0.073 &  0.071& & 0.090 & 0.104    \\
     &              &      & 1.0    &   & 0.073 &  & 0.040 &  & \textbf{0.032} & & 0.061 &  0.056& & 0.082 & 0.091    \\
     &              &      & 1.25   &   & 0.064 &  & 0.033 &  & \textbf{0.025} & & 0.053 & 0.047 & & 0.074 & 0.078    \\
     &              &      & 1.5    &   & 0.062 &  & 0.027 &  & \textbf{0.020} & & 0.047 & 0.039 & & 0.067 & 0.066    \\
     &              &      & 1.75   &   & 0.059 &  & 0.024 &  & \textbf{0.020} & & 0.041 & 0.033 & & 0.059 & 0.051    \\
     &              &      & 2.0    &   & 0.059 &  & \textbf{0.019} &  & 0.021 & & 0.026 & 0.026 & & 0.026 & 0.026    \\
     &              &      &    &   &   &  &  &  &  & & & & & &     \\
     & $\sim 50\%$  & 2400 & 0.1    &   & 0.079 &  & \textbf{0.057} &  & 0.061 & & 0.233 & 0.172 & & 0.237 & 0.259    \\
     &              &      & 0.25   &   & 0.114 &  & 0.103 &  & \textbf{0.088} & & 0.246 & 0.146 & & 0.257 & 0.231    \\
     &              &      & 0.5    &   & 0.109 &  & 0.130 &  & \textbf{0.071} & & 0.115 & 0.091 & & 0.130 & 0.146    \\
     &              &      & 0.75   &   & 0.090 &  & 0.098 &  & \textbf{0.054} & & 0.070 & 0.069 & & 0.092 & 0.112    \\
     &              &      & 1.0    &   & 0.078 &  & 0.075 &  & \textbf{0.040} & & 0.058 & 0.055 & & 0.079 & 0.095    \\
     &              &      & 1.25   &   & 0.070 &  & 0.057 &  & \textbf{0.031} & & 0.051 & 0.045 & & 0.072 & 0.079    \\
     &              &      & 1.5    &   & 0.068 &  & 0.046 &  & \textbf{0.025} & & 0.044 & 0.038 & & 0.065 & 0.068    \\
     &              &      & 1.75   &   & 0.066 &  & 0.038 &  & \textbf{0.022} & & 0.035 & 0.032 & & 0.056 & 0.059    \\
     &              &      & 2.0    &   & 0.066 &  & 0.023 &  & \textbf{0.022} & & 0.027 & 0.026 & & 0.026 & 0.026    \\
 \hline
  \multicolumn{15}{l}{\small\textit{DGP = Data generating process; N = Training sample size; CSFs = Causal Survival Forests; LSS = Local Survival Stacking;}} \\
  \multicolumn{15}{l}{\small\textit{pcox = penalized cox regression; gbm = gradient boosting machine.}}
\end{tabular}
}
\end{sideways}
\end{table}

\newpage
\begin{table}[ht]
\centering
\caption{Left truncated and right censored (LTRC) simulation study - Data generating process 2 (N=2400) - Difference in survival probabilities - Mean root mean squared error (RMSE) by time for the T-learner, CSFs and surv-iTMLE} 
\vspace{1em} 
\begin{sideways}
\resizebox{1.1\textwidth}{!}{%
\begin{tabular}{ccclcccccccccccc}
\hline
DGP & Truncation & N & Time & & T-learner & & CSFs & & surv-iTMLE &  & \multicolumn{2}{c}{ltrc-R} &  & \multicolumn{2}{c}{ltrc-DR}    \\
    &  level  &  &   &  &  &  &  & & &  &  \multicolumn{1}{c}{LSS/SuperLearner} & \multicolumn{1}{c}{pcox/gbm} &  & \multicolumn{1}{c}{LSS/SuperLearner} & \multicolumn{1}{c}{pcox/gbm}    \\\hline
2    & $\sim 25\%$  & 2400 & 0.1    &   & 0.068 &  & 0.064 &  & \textbf{0.055} & & 0.206 & 0.218 & & 0.211 & 0.248     \\
     &              &      & 0.2    &   & 0.070 &  & 0.054 &  & \textbf{0.051} & & 0.181 & 0.185 & & 0.193 & 0.216     \\
     &              &      & 0.3    &   & 0.069 &  & 0.049 &  & \textbf{0.048} & & 0.161 & 0.158 & & 0.179 & 0.195     \\
     &              &      & 0.4    &   & 0.066 &  & 0.047 &  & \textbf{0.046} & & 0.151 & 0.137 & & 0.171 & 0.185     \\
     &              &      & 0.5    &   & 0.065 &  & 0.045 &  & \textbf{0.043} & & 0.137 & 0.120 & & 0.164 & 0.174     \\
     &              &      & 0.6    &   & 0.063 &  & 0.043 &  & \textbf{0.042} & & 0.130 & 0.108 & & 0.157 & 0.161     \\
     &              &      & 0.7    &   & 0.061 &  & 0.042 &  & \textbf{0.040} & & 0.118 & 0.097 & & 0.151 & 0.150     \\
     &              &      & 0.8    &   & 0.060 &  & 0.041 &  & \textbf{0.039} & & 0.109 & 0.087 & & 0.142 & 0.137     \\
     &              &      & 0.9    &   & 0.060 &  & 0.041 &  & \textbf{0.037} & & 0.100 & 0.079 & & 0.127 & 0.115     \\
     &              &      & 1      &   & 0.060 &  & 0.071 &  & \textbf{0.036} & & 0.070 & 0.070 & & 0.070 & 0.070     \\
     &              &      &    &   &   &  &  &  &  & & & & & &     \\
     & $\sim 50\%$  & 2400 & 0.1    &   & \textbf{0.091} &  & 0.129 &  & 0.118 & & 0.361 & 0.356 & & 0.346 & 0.391     \\
     &              &      & 0.2    &   & 0.094 &  & 0.114 &  & \textbf{0.088} & & 0.280 & 0.291 & & 0.280 & 0.333     \\
     &              &      & 0.3    &   & 0.093 &  & 0.083 &  & \textbf{0.067} & & 0.235 & 0.240 & & 0.238 & 0.282     \\
     &              &      & 0.4    &   & 0.089 &  & 0.067 &  & \textbf{0.059} & & 0.205 & 0.199 & & 0.217 & 0.251     \\
     &              &      & 0.5    &   & 0.087 &  & 0.058 &  & \textbf{0.054} & & 0.186 & 0.168 & & 0.203 & 0.223     \\
     &              &      & 0.6    &   & 0.085 &  & 0.053 &  & \textbf{0.051} & & 0.166 & 0.144 & & 0.189 & 0.205     \\
     &              &      & 0.7    &   & 0.083 &  & 0.050 &  & \textbf{0.048} & & 0.149 & 0.123 & & 0.176 & 0.185    \\
     &              &      & 0.8    &   & 0.081 &  & 0.049 &  & \textbf{0.047} & & 0.137 & 0.106 & & 0.166 & 0.163     \\
     &              &      & 0.9    &   & 0.079 &  & 0.048 &  & \textbf{0.045} & & 0.117 & 0.091 & & 0.147 & 0.141     \\
     &              &      & 1      &   & 0.078 &  & 0.068 &  & \textbf{0.045} & & 0.070 & 0.070 & & 0.070 & 0.070     \\
 \hline
  \multicolumn{15}{l}{\small\textit{DGP = Data generating process; N = Training sample size; CSFs = Causal Survival Forests; LSS = Local Survival Stacking;}} \\
  \multicolumn{15}{l}{\small\textit{pcox = penalized cox regression; gbm = gradient boosting machine.}}
\end{tabular}
}
\end{sideways}
\end{table}

\newpage
\begin{table}[ht]
\centering
\caption{Left truncated and right censored (LTRC) simulation study - Data generating process 3 (N=2400) - Difference in survival probabilities - Mean root mean squared error (RMSE) by time for the T-learner, CSFs and surv-iTMLE} 
\vspace{1em} 
\begin{sideways}
\resizebox{1.1\textwidth}{!}{%
\begin{tabular}{ccclcccccccccccc}
\hline
DGP & Truncation & N & Time & & T-learner & & CSFs & & surv-iTMLE &  & \multicolumn{2}{c}{ltrc-R} &  & \multicolumn{2}{c}{ltrc-DR}    \\
    &  level  &  &   &  &  &  &  & & &  &  \multicolumn{1}{c}{LSS/SuperLearner} & \multicolumn{1}{c}{pcox/gbm} &  & \multicolumn{1}{c}{LSS/SuperLearner} & \multicolumn{1}{c}{pcox/gbm}    \\\hline
3    & $\sim 25\%$  & 2400 & 0.2    &   & 0.052 &  & \textbf{0.002} &  & 0.005 & & 0.094 & 0.055 & & 0.109 & 0.081     \\
     &              &      & 0.4    &   & 0.052 &  & \textbf{0.014} &  & \textbf{0.014} & & 0.157 & 0.116 & & 0.152 & 0.159     \\
     &              &      & 0.6    &   & 0.053 &  & 0.038 &  & \textbf{0.032} & & 0.180 & 0.141 & & 0.175 & 0.188     \\
     &              &      & 0.8    &   & 0.052 &  & 0.062 &  & \textbf{0.046} & & 0.169 & 0.149 & & 0.171 & 0.198     \\
     &              &      & 1.0    &   & 0.052 &  & 0.075 &  & \textbf{0.045} & & 0.150 & 0.147 & & 0.158 & 0.198     \\
     &              &      & 1.2    &   & 0.053 &  & 0.074 &  & \textbf{0.039} & & 0.137 & 0.141 & & 0.149 & 0.190     \\
     &              &      & 1.4    &   & 0.053 &  & 0.065 &  & \textbf{0.035} & & 0.125 & 0.129 & & 0.141 & 0.182     \\
     &              &      & 1.6    &   & 0.053 &  & 0.053 &  & \textbf{0.032} & & 0.112 & 0.115 & & 0.133 & 0.169     \\
     &              &      & 1.8    &   & 0.053 &  & 0.041 &  & \textbf{0.030} & & 0.086 & 0.090 & & 0.117 & 0.154     \\
     &              &      & 2.0    &   & 0.053 &  & 0.018 &  & 0.029 & & \textbf{0.000} & \textbf{0.000} & & \textbf{0.000} & 0.002     \\
     &              &      &    &   &   &  &  &  &  & & & & & &     \\
     & $\sim 50\%$  & 2400 & 0.2    &   & 0.056 &  & \textbf{0.002} &  & 0.012 & & 0.148 & 0.056 & & 0.151 & 0.103     \\
     &              &      & 0.4    &   & 0.056 &  & \textbf{0.008} &  & 0.011 & & 0.232 & 0.134 & & 0.216 & 0.191     \\
     &              &      & 0.6    &   & 0.057 &  & \textbf{0.026} &  & 0.029 & & 0.286 & 0.171 & & 0.270 & 0.247     \\
     &              &      & 0.8    &   & 0.056 &  & \textbf{0.051} &  & \textbf{0.051} & & 0.287 & 0.187 & & 0.274 & 0.268     \\
     &              &      & 1.0    &   & \textbf{0.056} &  & 0.076 &  & 0.060 & & 0.247 & 0.179 & & 0.242 & 0.260     \\
     &              &      & 1.2    &   & 0.056 &  & 0.094 &  & \textbf{0.053} & & 0.191 & 0.165 & & 0.193 & 0.245     \\
     &              &      & 1.4    &   & 0.056 &  & 0.097 &  & \textbf{0.045} & & 0.143 & 0.142 & & 0.154 & 0.223     \\
     &              &      & 1.6    &   & 0.056 &  & 0.087 &  & \textbf{0.037} & & 0.107 & 0.107 & & 0.131 & 0.199     \\
     &              &      & 1.8    &   & 0.056 &  & 0.069 &  & \textbf{0.034} & & 0.071 & 0.072 & & 0.105 & 0.170     \\
     &              &      & 2.0    &   & 0.056 &  & 0.021 &  & 0.033 & & \textbf{0.000} & \textbf{0.000} & & \textbf{0.000} & 0.001     \\
 \hline
  \multicolumn{15}{l}{\small\textit{DGP = Data generating process; N = Training sample size; CSFs = Causal Survival Forests; LSS = Local Survival Stacking;}} \\
  \multicolumn{15}{l}{\small\textit{pcox = penalized cox regression; gbm = gradient boosting machine.}}
\end{tabular}
}
\end{sideways}
\end{table}

\newpage
\begin{figure}[!htb]
\begin{center}
\includegraphics[scale=0.45]{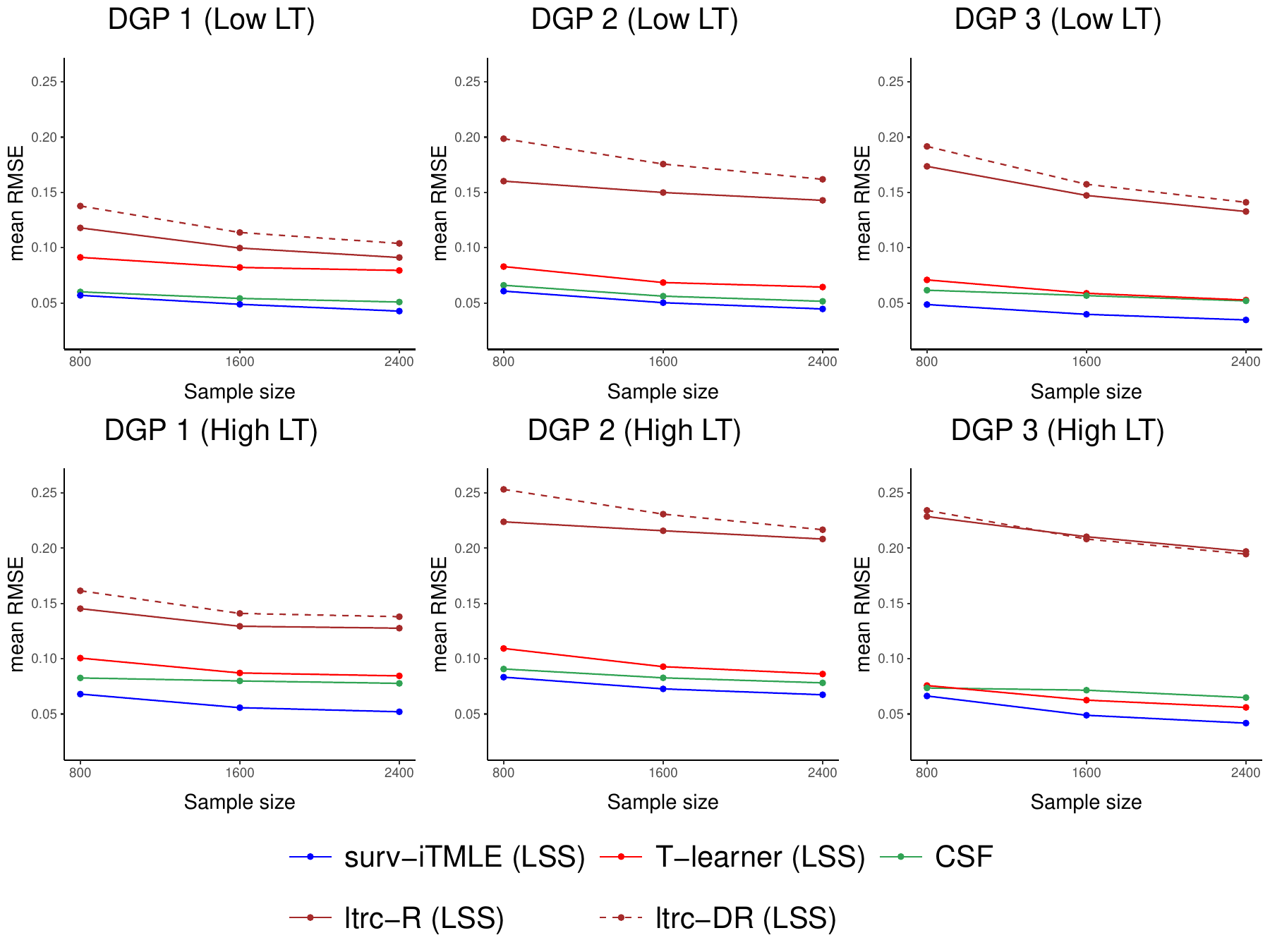} 
\end{center}
\caption{Overall mean root mean squared error (RMSE) by sample size for surv-iTMLE, CSFs, the T-learner and the ltrc R and DR learners when estimating the difference in conditional survival probabilities between two treatments. }
\label{Sim_plot1}
\end{figure}

\newpage
\begin{figure}[!htb]
\begin{center}
\includegraphics[scale=0.45]{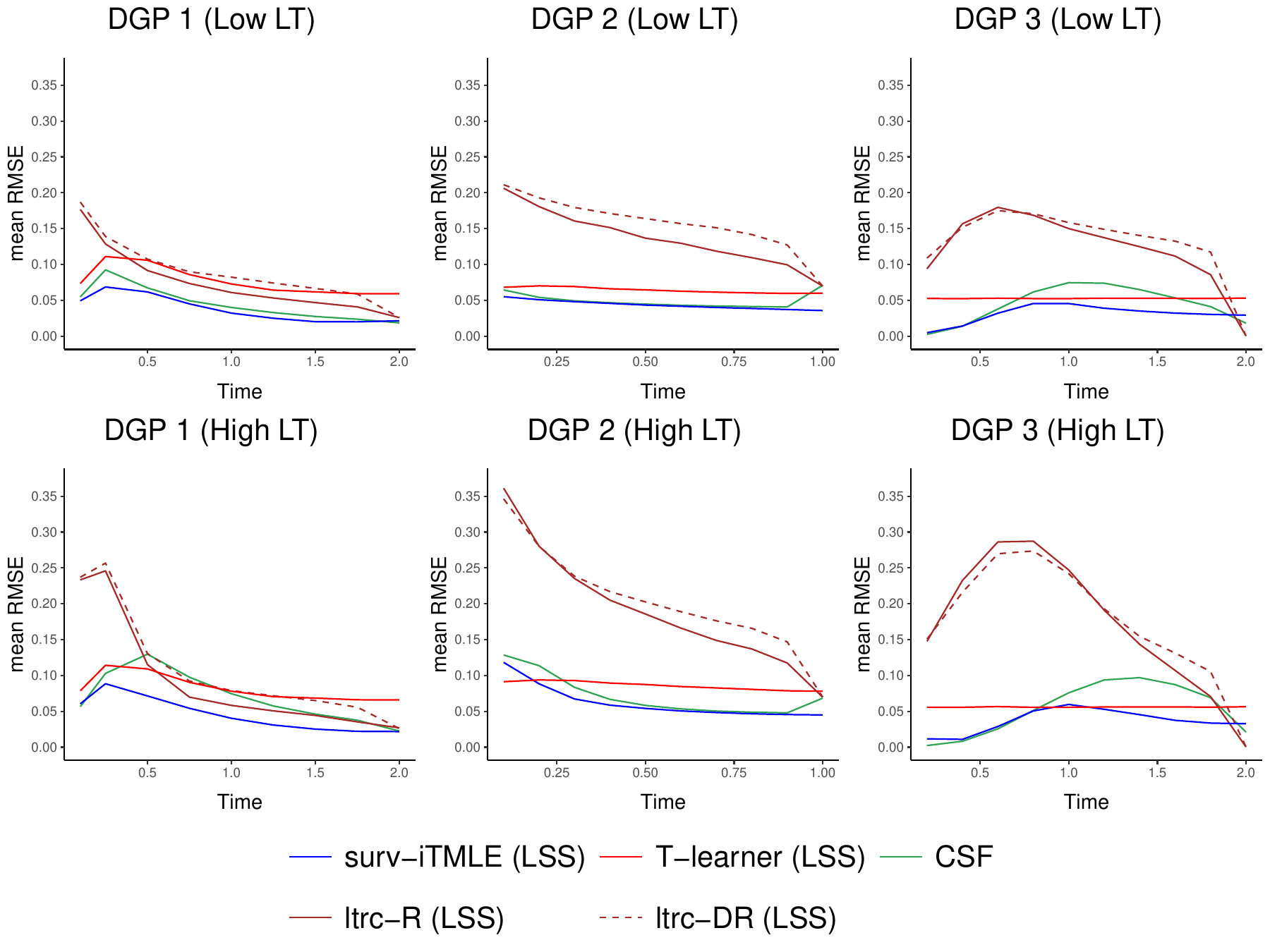} 
\end{center}
\caption{Mean root mean squared error (RMSE) by time point for surv-iTMLE, CSFs, the T-learner and the ltrc R and DR learners when estimating the difference in conditional survival probabilities between two treatments using training data with sample size N=2400. }
\label{Sim_plot2_ltrc}
\end{figure}

\newpage
\begin{figure}[!htb]
\begin{center}
\includegraphics[scale=0.45]{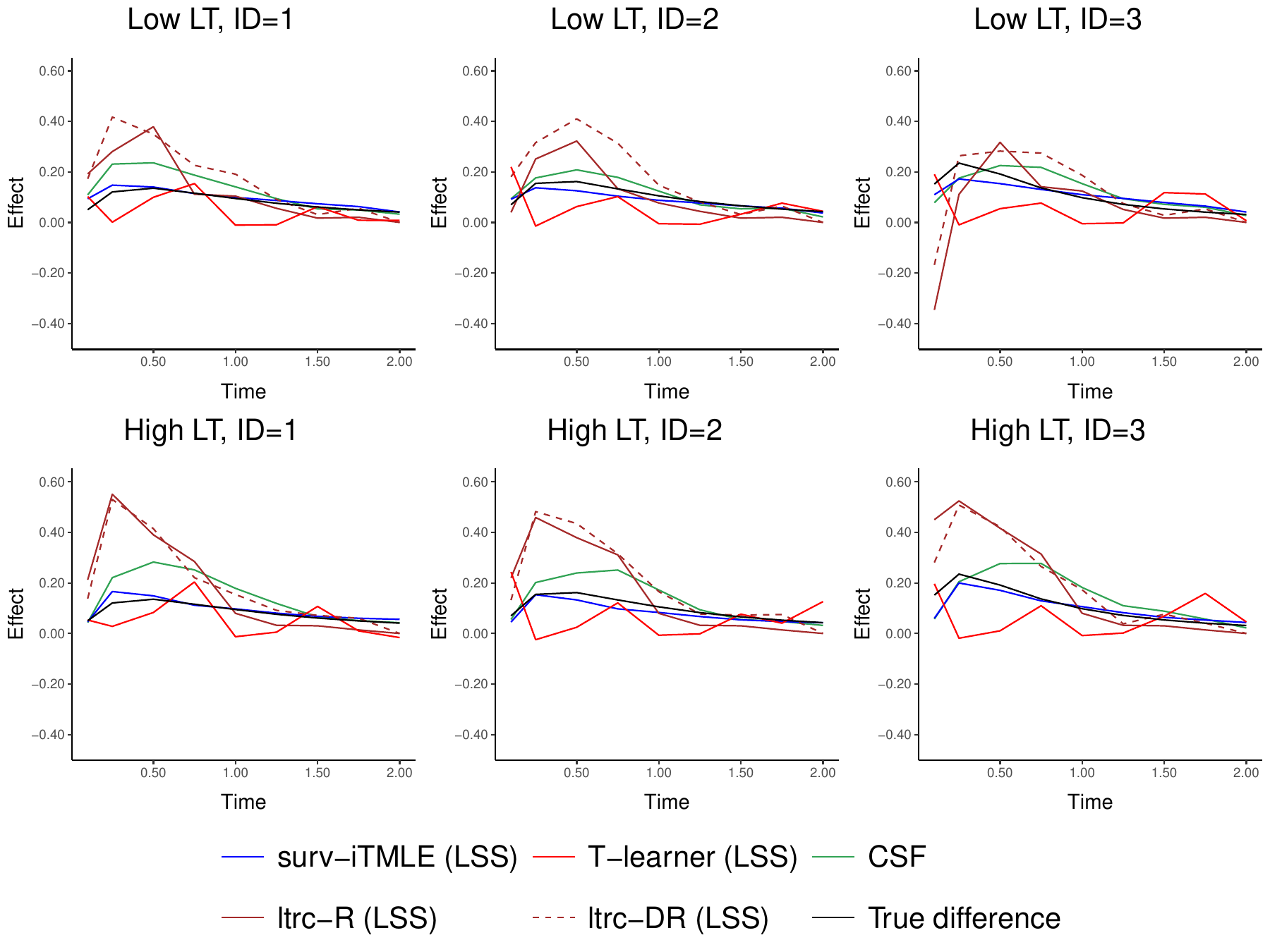} 
\end{center}
\caption{Grid of individual estimates for the difference in conditional survival probabilities between two treatments plotted over time for \textbf{data generating process 1}, with estimates obtained by surv-iTMLE, CSFs, the T-learner and the ltrc R and DR learners using training data of samples size N=2400.}
\label{Sim_plot3_DGP1}
\end{figure}

\newpage
\begin{figure}[!htb]
\begin{center}
\includegraphics[scale=0.45]{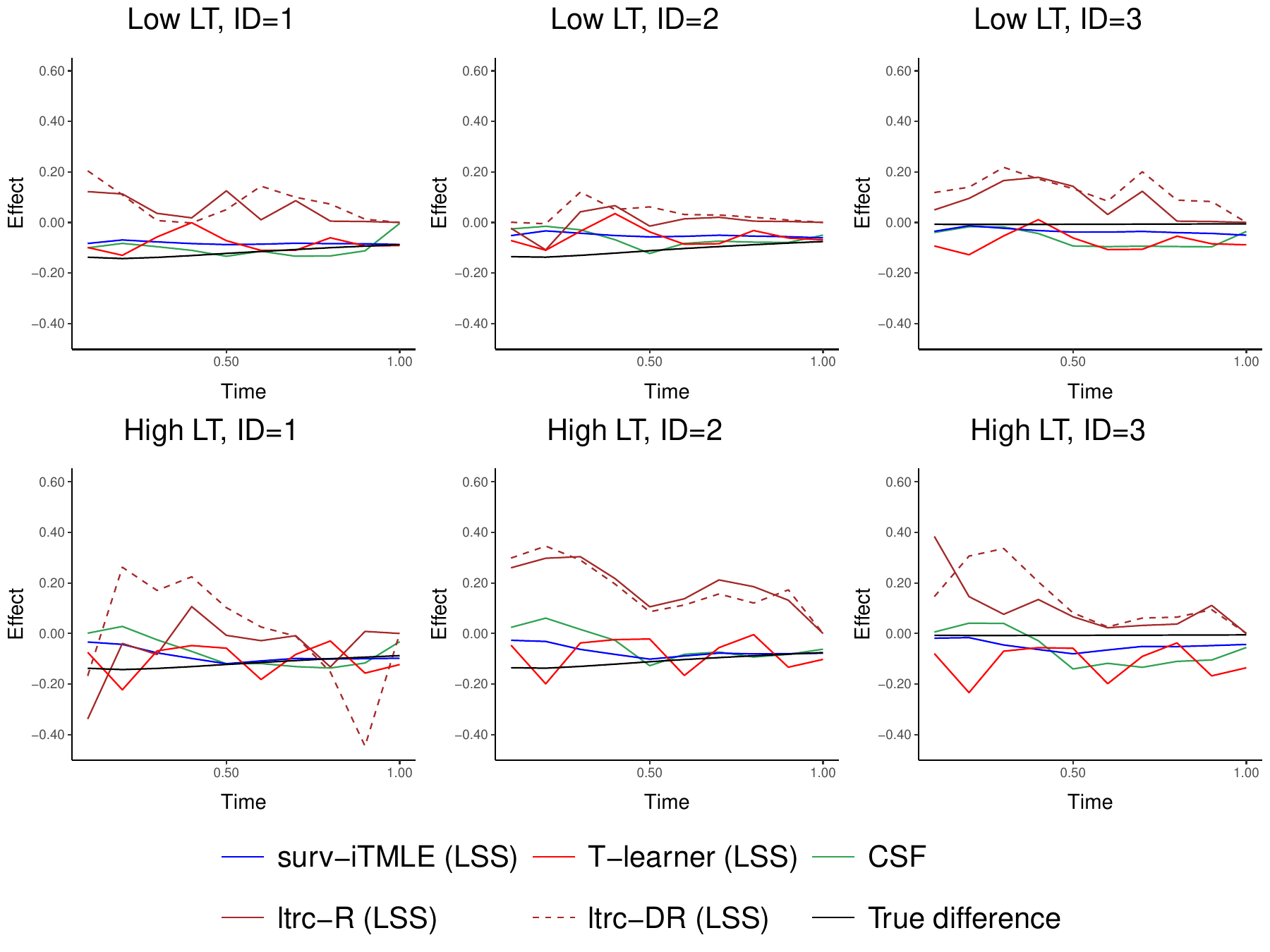} 
\end{center}
\caption{Grid of individual estimates for the difference in conditional survival probabilities between two treatments plotted over time for \textbf{data generating process 2}, with estimates obtained by surv-iTMLE, CSFs, the T-learner and the ltrc R and DR learners using training data of samples size N=2400.}
\label{Sim_plot3_DGP2}
\end{figure}

\newpage
\begin{figure}[!htb]
\begin{center}
\includegraphics[scale=0.45]{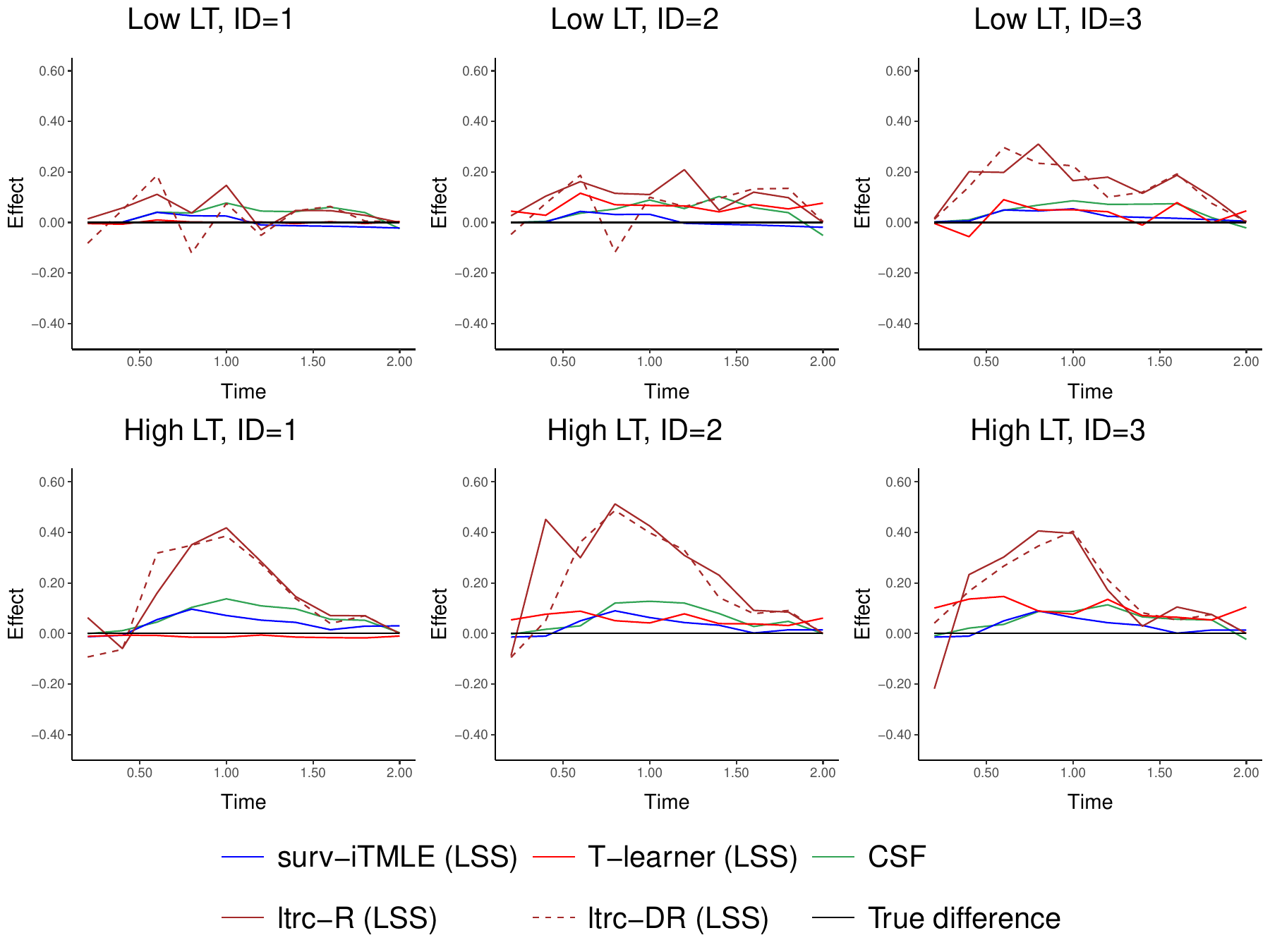} 
\end{center}
\caption{Grid of individual estimates for the difference in conditional survival probabilities between two treatments plotted over time for \textbf{data generating process 3}, with estimates obtained by surv-iTMLE, CSFs, the T-learner and the ltrc R and DR learners using training data of samples size N=2400.}
\label{Sim_plot3_DGP3}
\end{figure}

\newpage
\section*{S8. Right censored time-to-event data - Identification, Algorithm and Simulations}

\subsection*{S8.1. Identification}

When event times are right censored but not left truncated, a reduced set of identification assumptions are required. In this setting, five assumptions are required, positivity assumptions (A1)-(A2), a consistency assumption (A3) and conditional exchangeabilty assumptions (A4)-(A5): 
\begin{itemize}[leftmargin=4in]
    \item [(A1)]$~~P(A=a|Z)>0$, $P$-almost surely 
    \item [(A2)]$~~P(\tau \leq C| Z)>0$, $P_0$-almost surely
    \item [(A3)]~~When $A=a$, $T = T^{a}$.
    \item [(A4)]~~$T^aI(T^a\leq \tau) \indep A |Z$ 
    \item [(A5)]$~~T^aI(T^a\leq \tau) \indep CI(C\leq \tau) |A=a,Z$
\end{itemize}

Under (A1)-(A5), and using the product integral notation of \cite{RN8} we write our estimand in terms of observable data functions as:
\begin{align*}
    \theta(t|x) = \E\left[\left.\Prodi_{(0,t]}\{1-\Lambda(du|A=1,Z)\} - \Prodi_{(0,t]}\{1-\Lambda(du|A=0,Z)\}\right|X=x\right], 
\end{align*}
\noindent where $\Lambda(t|a,z)=\int_{(0,t]}\frac{F(du|a,z)}{R(u-|a,z)}$, for $a\in\{0,1\}$, $R(u-|a,z)=P(u\leq \tilde{T}|A=a,Z=z)$, and $F(u|a,z)=P(\tilde{T} \leq u,\Delta=1|A=a,Z=z)$. The identification proof in this setting is provided by \cite{RN2} in Section F of their supplementary material.

\subsection*{S8.2. surv-iTMLE algorithm}

As the RC setting is a special case of the LTRC setting, where $Q=0$ for each individual, surv-iTMLE can still be used in this setting. When this is the case, Algorithms 1 and 2 still follow the same structure, but no longer require truncation weights (i.e., $H(t|A,Z)=1$ for all $t\in (0,\tau]$ for all individuals). We provide code for the implementation of surv-iTMLE, in both the RC and LTRC settings at \href{https://github.com/Matt-Pryce/surv-iTMLE}{https://github.com/Matt-Pryce/surv-iTMLE}.

\subsection*{S8.3. Simulation study}

\subsubsection*{S8.3.1. Design}
To demonstrate surv-iTMLE's finite sample performance in the RC setting, we run a simulation study, comparing it to CSFs and the T-learner across the same three DGPs outlined in Section 6 of the main text, each with no left truncation introduced to the event times. The simulation study design is identical to that found in Section 6 of the main text, with a few key exceptions. Firstly, when fitting surv-iTMLE or the T-learner, we fit the time-to-event nuisance functions (i.e., the event and censoring models) using two different estimation techniques. The first being the same local survival stacking approach seen in the LTRC simulation study, and the second being an alternative survival ensemble, \textit{survSuperLearner} \citep{RN2}. By doing so, we can explore how choice of estimator for the nuisance functions impact model performance. Additionally, we note that the CSFs were designed specifically for this setting, and our first DGP comes from their paper. 

\newpage
\subsubsection*{S8.3.2. Findings}

Across each of the DGPs, both surv-iTMLE and CSFs consistently outperformed the T-learner, with each of these estimators providing comparable results. This can be seen for the difference in conditional survival probabilities in Figure 10. The performance between CSFs and surv-iTMLE fit using local survival stacking for its time-to-event nuisance functions can also be seen to be similar over time (See Figure 11). Meanwhile, surv-iTMLE fit using \textit{survSuperLearner} for these same nuisance functions performed best in DGP1, where the event/censoring times were generated using AFT/Cox models respectively, but struggled in DGP2 and DGP 3, where the approaches used to generate the event and censoring times were reversed. As a result, it seems that surv-iTMLE fit using local survival stacking for its time-to-event nuisance functions provides more consistent results, likely as this approach offers more flexibility when estimating the time-to-event curves. Finally, we note that the same smoothness properties discussed in the LTRC simulation study were observed, with surv-iTMLE present smoother treatment effect curves which closely track the truth (See Figure 8 in Section S8.3.3). 

\begin{figure}[h]
\begin{center}
\includegraphics[scale=0.45]{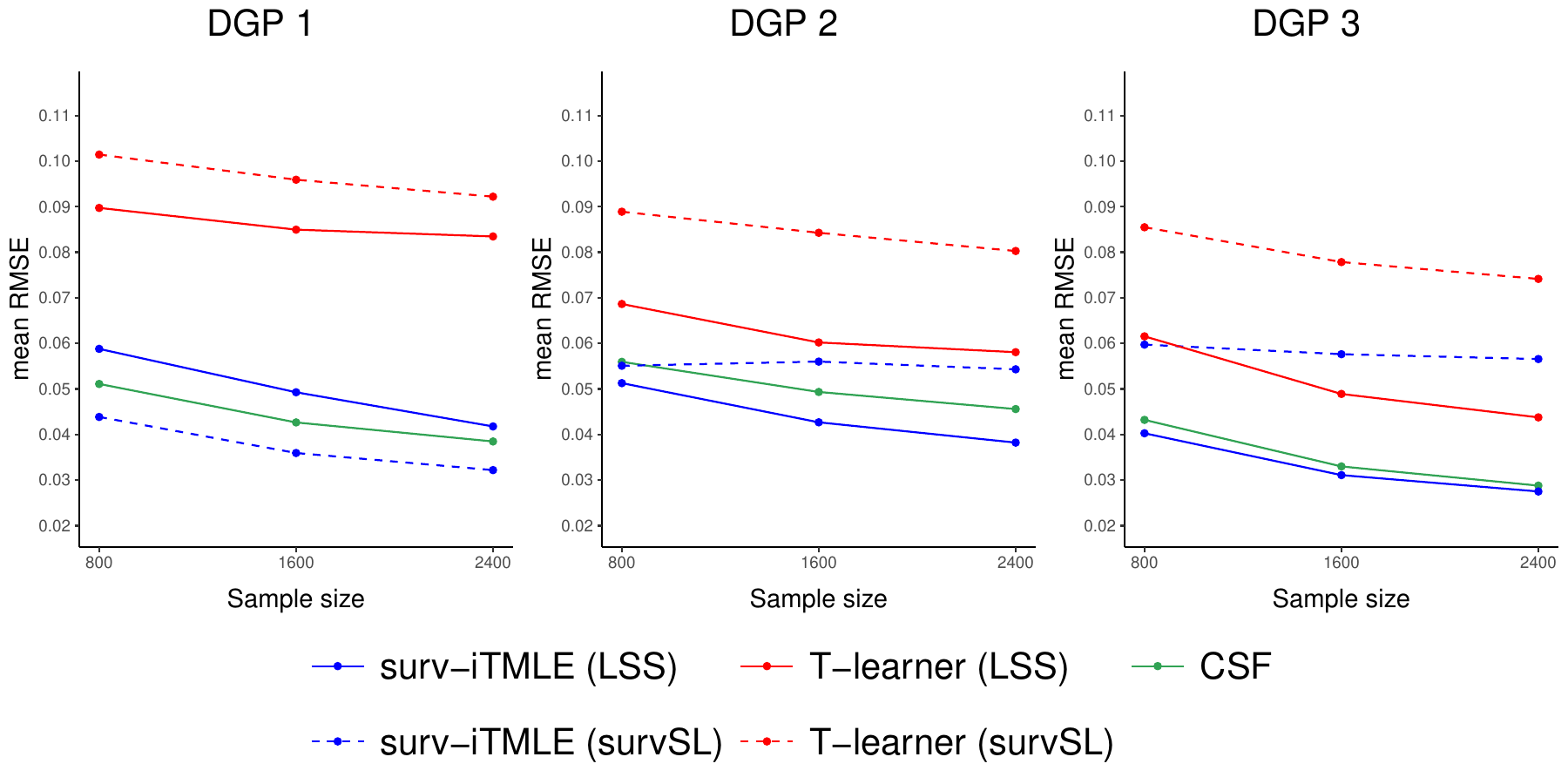} 
\end{center}
\caption{Overall mean root mean squared error (RMSE) by sample size for surv-iTMLE, CSFs and the T-learner when estimating the difference in conditional survival probabilities between two treatments.}
\label{RC_Sim_plot1}
\end{figure}

\begin{figure}[H]
\begin{center}
\includegraphics[scale=0.45]{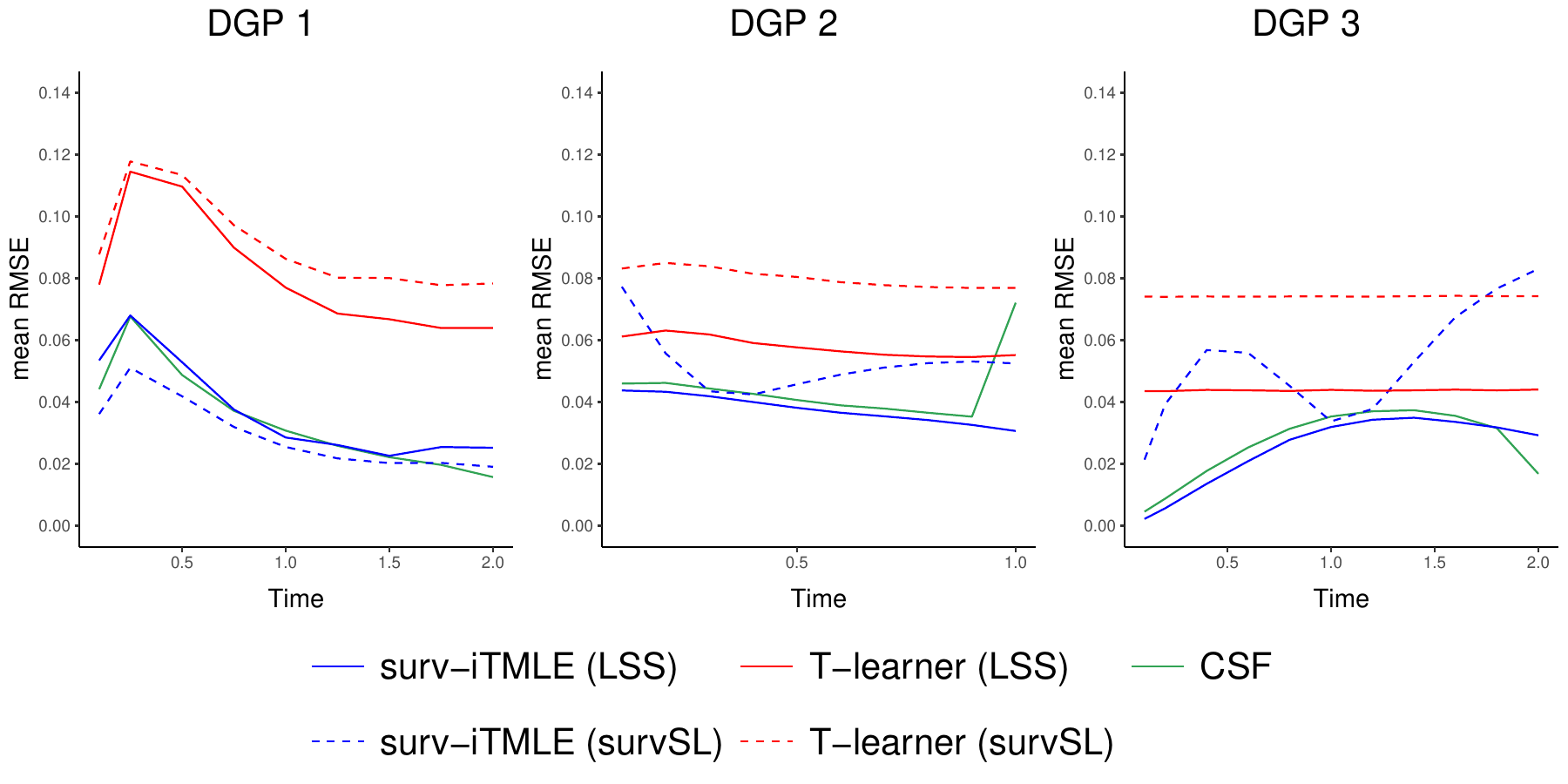} 
\end{center}
\caption{Mean root mean squared error (RMSE) by time point for surv-iTMLE, CSFs and the T-learner when estimating the difference in conditional survival probabilities between two treatments using training data with sample size N=2400.}
\label{RC_Sim_plot2}
\end{figure}

\subsubsection*{S8.3.3. Additional results}

\begin{figure}[H]
\begin{center}
\includegraphics[scale=0.45]{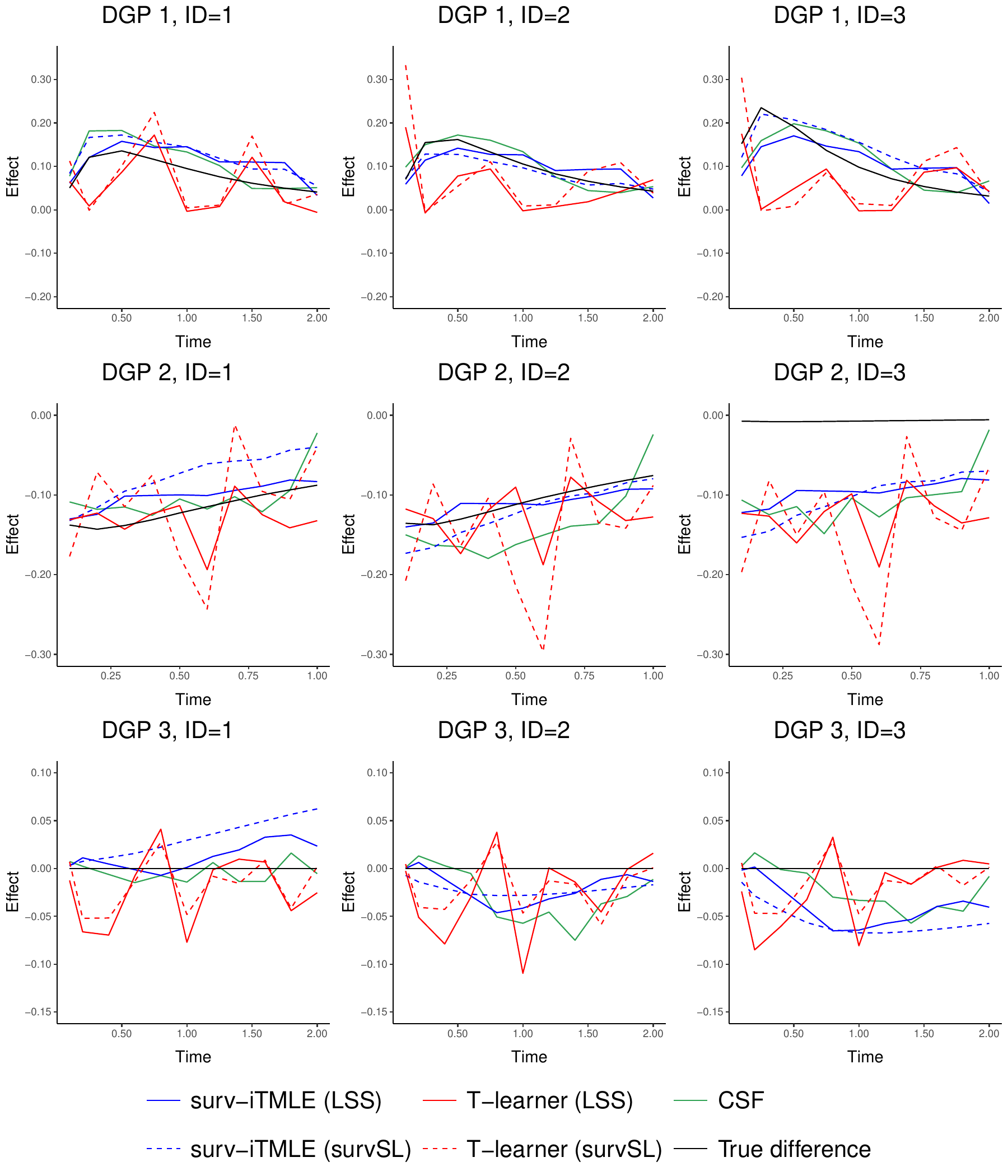} 
\end{center}
\caption{Grid of individual estimates for the difference in conditional survival probabilities between two treatments plotted over time with estimates obtained by surv-iTMLE, CSFs and the T-learner using training data of samples size N=2400.}
\label{RC_Sim_plot3}
\end{figure}

\newpage
\begin{table}[H]
\begin{center}
\caption{Right censored (RC) simulation study - Difference in survival probabilities - Mean root mean squared error (RMSE) for each learner by sample size} 
\begin{tabular}{cccccccccc}
\hline
DGP & N &  & \multicolumn{2}{c}{T-learner} & & CSFs & & \multicolumn{2}{c}{surv-iTMLE}           \\
 & &  & \multicolumn{1}{c}{LSS} & \multicolumn{1}{c}{survSL} & & & & \multicolumn{1}{c}{LSS} & \multicolumn{1}{c}{survSL}           \\
    \hline
1    & 800    &   & 0.090 & 0.101 &  & 0.051 &  & 0.059 & \textbf{0.044}   \\
     & 1600   &   & 0.085 & 0.096 &  & 0.043 &  & 0.049 & \textbf{0.036}    \\
     & 2400   &   & 0.083 & 0.092 &  & 0.038 &  & 0.042 & \textbf{0.032}    \\
     &        &   &  &  &  &  &  & &     \\   
2    & 800    &   & 0.069 & 0.089 &  & 0.056 &  & \textbf{0.051} & 0.055    \\
     & 1600   &   & 0.060 & 0.084 &  & 0.049 &  & \textbf{0.043} & 0.056    \\
     & 2400   &   & 0.058 & 0.080 &  & 0.046 &  & \textbf{0.038} & 0.054    \\
     &        &   &  &  &  &  &  & &     \\
3    & 800    &   & 0.062 & 0.085 &  & 0.043 &  & \textbf{0.040} & 0.060    \\
     & 1600   &   & 0.049 & 0.078 &  & 0.033 &  & \textbf{0.031} & 0.058    \\
     & 2400   &   & 0.044 & 0.074 &  & 0.029 &  & \textbf{0.028} & 0.057    \\
 \hline
  \multicolumn{10}{l}{\small\textit{DGP = Data generating process; N = Training sample size;}} \\
    \multicolumn{10}{l}{\small\textit{CSFs = Causal Survival Forests; LSS = Local Survival Stacking;}} \\
    \multicolumn{10}{l}{\small\textit{survSL = Survival SuperLearner.}} \\
\end{tabular}
\end{center}
\end{table}

\newpage
\begin{table}[H]
\begin{center}
\caption{Right censored (RC) simulation study - Difference in survival probabilities - Mean root mean squared error (RMSE) by time for each learner} 
\begin{tabular}{cccccccccccc}
\hline
DGP & N & time & & \multicolumn{2}{c}{T-learner} & & CSFs & & \multicolumn{2}{c}{surv-iTMLE}           \\
 & & & & \multicolumn{1}{c}{LSS} & \multicolumn{1}{c}{survSL} & & & & \multicolumn{1}{c}{LSS} & \multicolumn{1}{c}{survSL}           \\
    \hline
1    & 2400 & 0.1    &   & 0.078 & 0.088 &  & 0.044 & & 0.053 & \textbf{0.036}      \\
     &      & 0.25   &   & 0.114 & 0.118 &  & 0.068 & & 0.068 & \textbf{0.051}   \\
     &      & 0.5    &   & 0.110 & 0.113 &  & 0.049 & & 0.053 & \textbf{0.042}  \\
     &      & 0.75   &   & 0.090 & 0.097 &  & 0.037 & & 0.038 & \textbf{0.032}  \\
     &      & 1.0    &   & 0.077 & 0.086 &  & 0.031 & & 0.029 & \textbf{0.025}  \\
     &      & 1.25   &   & 0.069 & 0.080 &  & 0.026 & & 0.026 & \textbf{0.022}  \\
     &      & 1.5    &   & 0.067 & 0.080 &  & 0.022 & & 0.023 & \textbf{0.020}  \\
     &      & 1.75   &   & 0.064 & 0.078 &  & \textbf{0.020} & & 0.025 & \textbf{0.020}  \\
     &      & 2.0    &   & 0.064 & 0.078 &  & \textbf{0.016} & & 0.025 & 0.019  \\
     &      &        &   &  &  &  &  &  &  &   \\
2    & 2400 & 0.1   &   & 0.061 & 0.083 &  & 0.046 &  & \textbf{0.044} & 0.077     \\
     &      & 0.2   &   & 0.063 & 0.085 &  & 0.046 &  & \textbf{0.043} & 0.056  \\
     &      & 0.3   &   & 0.062 & 0.084 &  & 0.044 &  & \textbf{0.042} & 0.043  \\
     &      & 0.4   &   & 0.059 & 0.081 &  & 0.043 &  & \textbf{0.040} & 0.042  \\
     &      & 0.6   &   & 0.058 & 0.080 &  & 0.041 &  & \textbf{0.038} & 0.046  \\
     &      & 0.6   &   & 0.056 & 0.079 &  & 0.039 &  & \textbf{0.037} & 0.049  \\
     &      & 0.7   &   & 0.055 & 0.078 &  & 0.038 &  & \textbf{0.035} & 0.051  \\
     &      & 0.8   &   & 0.055 & 0.077 &  & 0.037 &  & \textbf{0.034} & 0.053  \\
     &      & 0.9   &   & 0.055 & 0.077 &  & 0.035 &  & \textbf{0.033} & 0.053  \\
     &      & 1.0   &   & 0.055 & 0.077 &  & 0.072 &  & \textbf{0.031} & 0.052  \\
     &      &     &   &  &  &  &  &  &  &   \\
3    & 2400 & 0.2   &   & 0.044 & 0.074 &  & 0.005 &  & 0.002 & \textbf{0.001}     \\
     &      & 0.4   &   & 0.044 & 0.074 &  & 0.009 &  & \textbf{0.006} & 0.039  \\
     &      & 0.6   &   & 0.044 & 0.074 &  & 0.018 &  & \textbf{0.014} & 0.057  \\
     &      & 0.8   &   & 0.044 & 0.074 &  & 0.025 &  & \textbf{0.021} & 0.056  \\
     &      & 1.0   &   & 0.044 & 0.074 &  & 0.031 &  & \textbf{0.028} & 0.045  \\
     &      & 1.2   &   & 0.044 & 0.074 &  & 0.035 &  & \textbf{0.032} & 0.034  \\
     &      & 1.4   &   & 0.044 & 0.074 &  & 0.037 &  & \textbf{0.034} & 0.038  \\
     &      & 1.6   &   & 0.044 & 0.074 &  & 0.037 &  & \textbf{0.035} & 0.053  \\
     &      & 1.8   &   & 0.044 & 0.074 &  & 0.036 &  & \textbf{0.034} & 0.067  \\
     &      & 2.0   &   & 0.044 & 0.074 &  & \textbf{0.032} &  & \textbf{0.032} & 0.077  \\
 \hline
  \multicolumn{11}{l}{\small\textit{DGP = Data generating process; N = Training sample size;}} \\
    \multicolumn{11}{l}{\small\textit{CSFs = Causal Survival Forests; LSS = Local Survival Stacking;}} \\
    \multicolumn{11}{l}{\small\textit{survSL = Survival SuperLearner.}} \\
\end{tabular}
\end{center}
\end{table}

\newpage
\section*{S9. NSCLC analysis - Additional model fitting information}

\subsection*{S9.1. Covariate adjustment sets}

When using the CGDB, a wide array of information is available for each patient, including information on their demographic factors, their health, their disease severity, and results from a range of lab/genomic tests. In this analysis, covariates which had $<30\%$ missingness were included in the adjustment sets, and genomic information was limited to biomarkers which had $>50$ positive mutations identified. A complete list of the covariate information which was assess for these criteria is provided in Table 9, and a complete list of covariate information that met these criteria is provided in Table 10. 

\begin{table}[ht] \label{cov_list}
\begin{center}
\caption{Covariates which were assessed for use in the clinico-genomic database (CGDB)} 
\begin{tabular}{|ll|}
\hline
Group           & Covariates \\ \hline
Demographic     & Age at time of treatment, Race, Gender, Smoking history  \\
                &            \\
Health/Disease  & BMI, Body weight, Body height, Heart rate, Systolic blood pressure, \\
                & Diastolic blood pressure, ECOG at time of treatment, \\
                & Time from diagnosis to treatment, COPD prior to diagnosis, Diabetes prior to diagnosis \\

                &    \\
Lab             & Hemoglobin, Urea nitrogen, Alkaline Phosphatase, Alanine Aminotransferase, \\
                & Aspartate Aminotransferase, Calcium, Creatinine, Total protein, Bilirubin, Albumin, \\
                & Hematocrit, Glucose, Platelet count, Lymphocyte count, \\
                & Monocyte count, Neutrophil count   \\
                &            \\
Genomic         &  EGFR mutation, BRAF mutation, KRAS mutation, HER2/ERBB2 mutation, \\
               & MET mutation, ROS1 fusion/rearrangement, ALK fusion/rearrangement \\
               & PDL1 expression level, RET fusion/rearrangement, NTRK1 fusion/rearrangement \\
               & NTRK2 fusion/rearrangement, NTRK3 fusion/rearrangement  \\ 
\hline
\end{tabular}
\end{center}
\end{table}

\newpage
\begin{table}[H] \label{cov_list2}
\begin{center}
\caption{Covariate which had sufficient available data from the clinico-genomic database (CGDB)} 
\begin{tabular}{|ll|}
\hline
Group           & Covariates \\ \hline
Demographic     & Age at time of treatment, Race, Gender, Smoking history  \\
                &            \\
Health/Disease  & Body weight, Body height, ECOG at time of treatment, \\
                & Time from diagnosis to treatment, COPD prior to diagnosis, Diabetes prior to diagnosis \\
                &  \\
Lab             & Hemoglobin, Urea nitrogen, Alkaline Phosphatase, Alanine Aminotransferase, \\
                & Aspartate Aminotransferase, Calcium, Creatinine, Total protein, Bilirubin, Albumin, \\
                & Hematocrit, Glucose, Platelet count, Lymphocyte count, \\
                & Monocyte count, Neutrophil count   \\
                &            \\
Genomic         &  EGFR mutation, KRAS mutation          \\ 
\hline
\end{tabular}
\end{center}
\end{table}

When using these covariates to estimate treatment effect heterogeneity, the sets of covariates which were used within each learner/model can be seen in Table 11, noting that information which came from genomic testing was not used within the truncation/treatment models as these would not have been available to inform these decisions. 

\begin{table}[H] \label{model_covs}
\begin{center}
\caption{Covariate adjustment sets for each learner/model} 
\begin{tabular}{|lll|}
\hline
Learner    & Model  & Covariates \\ \hline
surv-iTMLE & Outcome          & All \\
           & Censoring        & All \\
           & Truncation       & All except EGFR mutation and KRAS mutation  \\
           & Treatment        & All except EGFR mutation and KRAS mutation  \\
           & Pseudo-outcome   & (1) All and time, including interactions between each predictor and time\\
           &                  & (2) EGFR mutation, time and interactions between the two \\
           &                  & (3) Age at time of treatment, time and interactions between the two \\
           & &            \\
CSFs       & . & All \\
           & &  \\
T-learner  & Outcome & All          \\ \hline
\end{tabular}
\end{center}
\end{table}

\subsection*{S9.2. Imputation procedure}

In this study, covariates with less than 30\% missingness were used in the adjustment sets, as done by \cite{RN25}. The covariates which had missing values were: Race, ECOG, body mass index (BMI), Body weight, Hemoglobin, Urea nitrogen, Alkaline Phosphatase, Alanine Aminotransferase, Aspartate Aminotransferase, Calcium, Creatinine, Total protein, Bilirubin, Albumin, Hematocrit, Glucose, Platelet count, Lymphocyte count, Monocyte count, Neutrophil count. 

For race, missing values were coded at "Unknown", while for EGFR and KRAS mutations, missing values were categorized along with Negative test results. For the remaining covariates, multiple imputation by chained equations (MICE) was used to impute the missing values. We used the default options within the \textit{MICE} package in R, using predictive mean modelling to impute continuos variables, and logistic regressions to impute binary variables. All available covariates were used as predictors within these prediction models.

\subsection*{S9.3. ML estimation techniques used to implement surv-iTMLE, CSFs, the T-learner and the ltrc learners}

When estimating surv-iTMLE and the T-learner in our NSCLC analysis, local survival stacking approaches were taken to estimate the outcome functions, with the \textit{SuperLearner} used to estimate the binary outcome problems within these. This same approach was taken for censoring and truncation functions found in surv-iTMLE. The libraries defined for these binary outcomes problems can be found in Table 12. The propensity score in surv-iTMLE was estimated using the \textit{SuperLearner}, with the same library as used as in the local survival stacking models. For the pseudo-outcome regression within surv-iTMLE, a generalized additive model (GAM) was used, including factor–smooth interactions between $time$ and each variable in $X$, setting the the maximum basis dimension for each term to a maximum of four degrees of freedom ($K=4$), producing smooth treatment effect curves over time. CSF were then fit using the default options in the \textit{grf} package in R, and the ltrc learners fit using local survival stacking for the outcome, censoring and truncation models, while the super learner and cross-validated xgboost was used to estimate the propensity and final minimization respectively. Finally, we note that a 10-fold cross-fitting approach was used for the nuisance function estimation in each surv-iTMLE variation and for the ltrc learners.

\begin{table}[ht]
\caption{Super learner algorithm libraries used for binary outcome models (including those in local survival stacking (LSS) algorithms)}
\begin{center}
\begin{tabular}{ll}
\hline
\textbf{Algorithm}   & \textbf{Tuning parameters}     \\ \hline
Mean (SL.mean)     & .     \\
& \\
Linear model (SL.glm)     & .    \\
& \\
LASSO/Elastic net (SL.glmnet with logit link) & nlambda = (50,100,250)\\ 
 & alpha = (0.5,1)       \\
 & useMin = (False,True) \\
 & \\
Random forest (SL.ranger)     & mtry = (3,5)\\ 
& min.node.size = (10,20)      \\ 
& sample.fraction = (0.2,0.4,0.6) \\ 
& \\
Generalized additive models     & degrees of freedom = (2,3,4,5)   \\ \hline                          
\end{tabular}
\end{center}
\end{table}

\newpage
\section*{S10. NSCLC analysis results with ltrc learners}

\begin{figure}[!htb]
\begin{center}
\includegraphics[scale=0.5]{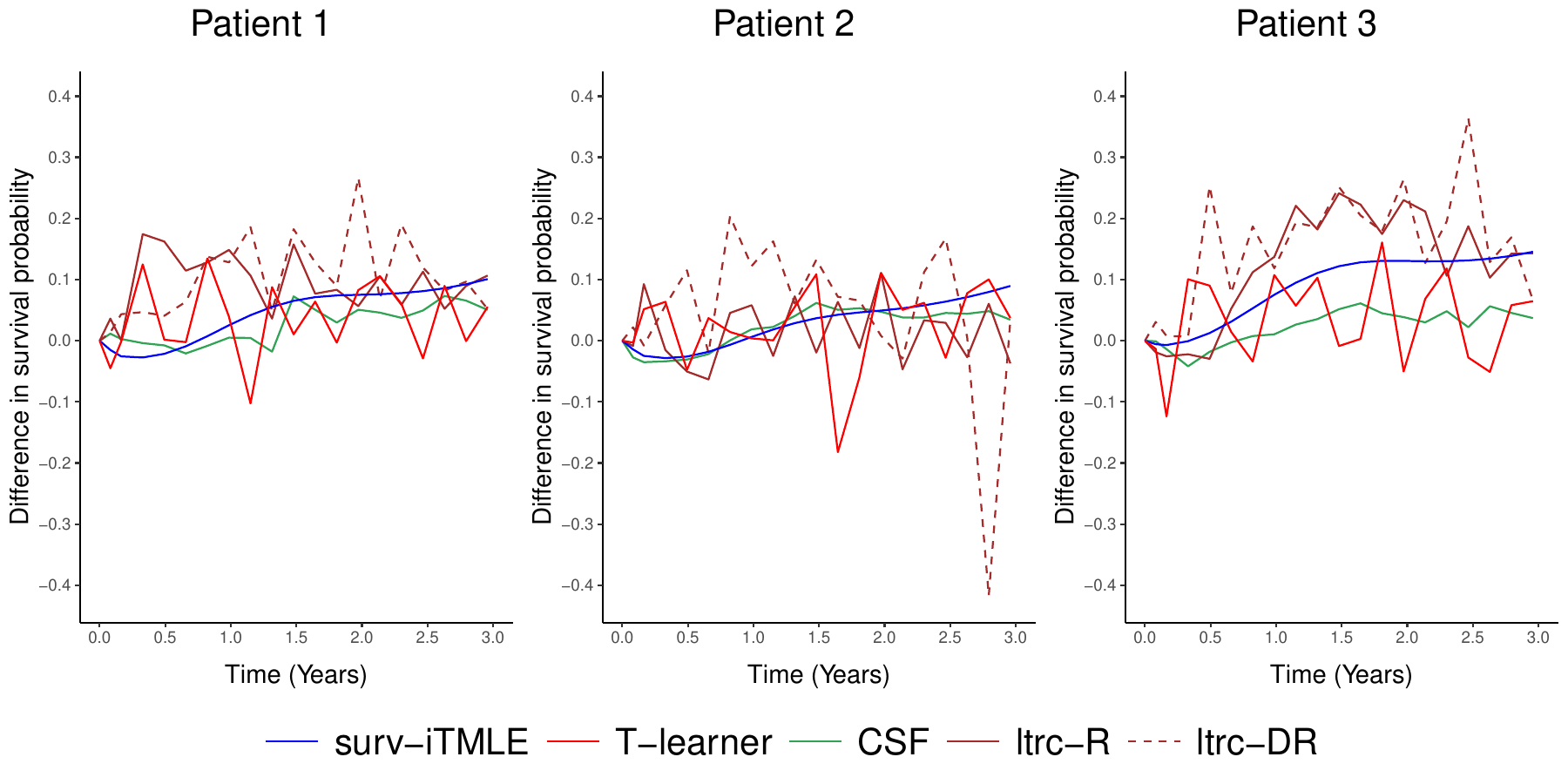} 
\end{center}
\caption{Individual difference in conditional survival probabilities over time for three example NSCLC patients if they were to initiate immunotherapy/chemotherapy after a NSCLC diagnosis, with treatment effects estimated by surv-iTMLE, CSFs, the T-learner and the ltrc DR and R learners. }
\label{NSCLC_plot1_ltrc}
\end{figure}

\end{document}